\documentclass[floatfix,aps,amsmath,prd,onecolumn,notitlepage,eqsecnum,showpacs,nofootinbib]{revtex4-1}
\usepackage{hyperref,url,color}
\usepackage[hyphenbreaks]{breakurl}
\usepackage{color,graphicx,graphics,amsfonts,amsmath,mathrsfs,amssymb,bm,psfrag,natbib,dcolumn,textcase}
\usepackage{enumerate}
\allowdisplaybreaks
\newcommand{\be}{\begin{equation}}
\newcommand{\ee}{\end{equation}}
\newcommand{\bea}{\begin{eqnarray}}
\newcommand{\eea}{\end{eqnarray}}

\newcommand{\bs}{\begin{subequations}}
\newcommand{\es}{\end{subequations}}
\def\no{\nonumber \\}

\begin{document}
\title{Gravitational-wave phasing for low-eccentricity inspiralling compact binaries to 3PN order}
\author{Blake Moore}
\email{mooreb4@mail.montclair.edu}
\author{Marc Favata}
\email{marc.favata@montclair.edu}
\affiliation{Mathematical Sciences Department, Montclair State University, 1 Normal Avenue, Montclair, New Jersey 07043, USA}
\author{K.~G.~Arun}
\email{kgarun@cmi.ac.in}
\affiliation{Chennai Mathematical Institute, Siruseri 603103, India}
\author{Chandra Kant Mishra}
\email{chandra@icts.res.in}
\affiliation{International Centre for Theoretical Sciences, Tata Institute of Fundamental Research, Bangalore 560012, India}
\date{Submitted 9 March 2016; published 24 June 2016}
\begin{abstract}
Although gravitational radiation causes inspiralling compact binaries to circularize, a variety of astrophysical scenarios suggest that binaries might have small but non-negligible orbital eccentricities when they enter the low-frequency bands of ground- and space-based gravitational-wave detectors.
If not accounted for, even a small orbital eccentricity can cause a potentially significant systematic error in the mass parameters of an inspiralling binary [M.~Favata, Phys.~Rev.~Lett.~{\bf 112}, 101101 (2014)]. Gravitational-wave search templates typically rely on the quasicircular approximation, which provides relatively simple expressions for the gravitational-wave phase to 3.5 post-Newtonian (PN) order.
Damour, Gopakumar, Iyer, and others have developed an elegant but complex \emph{quasi-Keplerian formalism} for describing the post-Newtonian corrections to the orbits and waveforms of inspiralling binaries with any eccentricity. Here, we specialize the quasi-Keplerian formalism to binaries with low eccentricity. In this limit, the nonperiodic contribution to the gravitational-wave phasing can be expressed explicitly as simple functions of frequency or time, with little additional complexity beyond the well-known formulas for circular binaries. These eccentric phase corrections are computed to 3PN order and to leading order in the eccentricity for the standard PN approximants. For a variety of systems, these eccentricity corrections cause significant corrections to the number of gravitational-wave cycles that sweep through a detector's frequency band. This is evaluated using several measures, including a modification of the \emph{useful cycles}. By comparing to numerical solutions valid for any eccentricity, we find that our analytic solutions are valid up to $e_0 \lesssim 0.1$ for comparable-mass systems, where $e_0$ is the eccentricity when the source enters the detector band. We also evaluate the role of periodic terms that enter the phasing and discuss how they can be incorporated into some of the PN approximants. While the eccentric extension of the PN approximants is our main objective, this work collects a variety of results that may be of interest to others modeling eccentric relativistic binaries. This includes a consistent eccentricity expansion of the Newtonian-order polarizations and a comparison of quasi-Keplerian results with numerical simulations. In addition to applications in gravitational-wave data analysis, the formulas derived here could be of use in comparing PN theory with numerical relativity or self-force calculations of eccentric binaries. They could also be useful in the construction of phenomenological inspiral-merger-ringdown waveforms that include eccentricity effects.
\end{abstract}
\maketitle
\section{\label{sec:intro}Introduction, Motivation, and Summary}
Shortly following the initial operations of a second generation of gravitational-wave (GW) interferometers  \cite{LIGO-stdrd-ref1,aVirgo-stdref,GEO-HF-stdref2010,KAGRA-stdrd-ref}, LIGO made the first direct detection of gravitational waves from a coalescing compact-object binary \cite{detectionPRL2016}. The detected waves were from a binary black hole merger, and the signal was consistent with black holes moving in circular orbits \cite{detection-PEpaper2016,detection-Astropaper2016ApJL} as predicted by General Relativity \cite{detection-TGRpaper2016}. The standard expectation is that future detections will be from binaries that have very small orbital eccentricities when they enter the LIGO frequency band ($f_{\rm low} \gtrsim 10$ Hz). This is due to the circularizing effect of gravitational radiation.\footnote{Note that test-particle calculations show that strong-gravity effects near the last stable orbit can cause the eccentricity to increase slightly before plunge \cite{cutler-kennefick-poisson,tanaka-etal-PThPh1993}.} However, one must be prepared for violations of standard expectations (as is often the case in science). As discussed in Sec.~\ref{subsec:astrophysics} below, there are some astrophysical scenarios that could produce binaries with eccentricities that are observationally relevant for ground-based detectors. Favata \cite{favata-PRL2014} has also shown that even very small eccentricities ($e_{0}\gtrsim 6\times 10^{-3}$)  can cause detectable systematic parameter biases in binary neutron stars (NSs) if eccentric corrections are not incorporated in waveform templates.

Post-Newtonian waveform models for circular, nonspinning binaries admit simple analytic expressions in the frequency domain, allowing computationally efficient data analysis. However, eccentric waveforms are much more complex, especially at high post-Newtonian (PN) orders. These waveforms are computed via a \emph{quasi-Keplerian} formalism \cite{damourderuelle,damourschafer,schafer-wex-PLA1993,*schafer-wex-PLA1993-erratum,DGI,quasikep3PN,quasikepphasing35PN} that provides a semianalytic description of conservative PN eccentric orbits supplemented by a set of ordinary differential equations (ODEs)  describing the radiative evolution of the orbital elements. For arbitrarily eccentric orbits, these waveforms must be computed by numerical evaluation of the ODEs supplemented with a root-finding procedure to solve the 3PN extension of Kepler's equation.\footnote{Note that, while we use the term ``eccentric'' throughout this paper, we consider only {\it elliptical} binaries in this work ($e<1)$. The formulas here are not applicable to hyperbolic or parabolic binaries ($e\geq 1$). We also note that by ``elliptical'' we refer to orbits that undergo periastron advance (which are not true ellipses in the context of Newtonian theory).} Fully analytic waveforms for eccentric binaries can be derived if one either ignores radiation reaction or other PN effects, or assumes that the eccentricity is small (see Sec.~\ref{subsec:eccwaveformreview} for further discussion). {\bf The primary purpose of this paper is to provide a \emph{simple} extension of the standard circular PN approximants that consistently incorporate the leading-order effects of eccentricity.} (PN \emph{approximants} provide different but related approaches for computing the phase and frequency evolution of a gravitational-wave signal.) This simple extension is possible because one can analytically solve for the evolution of the eccentricity as a function of frequency [$e(f)$] to 3PN order if one assumes that the eccentricity is small. 

The most important results of this paper are explicit formulas for the post-Newtonian approximants presented in Sec.~\ref{sec:approximants}.  
In the waveform phasing, these formulas are accurate to 3PN order [i.e., including relative corrections of $O(v^6)$ where $v$ is the relative orbital velocity] and to $O(e_0^2)$ (where $e_0$ is the eccentricity at a reference frequency $f_0$). For example, the orbital phase of the binary can be expanded as
\begin{multline}
\phi = \phi_c - \frac{1}{32 \eta v^5} \left\{ 1 + \left( \frac{3715}{1008} + \frac{55}{12}\eta \right) v^2 + \cdots + O(v^7) \right. \\
\left. - \frac{785}{272} e_0^2 \left(\frac{v_0}{v}\right)^{19/3} \left[ 1 + \left(\frac{6955261}{2215584} + \frac{436441 }{79128}\eta \right) v^2 + \left( \frac{2833}{1008} - \frac{197}{36}\eta  \right) v_0^2  + \cdots O(v^6) \right]  \right\} \,,
\end{multline}
where $v=(\pi M f)^{1/3}$, $v_0=(\pi M f_0)^{1/3}$, $f$ is the observed GW frequency, $M=m_1+m_2$ is the total binary mass, $\eta=m_1 m_2/M^2$ is the reduced mass ratio, and $\phi_c$ is the phase at coalescence. The above formula (and the other related PN approximants) are known to 3.5PN order [$O(v^7)$] in the circular terms (first line above). The low-eccentricity corrections (our main results) are listed schematically on the second line. The leading-order (Newtonian) term was computed in Ref.~\cite{krolak} and extended to 2PN order in Refs.~\cite{favata-phd,favata-PRL2014}. Here, we extend those derivations to 3PN order and to all the standard PN approximants (TaylorT1, TaylorT2, TaylorT3, TaylorT4, and TaylorF2). Readers wishing to get immediately to the main results can skip to Sec.~\ref{sec:approximants}. Of particular interest is the TaylorF2 approximant [Eq.~\eqref{eq:PsiFTecc}], which provides a fully analytic representation of the Fourier transform of the GW signal in the stationary phase approximation (SPA). Because there is no need to numerically solve ODEs or compute a Fourier transform, this formula is particularly useful for computationally intensive data analysis applications.  In Sec.~\ref{sec:numerical_comp}, we compare our leading-order eccentricity phasing with a numerical calculation of the phase evolution that does not assume small eccentricity. We estimate that our analytic formulas are valid for $e_0 \lesssim 0.1$ for comparable-mass binaries and $e_0 \lesssim 0.01$ for extreme-mass-ratio binaries. (The precise limits depend on the system masses. Note that for extreme-mass-ratio systems, the PN series converges slowly.)  
\begin{table}
	\caption{\label{tab:Ncyc-sum}Post-Newtonian contributions to the number of gravitational-wave cycles $\Delta N_{\rm cyc}$ for a NS/NS binary in the LIGO frequency band. This is computed from the orbital phase via $\Delta N_{\rm cyc} = [\phi(f_2)-\phi(f_1)]/\pi$. The first column lists the post-Newtonian order and the type of term: ``circ'' refers to the quasicircular contributions at that PN order, while ``ecc'' refers to the leading-order eccentric terms that are computed here. The binary enters the LIGO band at $f_1=10$ Hz. We truncate the signal at $f_2=1000$ Hz. The NS masses are $m_1=m_2=1.4\, M_{\odot}$. We assume an initial eccentricity of $e_0=0.1$ at a reference frequency $f_0=f_1=10$ Hz. These values can be scaled to other eccentricities by multiplying by $(e_0^{\rm new}/0.1)^2$. All numbers are rounded to at least three significant digits.}
	\begin{tabular}{|l|r|}
		\hline 
		\multicolumn{1}{|c|}{PN order} & \multicolumn{1}{|c|}{$\Delta N_{\rm cyc}$} \\
		\hline
		0PN(circ) & $16031$ \\
		0PN(ecc)  &  $-463$  \\
		1PN(circ) & $439$  \\
		1PN(ecc) & $-15.8$  \\
		1.5PN(circ) & $-208$ \\
		1.5PN(ecc) & $1.67$  \\
		2PN(circ) & $9.54$  \\
		2PN(ecc) & $-0.215$ \\
		2.5PN(circ) & $-10.6$  \\
		2.5PN(ecc) & $0.0443$  \\
		3PN(circ) & $2.02$  \\
		3PN(ecc) & $0.00200$  \\
		3.5PN(circ) & $-0.662$  \\
		\hline		
		Total & $15785$ \\
		\hline 
	\end{tabular}
\end{table}

To quantify the relative importance of the different PN correction terms, we compute in Sec.~\ref{sec:ncyc} several variants of the \emph{number of cycles} contributed from each PN term for different binary systems. For example, Table \ref{tab:Ncyc-sum} displays the contribution to the number of GW cycles $\Delta N_{\rm cyc}$ from circular and eccentric PN terms for a binary neutron-star system in the LIGO band (assuming $e_0 =0.1$ at $10$ Hz). Using the crude criterion that contributions $\Delta N_{\rm cyc}\sim O(1)$ are potentially significant, we see that eccentric corrections through 2PN order are significant for this system, while those at 2.5PN and 3PN orders are not.  Other measures, such as the number of ``useful'' cycles or the contribution to the phase of the Fourier transform are discussed in more detail in Sec.~\ref{sec:ncyc}.

Our objective is to provide waveforms that are only marginally more complex than circular ones yet consistently incorporate the effects of eccentricity. It is therefore important to understand the approximations that enter our analysis. Our approximants incorporate only the {\it secular} contribution to the phasing; there are also {\it oscillatory} contributions to the phasing that we do not include. These oscillatory contributions arise from two sources: (i) Even  Newtonian elliptical orbits have an instantaneous orbital frequency $\dot{\phi}$ that varies along an orbit. This is simply the statement that the binary phase angle evolves faster close to periastron and slower near apastron. (ii) In addition to slow secular changes of the orbital elements, the radiation reaction force also induces periodic oscillations in the orbital elements. We show in Sec.~\ref{sec:oscil} that (ii) does not affect the phasing until 5PN order. While (i) affects the phasing at 0PN order, we show in Sec.~\ref{sec:oscil} that it does not contribute more than $\sim O(1)$ cycle to the GW phase. We also briefly discuss how these oscillatory terms can be incorporated into the PN approximants. A more detailed treatment of this effect will be discussed in future work. Aside from these oscillatory terms (which we have shown to be small), all other PN effects are consistently incorporated into our phasing formulas at 3PN order and $O(e_0^2)$ in eccentricity.

Another important approximation arises from our treatment of the GW polarization amplitudes. Our amplitudes are accurate only to leading order in $v/c$; i.e., they are Newtonian-order accurate and contain no relative PN amplitude corrections. Furthermore, they contain no eccentric corrections to the amplitude. In other words, our polarizations have the form $h_{+,\times} = A_{+,\times} v^2 \cos2(\phi-\Phi_{+,\times})$, where $\Phi_{+}=0$, $\Phi_{\times}=\pi/4$, and $A_{+,\times}$ are constants depending on the masses, orbit inclination, and source distance. Our eccentric corrections only enter the waveform in the phasing $\phi(t)$ and the evolution of $v(t)$; our waveforms only oscillate at twice the azimuthal orbital frequency $\omega_{\phi}$. In Appendix \ref{app:polarization}, we provide a detailed derivation of how eccentricity affects the polarization amplitude at Newtonian order (but including the effects of periastron precession). In addition to $O(e^2)$ corrections to the functions $A_{+,\times}$, eccentricity also introduces terms of order $O(e)$ that oscillate at multiples of the radial orbit frequency $\omega_r$ and at frequencies $2\omega_{\phi} \pm j \omega_r$, where $j=1,2,3,\cdots$. In the context of our analysis, these eccentric amplitude corrections will be unimportant because our waveforms are already restricted to small eccentricities ($e_0 \lesssim 0.1$), and small corrections to the waveform amplitude are known to be much less important than small corrections to the phasing.

We emphasize that our objective is to obtain waveforms which are only \emph{marginally} more complex than circular waveforms, while incorporating eccentricity effects to the highest PN order available. We are further motivated to focus on the small-eccentricity limit for the following reasons: (i) Because GW emission tends to circularize binaries, it seems more likely than not that any class of GW sources will have more detectable events that are closer to very small eccentricities than to moderate or large eccentricities; (ii) while other studies have computed waveform corrections to higher orders in eccentricity than we provide, these calculations were not fully consistent in the PN approximation or did not include effects like periastron precession \cite{yunes-arun-berti-will-eccentric-PRD2009,huerta-etal-PRD2014}. Considering our limitation to binaries with small eccentricity, we envision our results as being applicable to the following situations:
\begin{enumerate}[(a)]
\item Studies that examine the systematic parameter bias of ignoring a small residual eccentricity (e.g., Ref.~\cite{favata-PRL2014}) or that wish to otherwise quantify (in a computationally efficient manner) the effect of small orbital eccentricity. While we are primarily concerned with applications to ground-based GW detection (LIGO/Virgo/Kagra/ET), our results could also be applied to studies concerned with sources for the Laser Interferometer Space Antenna (LISA, \cite{elisaweb}) or Pulsar Timing Arrays (PTAs, \cite{iPTA}).
\item To set limits on the orbital eccentricity of future candidate GW signals from nearly circularized binaries. 
\item As a tool to help reduce orbital eccentricity in numerical relativity (NR) simulations.
\item To provide formulas that may be of use in studies that compare NR or gravitational self-force (GSF) \cite{letiec-2014IJMPD,akcay-etal-GSFeccentricPRD2015,hopper-etal-eccentricGSFschw,forseth-etal-eccentricEMRIflux7PN,bini-damour-eccentricGSF-a,bini-damour-eccentricGSF,Akcay-vandeMeent-eccentricEOBpotential} calculations to analogous post-Newtonian results.
\item To extend phenomenological inspiral-merger-ringdown models \cite{ajith-etal-IMR-2007CQG,ajith-etal-BBHtemplates-PRD,ajith-etal-IMR-2011prl,santamaria-etal-phenom,Hannam:2013oca,Husa:2015iqa,Khan:2015jqa} to incorporate eccentricity effects.  
\end{enumerate}

In addition to the primary results summarized above, this paper also has the following secondary objectives and results:
\begin{enumerate}
\item While our focus is the low-eccentricity limit, we provide a clear discussion of how to apply the full quasi-Keplerian formalism to generate waveforms for arbitrary eccentricity. While some of this is discussed elsewhere in the literature, we feel that the transparency of our presentation will be useful to researchers and students who need to model PN corrections to eccentric orbits. By showing how the small-eccentricity limit arises from the full quasi-Keplerian formalism, the approximations implicit in our analysis are made more clear. Our presentation also provides a guide for extending our results to higher order. In particular, Sec.~\ref{sec:polarization} and Appendix \ref{app:polarization} discuss our notation and derive the waveform polarizations at Newtonian order and to $O(e^3)$ in the eccentricity (properly accounting for precessing orbits which display two fundamental orbital frequencies). Via a 3PN accurate inversion of Kepler's equation, we show how the polarizations can be expressed as explicit functions of time in an expansion in eccentricity [see e.g., Eqs.~\eqref{eq:lexpand} and \eqref{eq:hNl}]. A detailed description of the orbital motion and phasing (in the absence of radiation reaction) is discussed in Sec.~\ref{sec:quasikep}. This includes a reduction to the Newtonian case (Sec.~\ref{subsec:quasikep-Newt}), which helps elucidate the meaning of many of the quantities that enter the quasi-Keplerian formalism. Section \ref{sec:evolvebinaries} extends this to the case where radiation reaction is present, including an explicit evaluation of the periodic oscillations that are induced in the orbital elements (Sec.~\ref{sec:periodicsolns}) along with their secular variations (see Appendix \ref{app:ndot-edot} for the general eccentricity case and Sec.~\ref{sec:xiphi-define} for the low-eccentricity limit). These equations are then used to analytically determine how the eccentricity secularly evolves with frequency in the small-eccentricity limit (Sec.~\ref{sec:e_xi}). This result is used to derive (in Sec.~\ref{sec:expliciteqns}) several explicit formulas for the secular orbital phase and time to coalescence as a function of frequency, along with explicit functions of time for the frequency and eccentricity evolution. Most of the PN approximants can be read off of the results in that section (or independently derived from the orbital energy and GW luminosity as in Sec.~\ref{sec:approximants}). We also show in Sec.~\ref{sec:numerical_comp} and Appendix \ref{app:numericaleqns} how the secular piece of the orbital phasing as a function of the orbit frequency can be computed for arbitrary eccentricity via the numerical solution of two coupled ODEs. 
\item Section \ref{subsec:domain} provides a discussion of the region of validity of the quasi-Keplerian formalism. Earlier work \cite{DGI} (before the NR era) provided an argument for setting a particular upper frequency limit (or a minimum orbital separation) for which the formalism should be valid. By comparing with more recent NR and GSF calculations in Ref.~\cite{letiec-PRL2011}, we argue that the bound in Ref.~\cite{DGI} is too conservative. At least in the low-eccentricity limit, the quasi-Keplerian formalism should be valid over nearly the same range in dimensionless frequency ($Mf$) as circular waveforms. 
\item Although they do not enter our final results, we pay particular attention to the role of periodic terms in the waveform phasing. These terms are often neglected in other works. While the radiation-reaction induced oscillations in the orbital elements are shown in Sec.~\ref{sec:periodicsolns} to enter at 5PN order (and are hence negligible for our purposes), we derive in Sec.~\ref{sec:oscil} an explicit frequency-domain expression for the conservative oscillatory piece of the orbital phasing. We numerically evaluate its effect and find that, while small ($\lesssim 0.07$ GW cycles), it is comparable to the 2.5PN and 3PN-order eccentric secular corrections. We also briefly discuss in Sec.~\ref{sec:approximants} how this oscillatory correction can be added to the PN approximants. This will be explored in more detail in a future work. 
\end{enumerate}

In the remainder of this Introduction we first review the astrophysical expectations regarding binary eccentricity (with an emphasis on LIGO sources, Sec.~\ref{subsec:astrophysics}). Of particular note is an updated table of known NS/NS systems and their expected eccentricities when they enter the LIGO or LISA frequency band. In Sec.~\ref{subsec:eccwaveformreview}, we summarize the literature on modeling eccentric waveforms, emphasizing where our work differs from previous results. Throughout, we use units where $G=c=1$ and follow the conventions of Refs.~\cite{DGI,quasikepphasing35PN}.
\subsection{\label{subsec:astrophysics}Astrophysical expectations for eccentric binaries and implications for GW detection}
Since the early work of Peters and Mathews \cite{petersmathews,peters}, it has been understood that GW emission causes the eccentricity of a binary to decay. At Newtonian order, the orbital eccentricity $e_t$ of a binary emitting GWs at frequency $f_{\rm gw}$ is related to its earlier eccentricity $e_i$ (when the binary was wider and emitting GWs at frequency $f_{\rm gw,i}$) via\footnote{This formula follows from Eq.~(5.11) of Ref.~\cite{peters} where the semimajor axis $a$ is related to the ``fundamental'' GW frequency via $\pi f_{\rm gw}=\sqrt{M/a^3}$ for a binary with total mass $M$. Here, $f_{\rm gw}$ refers only to the frequency component of the GW signal that is emitted at twice the orbital frequency. This is the dominant frequency component when the eccentricity is small.}
\be
\label{eq:peters-circularize}
\frac{f_{\rm gw}}{f_{\rm gw, i}} = \left( \frac{e_i}{e_t} \right)^{18/19} \left(\frac{1-e_{t}^2}{1-e_{i}^2} \right)^{3/2} \left( \frac{304+121 e_i^2}{304+121 e_t^2} \right)^{1305/2299} \, .
\ee
To illustrate the circularizing efficiency of GWs, we consider the eccentricity evolution of double neutron-star systems. Table \ref{tab:knownbinaries} lists all such systems currently known. Using Eq.~\eqref{eq:peters-circularize}, we calculate the eccentricities of these binaries when they enter the LIGO and eLISA bands. The largest eccentricity at 10 Hz is $\approx 7 \times 10^{-6}$. This produces an entirely negligible correction to the GW phase in the frequency band of ground-based detectors. (For space-based detectors, the eccentricities are negligibly small in most cases, but potentially within the realm of detectability in others.)
\begin{table}
\caption{\label{tab:knownbinaries}Eccentricity evolution for confirmed or likely double neutron-star binaries. The columns indicate the source's name, orbital period $P_{\rm orb}$ (in days), current fundamental gravitational-wave frequency ($f_{\rm gw,i}=2/P_{\rm orb}$) in mHz, current eccentricity $e_i$, eccentricity at 5 mHz (eLISA band), and eccentricity at 10 Hz (LIGO band). The eccentricities $e_t(f_{\rm gw})$ are computed using Eq.~\eqref{eq:peters-circularize}. Most values for $P_{\rm orb}$ and $e_i$ are taken from the Australia Telescope National Facility (ATNF) pulsar catalog \cite{ATNF-catalogue}. The values for J0453+1559 are from Ref.~\cite{martinez-msthesis,martinez-etal2015}; those for J1807-2500B are from Ref.~\cite{lynch_etalApJL2012}. There is some uncertainty as to whether the systems denoted with an asterisk are double neutron-star binaries \cite{martinez-etal2015}.} 
\begin{tabular}{|l|r|r|r|r|r|}
\hline
\multicolumn{1}{|c|}{Source} & \multicolumn{1}{|c|}{$P_{\rm orb}$ (days)} & \multicolumn{1}{|c|}{$f_{\rm gw,i}$ (mHz)} & \multicolumn{1}{|c|}{$e_i$} & \multicolumn{1}{|c|}{$e_t(5 \text{ mHz})$} & \multicolumn{1}{|c|}{$e_t(10 \text{ Hz})$} \\
\hline
J0737-3039 & $0.10225156248$ & $0.226$ & $0.0877775$ & $0.00339$ & $1.11 \times 10^{-6}$ \\
J1906+0746$^{\ast}$ & $0.16599304683$ & $0.139$ & $0.0853028$ & $0.00198$ & $6.48\times 10^{-7}$ \\
J1756-2251 & $0.31963390143$ & $0.0724$ & $0.180594$ & $0.00220$ & $7.20 \times 10^{-7}$ \\
B1913+16 & $0.322997448911$ & $0.0717$ & $0.6171334$ & $0.0162$ & $5.32 \times 10^{-6}$ \\
B2127+11C & $0.33528204828$ & $0.0690$ & $0.681395$ & $0.0220$ & $7.23 \times 10^{-6}$ \\
B1534+12 & $0.420737298879$ & $0.0550$ & $0.27367752$ & $0.00270$ & $8.85 \times 10^{-7}$\\
J1829+2456 & $1.176027941$ & $0.0197$ & $0.1391412$ &  $4.17 \times 10^{-4}$ & $1.37 \times 10^{-7}$\\
J0453+1559 & $4.07246858$ & $0.00568$ & $0.11251832$ & $8.98 \times 10^{-5}$ & $2.94 \times 10^{-8}$ \\
J1518+4904 & $8.6340050964$ & $0.00268$ & $0.24948451$ & $9.89 \times 10^{-5}$ & $3.24 \times 10^{-8} $ \\
J1807-2500B$^{\ast}$ & $9.9566681588$ & $0.00232$ & $0.747033198$ & $9.32 \times 10^{-4}$ & $3.05 \times 10^{-7}$ \\
J1753-2240 & $13.6375668$ & $0.00170$ & $0.303582$ & $7.87 \times 10^{-5}$ & $2.58 \times 10^{-8}$ \\
J1811-1736 & $18.7791691$ & $0.00123$ & $0.828011$ & $9.29 \times 10^{-4}$ & $3.04 \times 10^{-7}$ \\
J1930-1852 & $45.0600007$ & $0.00514$ & $0.39886340$ & $3.36 \times 10^{-5} $ & $1.10 \times 10^{-8}$  \\
\hline
\end{tabular}
\end{table}

Despite the fact that these projected eccentricities are small, there are still several reasons why the consideration of eccentric gravitational waveforms might be important: (i) while orbits have time to circularize before they enter the frequency band of ground-based detectors, detectors that operate at lower frequencies (such as eLISA \cite{elisaweb} or pulsar timing arrays \cite{nanograv,PPTA,EPTA}) can observe sources that have not yet circularized (wide stellar-mass binaries, supermassive black hole (BH) binaries, extreme-mass-ratio inspirals); (ii) while not observed, astrophysical arguments suggest that there may be compact binaries in the frequency band of ground-based detectors ($\gtrsim 10$ Hz) that have not yet circularized. (These scenarios are discussed below.); (iii) lastly, it is  possible that eccentric binaries are produced by formation channels that have not yet been considered, so it is prudent to use the most general waveforms possible when analyzing GW data.

Dense stellar environments such as galactic nuclei and globular clusters are suspected to create binaries with significant eccentricities. This is partly due to hierarchical three-body interactions where the Kozai-Lidov mechanism can drive oscillations in the eccentricity of the inner binary of the triplet \cite{wen-eccentricity-ApJ2003}.\footnote{Note that Ref.~\cite{naoz-kocsis-loeb-yunes-PNkozai-2012} has shown that PN effects can resonantly enhance eccentricity in hierarchical triples beyond the Newtonian Kozai-Lidov effect.} In globular clusters, around 30\%\mbox{--}50\% \cite{antonini-murray-mikkola-ApJ2014} of coalescing BH binaries driven by the Kozai-Lidov mechanism will have $e \gtrsim 0.1$ when entering the LIGO frequency band at $10$ Hz.\footnote{Similar results were also found in Ref.~\cite{antognini-etalMNRAS2014}. While earlier studies \cite{gultekin-miller-hamilton2004ApJ,oleary-etal-BHmergersGC-ApJ2006} indicated that globular cluster BH binaries will have mostly circularized when they enter the LIGO band, those works relied on orbit-averaged equations of motion (which were shown to be inaccurate in Refs.~\cite{antonini-peretsApJ2012,antonini-murray-mikkola-ApJ2014}).} Recent work in Ref.~\cite{antonini-etalApJ2015} (which does not rely on orbit-averaged equations) indicates that $\sim 20\%$ of merging BHs formed via dynamical interactions in globular clusters will have eccentricities greater than $0.1$ when they enter the LIGO band at $10$ Hz. They estimate that mergers with eccentricities greater than this value will be detected by LIGO at rates between $0.05/\text{yr}$ to $3/\text{yr}$, with a ``realistic'' estimate of $0.4/\text{yr}$. More recent work in Ref.~\cite{rodriguez-chatterjee-rasio2016} indicates that $\sim 1\%$ of binary black holes formed in globular clusters will have eccentricities at $10$ Hz that exceed $10^{-3}$.  Reference \cite{samsing-macleod-ramirez2013} also found that some NS/NS binaries can be dynamically formed in the LIGO band with high eccentricity. 

Galactic nuclei are another potentially significant source of eccentric compact-object mergers. Reference \cite{oleary-kocsis-loeb-MNRAS2009} showed that BH/BH binaries in dense galactic nuclei can be formed in the LIGO band with high eccentricities ($90\%$ with $e>0.9$). 
Reference \cite{antonini-peretsApJ2012} found that $10\%$ of BH/BH binaries merging near supermassive BHs will have $e>0.1$ when they enter the LIGO band, with $\sim 2\%$ to $5\%$ of binaries having eccentricities $\sim 0.05$ and $\sim 10\%$ with $e \approx 0.001$ (see their Fig.~7; see also more recent work in Ref.~\cite{hong-lee_MNRAS2015}).  

Using a population synthesis code, Ref.~\cite{kowalska-etal-eccentricity-distribution-AA2011} computed the fraction of binaries with eccentricities exceeding $e=0.01$ at $30$ Hz (this is roughly the value where eccentricity will cause a systematic parameter bias \cite{favata-PRL2014}). For BH/BH binaries, $\lesssim 0.3\%$ will exceed this value at $30$ Hz. For BH/NS binaries, this fraction increases slightly to $<0.7\%$. For NS/NS binaries, this value is $<2\%$. The fraction of binaries that exceed $e=0.01$ at $3$ Hz (the ET band) is roughly double the numbers quoted above (see Table 3 of Ref.~\cite{kowalska-etal-eccentricity-distribution-AA2011}). 

For supermassive BH (SMBH) binaries that merge in galactic nuclei, Ref.~\cite{merrit-spurzem} found that these systems generally form with high eccentricities. When they enter the LISA band, a significant fraction of these binaries have eccentricities $e\sim 0.05 \mbox{--} 0.2$. (For a brief review of the literature on eccentric LISA sources, see Appendix A of Ref.~\cite{yunes-arun-berti-will-eccentric-PRD2009}.) While it is well known that extreme-mass-ratio inspirals (EMRIs) will be highly eccentric in the LISA band (see, e.g., Ref.~\cite{pau-EMRIreview2015JPhCS} for a recent review),  intermediate-mass-ratio inspirals (IMRIs) are likely to be nearly circular when they enter the LIGO band \cite{mandel-brown}.\footnote{Reference \cite{mandel-brown} found that IMRIs that harden via three-body interactions should have $e<10^{-4}$ at $10$ Hz, while $10\%$ of those formed by direct capture will have $e>0.1$ at $10$ Hz.}  

For binaries above a particular eccentricity threshold, the use of circular search templates results in a potential decrease in the signal-to-noise ratio (SNR). This issue has been investigated in a variety of studies \cite{mandel-brown,martel-poisson-eccentric-PRD1999,tessmer-gopu-PRD2008,cokelaer-pathak-detecteccentric-CQG2009,brown-zimmerman-eccentric-PRD2010,huerta-brown2013PRD}. The punch line of these analyses is that circular templates are sufficient for {\it detecting} compact binary inspirals with initial eccentricities (when entering the band of ground-based detectors)  $e\lesssim 0.02\mbox{--}0.05$ (for fitting factors $\gtrsim 0.95$).\footnote{The eccentricity threshold depends on the binary mass. The  range in eccentricity quoted above is most applicable to NS/NS binaries; for stellar-mass BH binaries, the eccentricity threshold for detection with circular templates is closer to $e\sim 0.15$. We do not quote precise values because the various studies in Refs.~\cite{martel-poisson-eccentric-PRD1999,tessmer-gopu-PRD2008,cokelaer-pathak-detecteccentric-CQG2009,brown-zimmerman-eccentric-PRD2010,huerta-brown2013PRD} differ in their details. Interestingly, the threshold for detecting IMRI systems in the LIGO band with circular templates is also $e\lesssim 0.05$ \cite{mandel-brown}, although estimates suggest there are not likely to be many IMRIs with eccentricities much higher than this. The inadequacy of circular templates for supermassive BH binaries in the LISA band is discussed in Ref.~\cite{porter-sesana2010}.} Although the waveform model developed here is not relevant to binaries with moderate to high eccentricities, the detectability of such binaries is considered in Refs.~\cite{kocsis-levinPRD2012,east-etalPRD2013,tai-mcwilliams-pretoriusPRD2014,coughlin-etalPRD2015}. 

We emphasize that currently understood astrophysical scenarios relevant for LIGO imply that if binaries enter the detector band with any eccentricity, the observed eccentricity is more likely to be small than large. This justifies our focus on the small-eccentricity limit. In addition, our formulas might be of use for parameter estimation or search template studies wishing to examine the general behavior of including eccentricity (sacrificing accuracy in the moderate eccentricity limit in exchange for simplicity and computational efficiency).

\subsection{\label{subsec:eccwaveformreview}Previous work on eccentric waveform models}
As circular binaries are thought to be more likely, circular-orbit waveforms are more developed than elliptical ones.\footnote{Recall that we take the word ``eccentric'' to refer to elliptical orbits ($e<1$) in this work. We do not review waveforms relevant to parabolic or hyperbolic binaries in detail.} Nonetheless, a significant body of work has explored eccentric waveform models. Much of this is reviewed in Sec.~10 of Ref.~\cite{blanchet-LRR2014}. Waveform polarizations for elliptical binaries are implicit in Peters and Mathews \cite{petersmathews} (who focus on computing the radiated power); they were first given explicitly in Wahlquist \cite{wahlquist} and later in Refs.~\cite{moreno-garrido-etal-eccentric-MNRAS1994,moreno-etal-eccentric-MNRAS1995,pierro-pinto-etal-MNRAS2001} to leading (0PN) order. These elliptical waveforms have since been extended to 1PN order in amplitude by Junker and Sch\"{a}fer \cite{junker-schafer} (see also Refs.~\cite{tessmergopamnras,majar-vasuth-unbound1pn-PRD2010}), to 1.5PN order in Blanchet and Sch\"{a}fer \cite{blanchetschafertails}, and to 2PN order (neglecting tails) in Gopakumar and Iyer \cite{GI2}. The 3PN (nonhereditary) contribution to the waveform was recently computed in Ref.~\cite{mishra-arun-iyer-3PNwaveform-PRD2015}. The nonlinear memory corrections to the polarizations (which enter at 0PN order) were computed in Ref.~\cite{favata-eccentricmemory}.  Information to compute the GW \emph{phasing} for elliptical binaries is currently known to relative 3PN order in the conservative \cite{damourderuelle,damourschafer,schafer-wex-PLA1993,*schafer-wex-PLA1993-erratum,DGI,quasikep3PN,quasikepphasing35PN} and dissipative parts \cite{wagoner-will,*wagoner-will-erratum,blanchetschafer1PN,junker-schafer,blanchetschafertails,riethschafer,GI1,arun-eccentrictailsEflux,arun-eccentricEflux3PN,arun-etal-eccentric-orbitalelements-PRD2009}. Time-domain waveforms that incorporate both spin and eccentricity via a direct numerical solution of the PN equations of motion are discussed in Ref.~\cite{cornish-key-spineccentricwaveforms-PRD2010,*cornish-key-spineccentricwaveforms-PRD2010-erratum,gopu-schafer-spinningeccentricPRD2011,levin-mcwilliams-contrerasCQG2011,CBwavesCQG2012}.

Especially useful for GW data analysis applications is the development of frequency-domain eccentric waveforms.
References \cite{tessmer-schafer-1pnfreqecc-PRD2010,mikoczi-etalHansencoeff2015} express the waveform amplitude to arbitrary eccentricity at 1PN order in the frequency domain but do not express the phasing explicitly as a function of frequency. This is extended to 2PN order in Ref.~\cite{tessmer-schafer-2pnfreqecc} with the phasing expressed as a hypergeometric function of the eccentricity. Explicit eccentric corrections to the phase of the Fourier transform of the GW signal (expressed as a function of frequency) were first computed in the stationary phase approximation (SPA) in Ref.~\cite{krolak}. They expressed the waveform amplitude at Newtonian order and without any eccentric corrections. The phase contained leading-order eccentric corrections [$O(e_0^2)$] at 0PN order. In Refs.~\cite{favata-phd,favata-PRL2014}, these $O(e_0^2)$ phase corrections were extended to 2PN order. The details of that computation, along with their extension to 3PN order, are the focus of this paper. The ``postcircular'' approximation in Ref.~\cite{yunes-arun-berti-will-eccentric-PRD2009} computed the amplitude and SPA phasing to Newtonian order as an expansion to $O(e^8)$. Post-Newtonian corrections to that work were recently incorporated in Ref.~\cite{huerta-etal-PRD2014}; however, these were not computed via a fully consistent PN approach but rather through the choice of a particular {\it ansatz} (as acknowledged in that work). For example, their results for the $O(e_0^2)$ corrections to the SPA phasing disagree with ours beginning at 1PN order. 
As this manuscript was nearing completion, we learned of related work in Ref.~\cite{tanay-haney-gopu-2015}. There, the postcircular approximation of Ref.~\cite{yunes-arun-berti-will-eccentric-PRD2009} is extended to 2PN order and $O(e^6)$,  with only Newtonian effects accounted for in the amplitude.  

The merger/ringdown portion of eccentric binary coalescence must be treated via numerical relativity. Eccentric merger simulations have been performed by multiple groups \cite{sperhake-etal-eBHBH2008PRD,hinder-etal-eBHBH2008PRD,vaishnav-etal-CQG2009,abdul-etal-eBHBH2010PRD,stephens-east-pretorius-eBHNS2011ApJ,east-pretorius-stephens-eBHNS2012PhRvD,gold-etal-eNSNS2012PRD,gold-brugmann-eBHBHPRD2013,east-paschalidis-pretorius-eBHNS2015ApJL,dietrich-etal2015,paschalidis-etal2015,east-etal2015}. Comparisons between eccentric post-Newtonian waveforms and NR simulations were performed in Refs.~\cite{gopu-etal-EtNRcomparePRD2008,hinder-etal-xmodelPRD2010}. More recent comparisons between PN, NR, and gravitational self-force (GSF) results for the periastron advance rate are discussed in Refs.~\cite{letiec-PRL2011,letiecPRD2013,Hinderer-etalPRD2013}; comparisons between PN and GSF calculations of the redshift invariant are found in Refs.~\cite{akcay-etal-GSFeccentricPRD2015,hopper-etal-eccentricGSFschw,bini-damour-eccentricGSF-a,bini-damour-eccentricGSF}. Recent attempts to extend the effective-one-body (EOB) formalism to handle eccentric orbits are discussed in Refs.~\cite{bini-damour-EOBeccentricRR-PRD2012,bini-damour-eccentricGSF-a,Akcay-vandeMeent-eccentricEOBpotential}.
\section{\label{sec:polarization}Gravitational wave polarizations}
We begin by defining our conventions and expressions for the GW polarizations. We work at leading (Newtonian) order in the amplitude of the polarizations but initially make no assumptions about the PN order of the phasing (that will be specified in later sections). Starting from an expression that is valid for general planar orbits, we then specialize to elliptical orbits (with the details relegated to Appendix \ref{app:polarization}). These expressions are further simplified to the case where eccentricity provides a negligible correction to the amplitude (but not to the phasing).

Consider a nonspinning eccentric binary with component masses $m_1$ and $m_2$, total mass $M=m_1+m_2$, and reduced mass $\mu=m_1 m_2/M$. Following Sec.~II of Damour, Gopakumar, and Iyer \cite{DGI}, introduce an orthonormal triad ${\bm p}$, ${\bm q}$, ${\bm N}$ in which ${\bm p}$ points toward a suitably defined ascending node, ${\bm N}$ points from the source to the observer, and ${\bm q}={\bm N} \times {\bm p}$. The relative separation vector of the binary ${\bm x}$, which has a magnitude $r$ and makes an angle $\phi$ with respect to ${\bm p}$, is given by
\be
\label{eq:x}
{\bm x} = {\bm p} r \cos \phi + ({\bm q} \cos \iota + {\bm N} \sin \iota) r \sin \phi \;,
\ee
where $\iota$ is the orbit inclination angle (the angle between ${\bm N}$ and the orbital angular momentum).
The plus ($+$) and cross ($\times$) polarizations of the gravitational-wave field can be expanded in a post-Newtonian series. The leading (``Newtonian'')-order piece of that expansion is given in terms of $r$, $\phi$, and their first time derivatives by Eq.~(6) of Ref.~\cite{DGI},
\bs
\label{eq:hN}
\begin{align}
\label{eq:hplusN}
h_{+}^{\rm N} &= -\frac{\eta M}{D} \bigg\{ (1+C^2) \bigg[ \bigg( \frac{M}{r} + r^2 {\dot{\phi}}^2 - {\dot{r}}^2 \bigg) \cos 2\phi + 2 \dot{r} r \dot{\phi} \sin 2\phi \bigg] + S^2 \bigg[ \frac{M}{r} - r^2 {\dot{\phi}}^2 - {\dot{r}}^2 \bigg] + O(v) \bigg\}   \, , \\
\label{eq:hcrossN}
h_{\times}^{\rm N} &= -2\frac{\eta M C}{D} \bigg[ \bigg( \frac{M}{r} + r^2 {\dot{\phi}}^2 - {\dot{r}}^2 \bigg) \sin 2\phi + 2 \dot{r} r \dot{\phi} \cos 2\phi + O(v) \bigg]  \, ,
\end{align}
\es
where $D$ is the distance to the binary, $\eta \equiv \mu /M$ is the reduced mass ratio, $C\equiv \cos \iota$, and $S\equiv \sin \iota$.
Corrections to these polarizations enter at $0.5$PN order [$O(v)$, where $v$ is the relative orbital speed].\footnote{Throughout, we denote $n$PN-order relative corrections as $O(v^{2n})$, denoting the appropriate power of $v/c$.} For eccentric binaries these amplitude corrections have been computed to 2PN-order by Gopakumar and Iyer \cite{GI2} and to 3PN order in Ref.~\cite{mishra-arun-iyer-3PNwaveform-PRD2015} (neglecting hereditary corrections). Nonlinear memory corrections are computed in Ref.~\cite{favata-eccentricmemory}. Only leading-order amplitude corrections are considered here. 

To further simplify Eqs.~\eqref{eq:hN}, we use the quasi-Keplerian parametrized solution for $r$ and $\phi$, which provides an analytic solution to the conservative part of the PN equations of motion (Sec.~\ref{sec:quasikep}). Using this quasi-Keplerian solution, we show in Appendix \ref{app:polarization} how the polarizations can be expressed in terms of the eccentric anomaly $u$ for arbitrary elliptical orbits [Eq.~\eqref{eq:hNu}]. Those expressions are further simplified by writing them as a series expansion in eccentricity [carried to $O(e^3)$] and in terms of the phase variable $\phi$ and the mean anomaly $l$ (which are straightforwardly expressed as functions of time for conservative orbits). The resulting expansion [Eqs.~\eqref{eq:hNl}] indicates how different frequency harmonics enter the waveform. Since we are focused on the small-eccentricity limit---in which most of the radiated power is concentrated at a GW frequency equal to twice the orbital frequency---we take the $e\rightarrow 0$ limit of our result and arrive at
\bs
\label{eq:hNcirc}
\begin{align}
\label{eq:hplusNcirc}
h_{+}^{\rm N} &= -2\frac{\eta M}{D} \left(\frac{M}{r} \right) (1+C^2) \cos 2\phi \;, \\
\label{eq:hcrossNcirc}
h_{\times}^{\rm N} &= -4\frac{\eta M}{D} \left(\frac{M}{r} \right) C \sin 2\phi \;.
\end{align}
\es
These are the expressions that we ultimately use for our polarizations. 
Note that this is equivalent to taking the circular orbit limit ($\dot{r}=0$ and $r^2 \dot{\phi}^2 = M/r$) of Eqs.~\eqref{eq:hN}. Eccentricity introduces $O(e)$ and higher corrections to these expressions.  However, we ignore them since amplitude corrections are less important than phase corrections (and since we are also assuming that eccentricity is small). We note that the expressions \eqref{eq:hNcirc} depend on only a single phase variable while the more general expressions \eqref{eq:hNl} depend on two phase variables. There are thus only three initial conditions [$r(t_0)$, $e(t_0)$, and $\phi(t_0)$] associated with the orbital motion that need to be specified in our waveforms. For general elliptical orbits, an additional constant associated with a second phase variable---and corresponding to the initial argument of pericenter $\varpi(t_0)$---would also need to be specified. However, because we neglect $O(e)$ corrections to the waveform amplitude \emph{and} oscillatory contributions to the waveform phase $\phi$, this dependence on a fourth initial condition drops out of our expressions. The extended discussion in Appendix \ref{app:polarization} further clarifies this point and the nature of the approximations leading to Eq.~\eqref{eq:hNcirc}. 
\section{\label{sec:quasikep}Quasi-Keplerian Parametrization}
First introduced in Refs.~\cite{damourschafer,schafer-wex-PLA1993,*schafer-wex-PLA1993-erratum}, the quasi-Keplerian formalism provides an analytic (parametric) solution to the conservative pieces of the PN equations of motion. It provides the orbital variables $(r,\phi)$ and their derivatives as a function of a parametric angle (the eccentric anomaly $u$). Combined with a numerical solution of the PN extension of Kepler's equation, this formalism allows one to determine the orbital evolution as a function of time without the need to solve ODEs. In this section, we first review the more familiar Newtonian case, expressing the Keplerian solution in a form that will be similar to its PN generalization (Sec.~\ref{subsec:quasikep-Newt}). We then provide the related generalization to the PN case (quasi-Keplerian), listing the relevant equations from the literature that one needs to model eccentric PN orbits (Sec.~\ref{quasikep2}). In Sec.~\ref{subsec:convertfreq}, we perform a change of variables that more naturally connects with the circular limit. The overall domain of validity of the quasi-Keplerian formalism is discussed in Sec.~\ref{subsec:domain}. The effects of incorporating radiation reaction are then discussed in Sec.~\ref{sec:evolvebinaries} and the remainder of the paper. 
\begin{figure}[t]
\includegraphics[angle=0, width=0.45\textwidth]{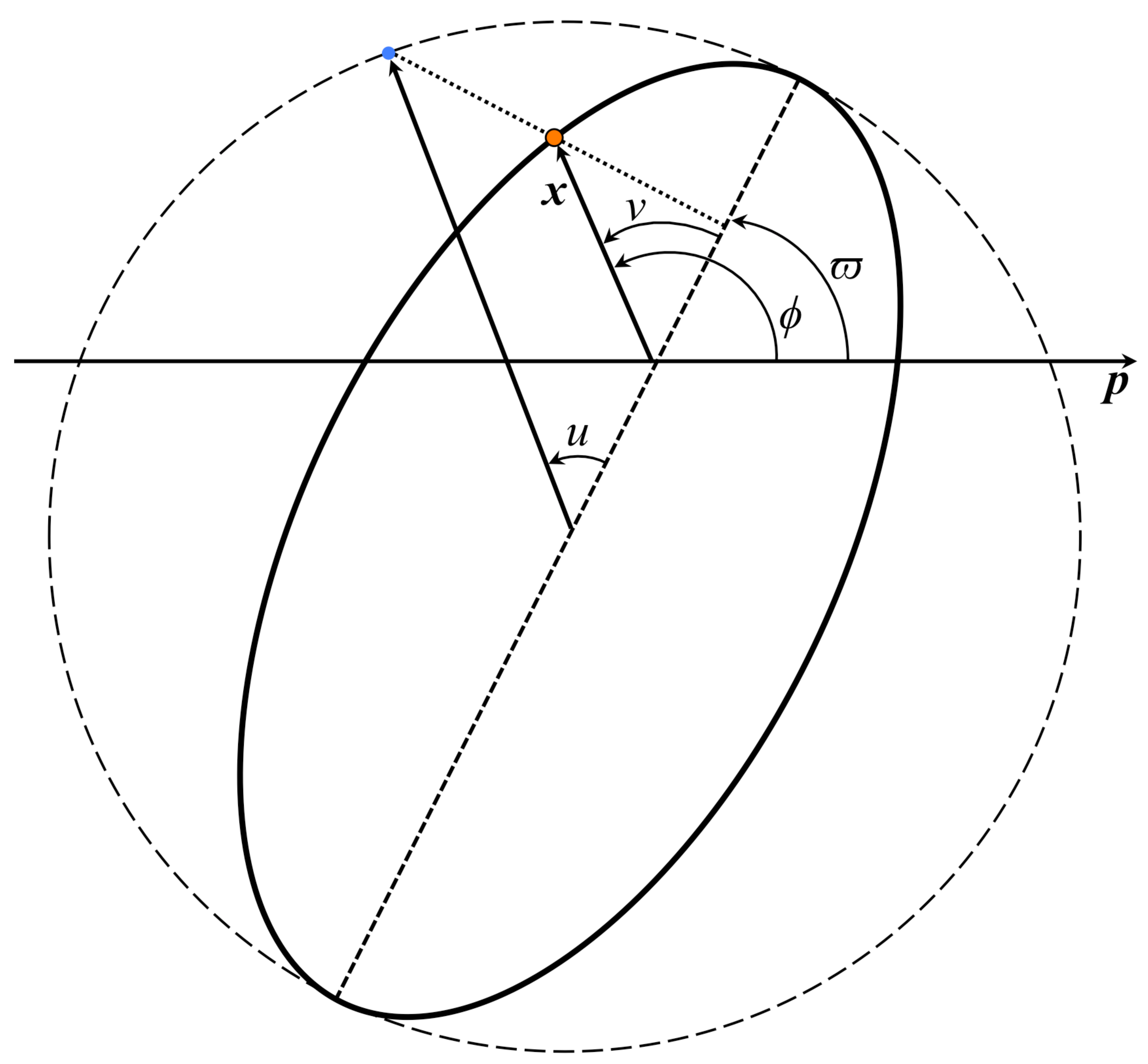}
\caption{\label{fig:ellipse} Angles describing elliptical orbits. The vector ${\bm x}$ is the relative separation vector which points from the binary center of mass to the position of the reduced mass. The reduced mass traces out a curve, and we show the ellipse which is momentarily tangent to that curve. The vector ${\bm x}$ makes an angle $\phi$ (the orbital phase angle) from the unit vector ${\bm p}$ and an angle $v$ (the true anomaly) from the pericenter point of the ellipse. The pericenter itself makes an angle $\varpi$ (the argument of pericenter) from the direction ${\bm p}$. The eccentric anomaly $u$ is the angle from the center of the ellipse to the projection (perpendicular to the major axis) of the particle's position on the circle circumscribing the ellipse. The mean anomaly $l$ (which does not have a geometric interpretation) is an angular parameter that varies along the orbit at a uniform rate, completing one cycle in a time $2\pi/n$. Note that $u$, $v$, and $l$ are measured from the pericenter direction.}
\end{figure}
\subsection{\label{subsec:quasikep-Newt}Newtonian-order (Keplerian) parametrization}
In the Newtonian limit, the radial and angular pieces of elliptical motion can be expressed via the following equations:
\bs
\label{eq:newteqns}
\begin{align}
\label{eq:rnewt}
r_{\rm N} &= S(l;n,e)=a_r(1-e \cos u) \, , \\
\label{eq:rdotnewt}
\dot{r}_{\rm N} &=n \frac{\partial S}{\partial l}(l;n,e)=\frac{\xi^{1/3} e \sin u}{(1-e \cos u)} \,, \\
\label{eq:phinewt}
\phi_{\rm N} &= \lambda_{\rm N} + W_{\rm N}(l;n,e)=\lambda_{\rm N} + v - u + e \sin u \, , \\
 &= v + \varpi = v + (c_{\lambda} - c_l) \, , \\
\label{eq:lambdanewt}
\lambda_{\rm N} &= n(t-t_0)+c_{\lambda} \, , \\
\label{eq:phidotnewt}
\dot{\phi}_{\rm N} &= n+n\frac{\partial W_{\rm N}}{\partial l}(l;n,e) = \frac{n \sqrt{1-e^2}}{(1-e \cos u)^2} \, , \\
\label{eq:keplereqnewt}
l_{\rm N} &= n(t-t_0)+c_l = u- e \sin u \, , \\
\label{eq:veqnewt}
v &= V_{\rm N}(u) \equiv 2 \arctan \left[ \left( \frac{1+e}{1-e}\right)^2 \tan\Big(\frac{u}{2}\Big) \right] \;.
\end{align}
\es
The subscript N denotes that these are Newtonian quantities (to distinguish between their PN generalizations below). The angles $u$, $v$, and $l$ are the eccentric, true, and mean anomalies. Figure~\ref{fig:ellipse} discusses their geometrical interpretation (which carries over to the PN case). The semimajor axis of the ellipse is $a_r=(M/n^2)^{1/3}$, and the \emph{mean motion} is $n\equiv 2\pi/P$, where $P$ is the radial orbital (periastron to periastron) period. (For Newtonian orbits, the radial and azimuthal frequencies are identical, but not for PN orbits). We also define the dimensionless radial angular orbit frequency  $\xi = M n$; this dimensionless variable will be used extensively in this work and serves as a PN expansion parameter. The definitions of the functions $S$ and $W_{\rm N}$ can be read off of the above equations. The phase angle $\lambda_{\rm N}$ provides the linearly accumulating piece of the orbital phase. In the Newtonian limit, it is equivalent (up to a constant shift) to the mean anomaly $l_{\rm N}$. The constants $c_l$ and $c_{\lambda}$ provide the values of $l_{\rm N}$ and $\lambda_{\rm N}$ at some instant $t_0$. They are related to the argument of pericenter $\varpi = c_{\lambda} - c_l$, which is the angle the pericenter makes with respect to ${\bm p}$. 

Note that a planar orbit requires the specification of four initial conditions or constants of the motion. For example, these might be the set $[r(t_0), \phi(t_0), \dot{r}(t_0), \dot{\phi}(t_0)]$. Equivalently (and of greater geometrical meaning), one can specify $[a_r, e, \varpi, \phi(t_0)]$. The constant $a_r$ could be replaced by $n$ or a related frequency variable. In Eqs.~\eqref{eq:newteqns} above the constants that must be specified are $[n, e, c_{\lambda}, c_l]$. Given $\varpi$ and $\phi(t_0)$, $c_l$ is determined by first solving for the initial value of $v$ via $\phi(t_0) = \varpi + v_0$ and then inverting Eq.~\eqref{eq:veqnewt} [$v_0=V_{\rm N}(u_0)$] for $u_0$. The value $u_0$ is substituted in Eq.~\eqref{eq:keplereqnewt} (with $t=t_0$) to give $c_l$.
\subsection{\label{quasikep2}Quasi-Keplerian parametrization}
When post-Newtonian effects are considered, the resulting orbits are no longer \emph{Keplerian} ellipses, but the parametric equations for $r$, $\phi$, $\dot{r}$, and $\dot{\phi}$ take a form similar  to Eqs.~\eqref{eq:newteqns} (but are much more complicated). The resulting solution is thus referred to as  \emph{quasi-Keplerian}. These explicit analytic expressions can only be obtained for the conservative contributions to the equations of motion (i.e., one ignores the dissipative or radiation-reaction contributions to the equations of motion at 2.5PN, 3.5PN, and higher orders).  The extension of Eqs.~\eqref{eq:newteqns} is known to 3PN order \cite{quasikep3PN,quasikepphasing35PN} and has the following form [Eqs.~(7\mbox{--}12) of Ref.~\cite{quasikepphasing35PN}]:
\bs
\label{eq:quasikeeqns}
\begin{align}
\label{eq:req}
r &= S(l;n,e_t)=a_r(1-e_r \cos u) \, , \\
\label{eq:rdot}
\dot{r} &= n \frac{\partial S}{\partial l}(l;n,e_t) \, , \\
\label{eq:phi}
\phi &= \lambda + W(l;n,e_t) \, , \\
\label{eq:lambda}
\lambda &= (1+k) n (t-t_0)+c_{\lambda} \, , \\ 
\label{eq:W}
W(l;n,e_t)
&= ( 1 + k ) ( v - l ) 
+ \bigg( \frac{ f_{4\phi} }{c^4} + \frac{ f_{6\phi} }{c^6} \bigg) \sin 2 v
+ \bigg( \frac{ g_{4\phi} }{c^4} + \frac{ g_{6\phi} }{c^6} \bigg) \sin 3 v
+ \frac{ i_{6\phi} }{c^6} \sin 4 v + \frac{ h_{6\phi} }{c^6} \sin 5 v
\, , \\
\label{eq:phidot}
\dot{\phi} &= (1+k) n+n\frac{\partial W}{\partial l}(l;n,e_t) \, , \\
\label{eq:keplereq}
l &= n(t-t_0)+c_l = u - e_t \sin u
+ \bigg( \frac{g_{4t}}{c^4} + \frac{g_{6t}}{c^6} \bigg) (v - u)
+ \bigg( \frac{f_{4t}}{c^4} + \frac{f_{6t}}{c^6} \bigg) \sin v
+ \frac{i_{6t}}{c^6} \sin 2v + \frac{h_{6t}}{c^6} \sin 3v
\, , \\
\label{eq:veqn}
v &= V(u) \equiv 2 \arctan \left[ \left( \frac{1+e_{\phi}}{1-e_{\phi}}\right)^2 \tan\Big(\frac{u}{2}\Big) \right] \, . 
\end{align}
\es
The various symbols have the same interpretation as their Newtonian counterparts, but the function $W$ and the 3PN analog of Kepler's equation are significantly more complex. The functions $g_{4t}$, $g_{6t}$, $f_{4t}$, $f_{6t}$, $i_{6t}$, $h_{6t}$, $g_{4\phi}$, $g_{6\phi}$, $f_{4\phi}$, $f_{6\phi}$, $i_{6\phi}$, and $h_{6\phi}$  are given in Ref.~\cite{quasikep3PN} [in both Arnowitt-Deser-Misner (ADM) and harmonic gauges], while more explicit and useful expressions for $r$, $\dot{r}$, $\phi$, and $\dot{\phi}$ are given in harmonic gauge in Ref.~\cite{quasikepphasing35PN} (to 3PN order) or in ADM gauge to 2PN order in Ref.~\cite{DGI}. For brevity, we list here only expressions in {\it harmonic gauge}\footnote{Note that the harmonic and ADM gauge expressions differ only at 2PN and higher orders; at 1PN order, they are identical.} and to 2PN order (we list some results to 3PN order if they are crucial for later steps in our analysis). As in the Newtonian case, the functions $S(l)$ and $W(l)$ are $2\pi$ periodic in $l$.

Besides the overall increase in complexity, the quasi-Keplerian case introduces some additional new features. There now appear three eccentricities $e_t$, $e_r$, and $e_{\phi}$ instead of one in the Newtonian case. These eccentricities can be related to each other or to the orbital energy and angular momentum (see Refs.~\cite{DGI,quasikep3PN}).  The quasi-Keplerian equations also show the well-known periastron precession, a secular effect embodied in the constant $k=\Delta \Phi/(2\pi)$, where $\Delta \Phi$ is the advance of the periastron angle in the time interval $P$. The explicit expression for $k$ is given to 3PN order by Eq.~(25d) of Ref.~\cite{quasikepphasing35PN}:
\begin{multline}
\label{eq:k3pn}
k = \frac{ 3 \xi ^{2/3} }{ 1 - e_t^2 }
+ \left[ 78  - 28 \eta + ( 51 - 26 \eta ) e_t^2 \right]\frac{ \xi ^{4/3} }{ 4 ( 1 - e_t^2 )^2 }
+ \Big \{
18240 - 25376 \eta + 492 \pi^2 \eta + 896 \eta ^2
\\
+ ( 28128 - 27840 \eta + 123 \pi^2 \eta + 5120 \eta^2 ) e_t^2
+ ( 2496 - 1760 \eta + 1040 \eta^2  ) e_t^4
\\
+ \left[ 1920 - 768 \eta + ( 3840 - 1536 \eta ) e_t^2 \right] \sqrt{1 - e_t^2}
\Big \}\frac{ \xi^2 }{ 128 ( 1 - e_t^2 )^3 } \,.
\end{multline}

Making the above expressions more explicit requires choosing an appropriate set of constants of the motion. Several choices are possible for the principal constants of motion, including some combination of the orbital energy $E$, the magnitude of the reduced angular momentum $h$, the mean motion $n$, the semimajor axis $a_r$, or one of the three eccentricities $(e_t, e_r, e_{\phi})$. A convenient choice is to choose the mean motion $n$ and the time eccentricity $e_t$ \cite{DGI,quasikepphasing35PN}. In addition to the principal (intrinsic) constants of motion, two positional (extrinsic) constants of motion ($c_l$ and $c_{\lambda}$) determine the orientation of the orbit and the orbital phase at a reference time $t_0$. These constants ($n, e_t, c_l, c_{\lambda}$) are fixed for conservative orbits (no radiation reaction). When radiation reaction is considered, these ``constants'' now evolve---their initial values must be chosen, and a scheme for evolving them must be supplied. This is discussed in Sec.~\ref{sec:evolvebinaries}. In the present section, we ignore radiation reaction effects.

Given this choice of constants, we can now express $r$, $\dot{r}$, $\phi$, and $\dot{\phi}$ explicitly in terms of $n$, $e_t$, $c_l$, $c_{\lambda}$, and the eccentric anomaly $u$. The exact expressions are given to 3PN order and harmonic gauge in Eqs.~(23)--(26) of Ref.~\cite{quasikepphasing35PN} or to 2PN order and ADM gauge in Eqs.~(51)--(54) of Ref.~\cite{DGI}. For reference, we list the complete expressions up to 2PN order and in harmonic gauge [see Eqs.~(23)--(27) in Ref.~\cite{quasikepphasing35PN} for the lengthy 3PN terms]:
\bs
\label{eq:quasikep2pn}
\begin{multline}
\label{eq:r2pn}
r= M \xi^{-2/3} (1-e_t \cos u) \Bigg( 1 +  [ -18 +2\eta - (6-7\eta) e_t \cos u]\frac{\xi^{2/3}}{6 (1-e_t \cos u)} +  \bigg\{ -72(4-7\eta)
+ [ 72 + 30 \eta + 8 \eta^2 
\\
 - (72-231 \eta + 35 \eta^2) e_t \cos u ] (1-e_t^2) - 36 (5-2\eta)(2+e_t \cos u) \sqrt{1-e_t^2} \bigg\}\frac{\xi^{4/3}}{72 (1-e_t^2)(1-e_t \cos u)} + O(\xi^2) \Bigg) \, ,
\end{multline}
\begin{multline}
\label{eq:rdot2pn}
\dot{r} = \frac{ \xi^{1/3} }{ (1 - e_t \cos u) } e_t \sin u \Bigg\{ 1+  ( 6 - 7\eta )\frac{ \xi^{2/3} }{6} + 
\biggl [- 468 - 15 \eta + 35 \eta^2 + ( 135 \eta - 9 \eta^2 ) e_t^2 + ( 324 + 342 \eta 
\\
- 96 \eta^2 ) e_t \cos u 
+ ( 216 - 693 \eta + 105 \eta^2 ) ( e_t \cos u )^2 - ( 72 - 231 \eta + 35 \eta^2 ) ( e_t \cos u )^3
\\
+ \frac{ 36 }{ \sqrt{1 - e_t^2} } ( 1 - e_t \cos u )^2 ( 4 - e_t \cos u ) ( 5 - 2 \eta )
\biggr ]\frac{ \xi^{4/3} }{ 72 ( 1 - e_t \cos u)^3 } +O(\xi^2) \Bigg\} \, ,
\end{multline}
\be
\label{eq:phi2pn1}
\phi(\lambda,l) = \lambda + W(l) \, ,
\ee
\begin{multline}
\label{eq:phi2pn3}
W(l)= (v-u + e_t \sin u) +  (v-u + e_t \sin u)\frac{3\xi^{2/3}}{1-e_t^2}
+   \bigg ( 8 \left[ 78 - 28 \eta + ( 51 - 26 \eta) e_t^2 - 6 ( 5 - 2 \eta ) ( 1 - e_t^2 )^{3/2} \right] 
\\
( v - u ) ( 1 - e_t \cos u )^3
+ \Big \{624 - 284 \eta + 4 \eta^2 + ( 408 - 88 \eta - 8 \eta^2 ) e_t^2 - ( 60 \eta - 4 \eta^2 ) e_t^4 + \left[ - 1872 + 792 \eta 
\right.  \\ \left.
 - 8 \eta^2 - ( 1224 - 384 \eta
- 16 \eta^2 ) e_t^2
 + ( 120 \eta - 8 \eta^2 ) e_t^4 \right] e_t \cos u + \left[ 1872 - 732 \eta + 4 \eta^2 + ( 1224 - 504 \eta - 8 \eta^2 ) e_t^2 
\right. \\ \left.
- ( 60 \eta - 4 \eta^2) e_t^4 \right] ( e_t \cos u )^2
+ \left[ - 624 + 224 \eta - ( 408 - 208 \eta ) e_t^2 \right] ( e_t \cos u )^3 \Big \} e_t \sin u 
\\
+ \Big \{-(8 + 153 \eta - 27 \eta^2 ) e_t^2 + ( 4 \eta - 12 \eta^2 ) e_t^4
+ \left[ 8 + 152 \eta - 24 \eta^2 
+ ( 8 + 146 \eta - 6 \eta^2 ) e_t^2 \right] e_t \cos u
\\
 + \left[ - 8 - 148 \eta + 12 \eta^2 - ( \eta - 3 \eta^2 ) e_t^2 \right] ( e_t \cos u )^2 \Big \} e_t \sin u \sqrt{1 - e_t^2} \Bigg) \frac{ \xi ^{4/3} }{ 32 ( 1 - e_t^2 )^2( 1 - e_t \cos u )^3 } + O(\xi^2)
\,,
\end{multline}
\begin{multline}
\label{eq:phidot2pn}
\dot{\phi} = \frac{ (\xi/M) \sqrt{1 - e_t^2} }{ (1 - e_t \cos u )^2 } \Bigg( 1+  \left[ 3  - ( 4 - \eta ) e_t^2  + ( 1 - \eta ) e_t \cos u \right]\frac{ \xi^{2/3} }{ (1 - e_t^2) ( 1 - e_t \cos u) } \\
+  \biggl \{ 144 - 48 \eta - ( 162 + 68 \eta - 2 \eta^2 ) e_t^2 + ( 60 + 26 \eta - 20 \eta^2 ) e_t^4 + ( 18 \eta + 12 \eta^2 ) e_t^6 + \left[ - 216 + 125 \eta + \eta^2 
\right. \\ \left.
+ ( 102 + 188 \eta + 16 \eta^2 ) e_t^2 - ( 12 + 97 \eta - \eta^2 ) e_t^4 \right] e_t \cos u
+ \left[ 108- 97 \eta - 5 \eta^2
+ ( 66 - 136 \eta + 4 \eta^2 ) e_t^2 
\right. \\ \left.
 - ( 48 - 17 \eta + 17 \eta^2 ) e_t^4 \right] ( e_t \cos u )^2
+ \left[ - 36 + 2 \eta - 8 \eta^2 - ( 6 - 70 \eta - 14 \eta^2 ) e_t^2 \right] ( e_t \cos u )^3 
 \\ 
+ 18 ( 1 - e_t \cos u )^2 ( 1 - 2 e_t^2 + e_t \cos u ) ( 5 - 2 \eta ) \sqrt{1 - e_t^2} \biggr \}\frac{ \xi^{4/3} }{ 12 (1 - e_t^2)^2 ( 1 - e_t \cos u)^3 } + O(\xi^2) \Bigg)
\;.
\end{multline}
\es
Some of these equations depend on the true anomaly $v$ in the combination $v-u$. Rather than using the discontinuous function in Eq.~\eqref{eq:veqn}, this combination is more easily described by the smooth function \cite{DGI,quasikepphasing35PN}
\be
\label{eq:vmu}
v-u = 2 \tan^{-1} \left( \frac{\beta_{\phi} \sin u}{1-\beta_{\phi} \cos u} \right) \;,
\ee
where $\beta_{\phi}=(1-\sqrt{1-e_{\phi}^2})/e_{\phi}$. An explicit expression for $e_{\phi}$ in terms of $e_t$ and $n$ is not listed in the literature, but it can be derived by combining Eq.~(25r) of Ref.~\cite{quasikep3PN} with Eqs.~(21) of Ref.~\cite{quasikepphasing35PN}. The result to (3PN order and harmonic gauge) is
\begin{multline}
\label{eq:ephi}
\frac{e_{\phi}}{e_t} = 1 + \xi^{2/3} (4-\eta) +  \bigg[ 2016 - 260\eta-4\eta^2-(1152-659\eta+41\eta^2) e_t^2 + (720-288\eta) \sqrt{1-e_t^2} \bigg]\frac{\xi^{4/3}}{96 (1-e_t^2)} \\
+  \Bigg\{ 2553600 - 2719360\eta + 17220 \pi^2 \eta + 268800\eta^2 - (3494400 - 3203200 \eta + 17220 \pi^2 \eta  + 255360 \eta^2) e_t^2 \\
+ (940800 - 483840 \eta - 13440 \eta^2) e_t^4 + \bigg[ 4139520 - 3574960 \eta + 17220 \pi^2 \eta - 155680 \eta^2 
- (2419200 - 1290048 \eta + 4305 \pi^2 \eta
\\
+ 483420 \eta^2 - 18900 \eta^3) e_t^2 + (860160 - 786310 \eta + 134050 \eta^2 - 1050 \eta^3 ) e_t^4 \bigg] \sqrt{1-e_t^2} \Bigg\}\frac{\xi^2}{26880 (1-e_t^2)^{5/2}} \;.
\end{multline}

The above equations allow one to determine the functions $r$, $\dot{r}$, $\phi$, $\dot{\phi}$, and the waveform polarizations $h_{+,\times}$ entirely in terms of the eccentric anomaly $u$ and the chosen constants $e_t$, $n$, $c_l$, and $c_{\lambda}$ [see Eq.~\eqref{eq:hNu}]. With the addition of the 3PN extension of Kepler's equation [Eq.~(27) of Ref.~\cite{quasikepphasing35PN}],
\be
\label{eq:keplereqn}
l = u - e_t \sin u +  \left[ (15 \eta - \eta^2) e_t \sin u \sqrt{1-e_t^2} + 12 (5-2\eta) (v-u) (1- e_t \cos u) \right]\frac{\xi^{4/3}}{8\sqrt{1-e_t^2} (1-e_t \cos u)} + O(\xi^2) ,
\ee
one can parametrically obtain the orbit and waveform as a function of the mean motion angle or time. To do this Eq.~\eqref{eq:keplereqn} must be inverted. Techniques for doing this numerically are given in celestial mechanics texts (see, e.g., Sec.~6.6 of Danby \cite{danby}) and are discussed more recently in the context of eccentric compact binaries in Ref.~\cite{tessmergopamnras}. For Newtonian binaries, Bessel derived a series expansion that solves Kepler's equation (e.g., Appendix D of Ref.~\cite{danby}):
\be
\label{eq:keplerbessel}
u = l + 2 \sum_{n=1}^{\infty} \frac{1}{n} J_n(n e_t) \sin(n l) \;.
\ee
Since PN corrections to Kepler's equation do not enter until 2PN order, the above equation can also be applied to binaries at 1PN order. In Appendix \ref{app:polarization}, we analytically invert the 3PN version of Eq.~\eqref{eq:keplereqn}, obtaining $u(l)$ as a series expansion to order $O(e_t^3)$. This expansion can then be plugged into expressions for $r$, $\dot{r}$, $\phi$, $\dot{\phi}$, and $h_{+,\times}$, yielding explicit functions of time [e.g., Eqs.~\eqref{eq:quasikepsmallet}--\eqref{eq:hNl}].  
\subsection{\label{subsec:convertfreq}Converting between radial and azimuthal frequencies}
In the quasi-Keplerian formalism, PN expansions of elliptical orbit quantities are most naturally expressed in terms of the radial orbit angular frequency $\omega_r\equiv n\equiv \xi/M$ (i.e., the mean motion or periastron to periastron angular frequency).  However, circular orbit quantities are more naturally expanded in terms of the azimuthal or $\phi$-angular frequency $\omega_{\phi} \equiv \xi_{\phi}/M$ (this is related to the time to return to the same azimuthal angular position in the orbit). The relation between $\omega_r$ and the \emph{instantaneous} azimuthal angular frequency $\dot{\phi}$ is given by Eq.~\eqref{eq:phidot}. However, since we are often more concerned with secular variations, it is useful to derive the relationship between the orbit-averaged azimuthal frequency $\omega_{\phi} \equiv \langle \dot{\phi} \rangle$ and the radial frequency. Upon orbit averaging Eq.~\eqref{eq:phidot}, the periodic $dW/dl$ term vanishes, and we have 
\begin{multline}
\label{eq:xiphi-xi}
{\xi}_{\phi} = M \langle \dot{\phi} \rangle =M \frac{d{\lambda}}{dt}= (1+{k}) {\xi} \\
={\xi} \Bigg( 1 + \frac{ 3{\xi} ^{2/3} }{ 1 -{ e}_t^2 }
+ 
\left[ 78  - 28 \eta + ( 51 - 26 \eta ) {e}_t^2 \right]\frac{ {\xi} ^{4/3} }{ 4 ( 1 -{ e}_t^2 )^2 }
+ 
 \bigg\{
18240 - (25376 - 492 \pi^2) \eta + 896 \eta ^2
+ [ 28128 - (27840  - 123 \pi^2) \eta 
\\
+ 5120 \eta^2 ]{e}_t^2
+ ( 2496 - 1760 \eta + 1040 \eta^2  ) {e}_t^4
+ \left[ 1920 - 768 \eta + (3840 - 1536 \eta ) {e}_t^2 \right] \sqrt{1 - {e}_t^2}
\bigg\}\frac{{\xi}^2 }{ 128 ( 1 -{ e}_t^2 )^3 }  \Bigg).
\end{multline}
This equation can be inverted to give
\begin{multline}
\label{eq:xi-xiphi}
{\xi} ={\xi}_{\phi} \Bigg(  1 - \frac{ 3 {\xi}_{\phi}^{2/3} }{ 1 - {e}_t^2 }
- \left[ 18  - 28 \eta + ( 51 - 26 \eta ) {e}_t^2 \right]\frac{ {\xi}_{\phi}^{4/3} }{ 4 ( 1 - {e}_t^2 )^2 } -  \bigg\{ -192-(14624 -492 \pi^2)\eta + 896\eta^2
+[8544-(17856
\\
-123\pi^2)\eta
+5120\eta^2] {e}_t^2 + (2496-1760\eta+1040\eta^2) {e}_t^4 + \left[ 1920-768\eta + (3840-1536\eta) {e}_t^2 \right] \sqrt{1-{e}_t^2} \bigg\}\frac{\xi_{\phi}^2}{128(1-{e}_t^2)^3} \Bigg).
\end{multline}
This relation allows us to replace the constant of the motion $\xi = M n$ in the quasi-Keplerian formalism with the equivalent constant $\xi_{\phi} = M \dot{\lambda}$. 
The advantage in doing so is that $\xi_{\phi}$ is directly related to the standard PN expansion parameters $x$ and $v$ that appear in PN expansions for quasicircular orbits. Specifically, we make the following identifications between the various PN expansion parameters:
\be
\label{eq:xi_x_v_relate}
\xi_{\phi} = M \omega_{\phi} = x^{3/2} = v^{3} \, .
\ee
In the eccentric case, $x$ and $v$ are defined by the above relation and reduce to their standard meanings in the circular limit. 
As an application of these relations, it will be useful (in the next section) to express $k$ in terms of $\xi_{\phi}$ via substitution of \eqref{eq:xi-xiphi} into \eqref{eq:k3pn} and expanding to 3PN order: \begin{multline}
\label{eq:k3pnxiphi}
k(\xi_{\phi}) =  \frac{ 3 \xi_{\phi}^{2/3} }{ 1 - e_t^2 }
+ \left[ 54  - 28 \eta + ( 51 - 26 \eta ) e_t^2 \right]\frac{ \xi_{\phi}^{4/3} }{ 4 ( 1 - e_t^2 )^2 }
+ \bigg\{
6720 - (20000 - 492 \pi^2) \eta + 896 \eta ^2
\\
+ [ 18336 - (22848 - 123 \pi^2) \eta + 5120 \eta^2 ] e_t^2
+ ( 2496 - 1760 \eta + 1040 \eta^2  ) e_t^4
\\
+ \left[ 1920 - 768 \eta + ( 3840 - 1536 \eta ) e_t^2 \right] \sqrt{1 - e_t^2}
\bigg\}\frac{ \xi_{\phi}^2 }{ 128 ( 1 - e_t^2 )^3 }.
\end{multline}

Note that in the $e_t \rightarrow 0$ limit the frequency variables $\xi$ and $\xi_{\phi}$ do not agree. This arises from the fact that the well-known formula for the periastron advance angle is independent of eccentricity in the $e_t \rightarrow 0$ limit. Readers should keep in mind that it is the quantity ${\xi}_{\phi}$ that is equivalent to the circular orbit frequency in the $e_t \rightarrow 0$ limit. When PN quantities for eccentric orbits are specialized to the $e_t \rightarrow 0$ limit, they generally will reduce to the standard circular expressions only if they are expressed in terms of ${\xi}_{\phi}$ (not ${\xi}$). This is an important point and motivates much of the approach that we follow throughout this paper. 
\subsection{\label{subsec:domain}Domain of validity of the quasi-Keplerian formalism}
In Sec.~III of Ref.~\cite{DGI}, an estimate was provided for the range of validity of the quasi-Keplerian formalism. We review this constraint here, illustrate its problems, and then argue that recent comparisons with numerical relativity and gravitational self-force  calculations suggest that the constraint (in the small $e_t$ limit) should be relaxed.  

Reference \cite{DGI} argues that the parameters of elliptical motion in the quasi-Keplerian formalism should be chosen such that i) orbits are not plunging and ii) rapid periastron precession is avoided. Section III of Ref.~\cite{DGI} adequately discusses the first point. In the small $e_t$ limit, this constraint is simply that one stays outside the last stable orbit, which we will take to be the innermost stable circular orbit (ISCO; there are eccentric corrections to this, but they will be negligible for our purposes as any small eccentricity will have rapidly decayed at frequencies near plunge). The second point is a more conservative constraint---effectively requiring the binary separation (for a given eccentricity) to be sufficiently far from the plunge boundary that the periastron advance rate is ``slow.'' (This also eliminates the possibility of zoom-whirl orbits.)  Using a particular (and somewhat arbitrary) choice for specifying the smallness of the periastron advance rate in the Schwarzschild spacetime,  Ref.~\cite{DGI} proposes that the following condition be satisfied:
\be
\label{eq:quasikepapprox}
\frac{\xi}{(1-e_t^2)^{3/2}} < 0.0030.
\ee
Rearranging the above and approximating (at Newtonian order) $\xi 
\approx \pi M f$ yields the following constraint on the eccentricity parameter,
\be
\label{eq:etconstraint}
e_t < \sqrt{1- \left( \frac{\pi M f}{\alpha} \right)^{2/3}},
\ee
where $f$ is the GW frequency of the fundamental harmonic (twice the orbital frequency) and $\alpha = 0.0030$. 
This equation implies that (for the quasi-Keplerian formalism to be valid) a NS/NS system should have $e_t \lesssim 0.85$ at $10$ Hz, while a $M=10 M_{\odot}$ BH/BH binary should have $e_t \lesssim 0.6$. Similarly a BH/BH binary  with $M \gtrsim 19.3 M_{\odot}$ is constrained to be fully circularized before entering the LIGO band.

An immediate problem arises when one considers the decay of eccentricity with frequency, $e_t(f) \approx e_0 (f_0/f)^{19/18}$ for small $e_t$. Substituting this relation into Eq.~\eqref{eq:etconstraint} implies that the GW frequency must satisfy the following inequality throughout the inspiral:
\be
\label{eq:f-constraint}
e_0^2 \left( \frac{f_0}{f} \right)^{19/9} + \left( \frac{\pi M f}{\alpha}\right)^{2/3} < 1.
\ee 
If a binary enters the LIGO band ($f_0=10$ Hz) with $e_0=0.1$, this implies that the formalism is valid only for $f\lesssim 69$ Hz for a NS/NS binary or $f \lesssim 19$ Hz for a $M=10 M_{\odot}$ BH/BH binary. For $e_0 \lesssim 0.1$, the first term in \eqref{eq:f-constraint} hardly affects the result, and one can simplify the criterion to 
\be
\label{eq:f-constraint2}
f < \frac{\alpha}{\pi M} = 194 \, {\rm Hz} \left( \frac{\alpha}{0.0030} \right) \left( \frac{1 M_{\odot}}{M} \right).
\ee
In comparison, the Schwarzschild ISCO frequency is
\be
\label{eq:isco}
f_{\rm isco} = \frac{6^{-3/2}}{\pi M} = 4397 \, {\rm Hz} \left( \frac{1 M_{\odot}}{M} \right). 
\ee
Equation \eqref{eq:f-constraint2} is vastly more constraining than the ISCO limit and suggests that quasi-Keplerian waveforms would be essentially useless for LIGO data analysis. This is surprising. While circular PN waveforms certainly breakdown before the ISCO limit above, they can often be pushed much closer to the ISCO frequency than Eq.~\eqref{eq:f-constraint2} would imply (especially in the comparable-mass limit). In the low-eccentricity limit, the quasi-Keplerian waveforms we derive here smoothly approach the circular limit, so we would expect that they would have a similar range of validity. For example, for circular binaries, $\dot{\xi_{\phi}}$ agrees with equal-mass NR calculations to within a few percent at $\xi_{\phi} = 0.1$ (see Fig.~13 of Ref.~\cite{boyle-etal-Efluxcomparison}). We therefore expect our low-eccentricity expressions to remain accurate to comparable frequencies.

Independently of the approach in Sec.~III of Ref.~\cite{DGI}, the constraint \eqref{eq:quasikepapprox} can be derived by specifying a condition for the slowness of the periastron advance rate. Recall that the angular frequency $\omega_p \equiv d\Phi/dt$ of the periastron angle $\Phi=\varpi$ is given by $\omega_p = \omega_{\phi} - \omega_r$, where $\omega_{\phi} = \langle \dot{\phi} \rangle$ is the orbit averaged azimuthal orbit frequency and $\omega_r = n$ is the radial orbital frequency (mean motion). Since $\omega_{\phi} = (1+k) \omega_r$, $\omega_p = k \omega_r$.  Let us phrase our constraint on the ``slowness'' of the periastron advance rate in the following terms: we require that the periastron advance angle not exceed one cycle ($\Delta \Phi < 2\pi$) in $N$ radial orbital periods. This implies the constraint $k < 1/N$ or, using the leading-order expression for $k$ [cf., Eq.~\eqref{eq:k3pn}],
\be
\label{eq:quasikepapprox-v2}
\frac{\xi}{(1-e_t^2)^{3/2}} < \left( \frac{1}{3N} \right)^{3/2}.
\ee
One may then attempt to set a constraint by choosing an appropriate value for $N$ that assures that $k$ is sufficiently small. For example, the criterion in Ref.~\cite{DGI} for ``slowness'' \eqref{eq:quasikepapprox} corresponds to $N>16$ radial cycles for every cycle of the periastron [i.e., using $\alpha \rightarrow (3N)^{-3/2}$ and $\alpha =0.003$]. Zoom-whirl behavior corresponds to the condition $N<1$ (i.e., $k<1$ is required to forbid zoom-whirl orbits). However, ``sufficiently small'' is an arbitrary criterion on which reasonable people can differ. (E.g., is $N>1$ sufficient? Or $N>10$? Or $N>100$?) Unfortunately the precise choice is important as the upper bound on the frequency of validity depends sensitively on the chosen value of $N$ [or equivalently, the chosen $\alpha$ in Eq.~\eqref{eq:f-constraint2}]. 

We argue that this is not the appropriate way to determine the constraint. A better constraint comes from requiring that a specified PN quantity agrees with an exact (or at least ``better'') calculation to within a specified accuracy (e.g., within $\sim 1\%$ to $10\%$).\footnote{This accuracy level and the particular quantity that one chooses to constrain are also subjective elements, but we argue that the approach here provides a more natural way to set an appropriate constraint on the domain of validity of the quasi-Keplerian formalism.} As a proxy for an ``exact'' calculation, we use the relationship between the periastron advance parameter and orbital frequency given in Eq.~(7) of \cite{letiec-PRL2011}\footnote{Note that the notation here is related to that in Ref.~\cite{letiec-PRL2011} via $k=K-1$. That reference also uses $\nu$ for the reduced mass ratio ($\eta$ here).} 
\be
\label{eq:kGSFeta}
k_{\rm GSF}^{\eta} = \frac{1}{\sqrt{1-6 x}} \left[ 1 - \frac{\eta}{2} \frac{\rho(x)}{(1-6 x)} + O(\eta^2) \right] -1,
\ee
where 
\be
\rho(x) = \frac{14 x^2 (1+ 12.9906 x)}{1+4.57724 x- 10.3124 x^2} 
\ee
and $x=\xi_{\phi}^{2/3} = (M \omega_{\phi})^{2/3} = (\pi M f)^{2/3}$ is the standard circular PN expansion parameter.
This equation arises from an EOB calculation of $k$ in the $e_t \rightarrow 0$ limit \cite{barack-damour-sago_periastron}. The resulting expression automatically incorporates the Schwarzschild $e_t \rightarrow 0$ result and additionally incorporates an $O(q)$ correction to the Schwarzschild limit, where $q= m_1/m_2 \leq 1$ is the binary mass ratio. This correction arises from the first-order GSF and is determined by fitting the function $\rho(x)$ to accurate GSF calculations. The resulting linear-in-$q$ expression ($k_{\rm GSF}^{q}$) can then be ``improved'' by replacing $q \rightarrow \eta$, resulting in Eq.~\eqref{eq:kGSFeta} above. This replacement of the mass ratio $q$ with $\eta$ has the following remarkable effect: while the $q$ dependent expression $k_{\rm GSF}^{q}$ agrees with low-eccentricity NR simulations at small $q$ (as expected), the expression $k_{\rm GSF}^{\eta}$ agrees exceptionally well with NR simulations at {\it all} mass ratios, essentially lying within (or close to) the error bars of the NR simulations for the frequency range analyzed (see Figs.~1\mbox{--}3 of Ref.~\cite{letiec-PRL2011}). The figures there also clearly indicate good agreement between NR calculations of $k$ and the 3PN result we use here (e.g., less than $1\%$ error up to $\xi_{\phi} = 0.03$). 

\begin{figure}[t]
\includegraphics[angle=0, width=0.45\textwidth]{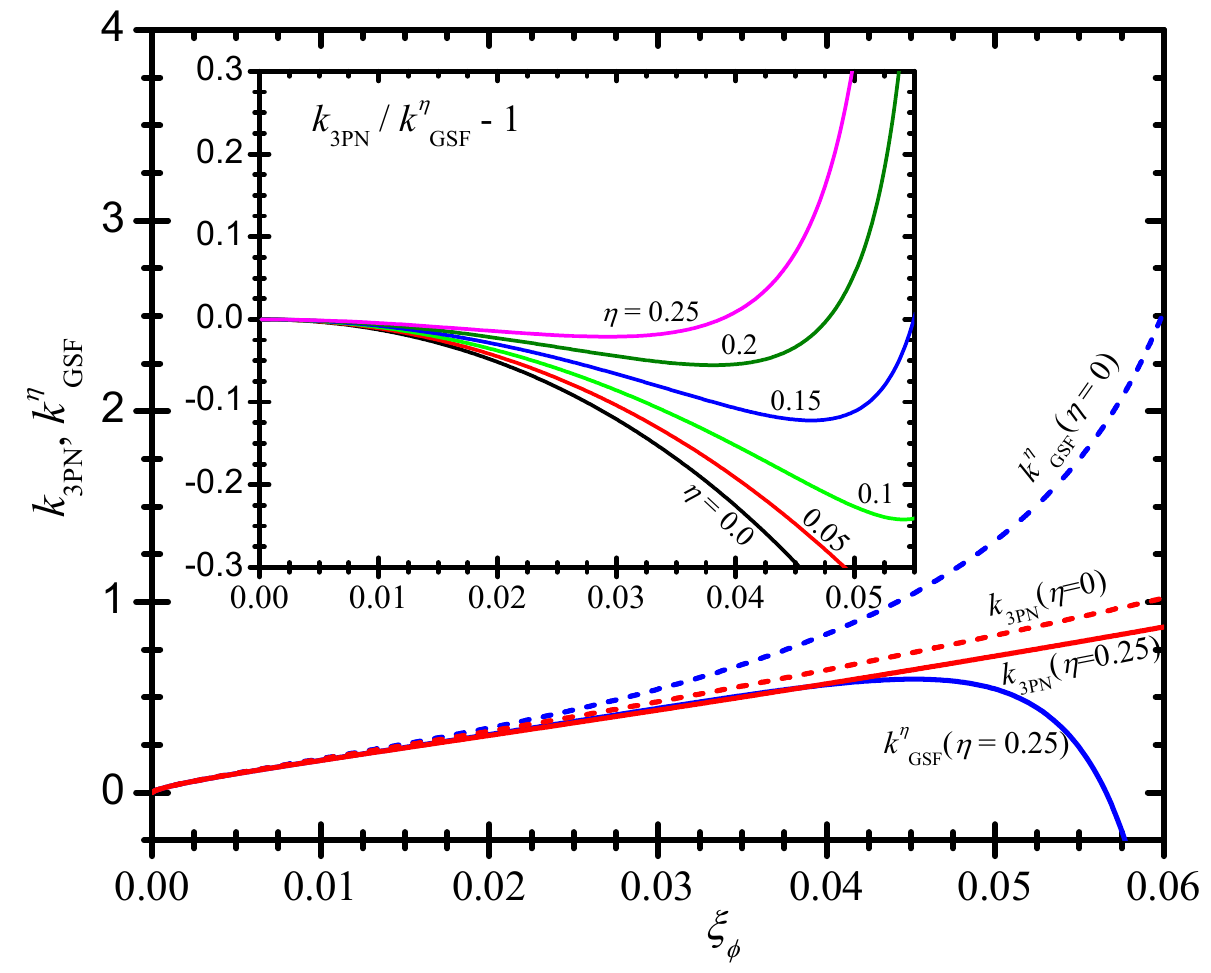}
\caption{\label{fig:k} Comparison of 3PN and ``exact'' calculations of the periastron advance constant $k$. We plot $k_{\rm GSF}^{\eta}$ [Eq.~\eqref{eq:kGSFeta}] and $k(e_t \rightarrow 0)$ [Eq.~\eqref{eq:ket0}] vs $\xi_{\phi} = M \omega_{\phi}$. The inset shows the fractional error between the quantities for different values of the reduced mass ratio $\eta$. As expected, agreement is better for comparable-mass binaries.}
\end{figure}

Considering the importance of understanding the realm of validity of the quasi-Keplerian formalism (and hence the results of this study), we further quantify the bounds on our PN expressions. Since it exhibits excellent agreement with NR results for all mass ratios, we take Eq.~\eqref{eq:kGSFeta} as our ``exact'' result and compare it with the 3PN expression for $k$ expressed in terms of $x=\xi_{\phi}^{2/3}$ [Eqs.~\eqref{eq:k3pnxiphi}] and specialized to the $e_t \rightarrow 0 $ limit. This reduces to
\be
\label{eq:ket0}
k (e_t\rightarrow 0) = 3 \xi_{\phi}^{2/3} +  ( 13.5 -7 \eta)\xi_{\phi}^{4/3} +  (67.5 -124.3137 \eta + 7 \eta^2)\xi_{\phi}^2.
\ee
In Fig.~\ref{fig:k} we see the behavior of these two functions as well as the fractional error between the 3PN formula and the ``exact'' result. The 3PN periastron advance constant shows good agreement with Eq.~\eqref{eq:kGSFeta} for nearly all mass ratios. Depending on the acceptable level of accuracy required and the mass ratio of the system, Fig.~\ref{fig:k} determines the frequency where the required accuracy threshold is violated. For example, if we require that the 3PN expression agrees with $k_{\rm GSF}^{\eta}$ to within $10\%$, then this implies that $\xi_{\phi}< 0.0274$ for $\eta =0$. This is a significantly larger upper limit than the $\approx 0.003$ value required by Eq.~\eqref{eq:quasikepapprox}. For $\eta=0$ and $\xi_{\phi}=0.003$, the fractional error from $k_{\rm GSF}^{\eta}$ is $0.1 \%$, a value that we argue is unnecessarily conservative. Furthermore, for larger mass ratios, the 3PN expression agrees even more closely with the NR/GSF result. A $<10\%$ error is achieved for $\eta=0.25$ so long as $\xi_{\phi} < 0.0458$. This upper bound increases to $\xi_{\phi} < 0.0511$ for $\eta=0.2$.  

In summary, we believe that the constraint \eqref{eq:quasikepapprox} suggested by Ref.~\cite{DGI} is too conservative, as it implies that the quasi-Keplerian formalism is not applicable to binaries in the frequency band of ground-based interferometers. Instead, more recent comparisons with NR and GSF simulations (summarized in Ref.~\cite{letiec-PRL2011} and discussed above) suggest that a more appropriate constraint for comparable-mass binaries is $\xi_{\phi} \lesssim 0.04$ or 
\be
\label{eq:f-constraint3}
f \lesssim \frac{\alpha}{\pi M} = 2585 \, {\rm Hz} \left( \frac{\alpha}{0.04} \right) \left( \frac{1 M_{\odot}}{M} \right).
\ee
This limit is nearly $60\%$ of the frequency of the Schwarzschild ISCO. This implies that our formalism should be accurate up to $f\approx 920$ Hz for NS/NS binaries or $f \approx 260$ Hz for a $M=10 M_{\odot}$ BH/BH binary.
\section{\label{sec:evolvebinaries}quasi-Keplerian phasing for evolving binaries}
In the previous section, we described the quasi-Keplerian formalism, which provides a parametric solution to the conservative pieces of the PN equations of motion. This analytic solution follows from the fact that conservative quasielliptical orbits admit four constants of motion: the principal (intrinsic) constants $n$ and $e_t$ which determine the shape of the orbit, plus two positional (extrinsic) constants (here taken to be $c_l$ and $c_{\lambda}$) that determine the orientation of the orbit and the initial binary configuration.
We now consider the inclusion of radiation reaction, both generally \cite{DGI,quasikepphasing35PN} and in the low-eccentricity limit. When dissipative terms are included, these four constants will generally evolve with time. A scheme for evolving the constants of the motion for nonspinning eccentric binaries has been detailed in Refs.~\cite{DGI,quasikepphasing35PN}. Here, we briefly summarize their results and then specialize them to low-eccentricity orbits.

As in the conservative case, the essential problem is to determine the functions $r(t)$, $\phi(t)$ and their derivatives as solutions to the full PN equations of motion. Rather than numerically solving the PN equations of motion (at say 3.5PN order), Ref.~\cite{DGI} employs a method of \emph{variation of constants} in which the functional form of the 3PN conservative solution [Eqs.~(\ref{eq:quasikeeqns})] is used as a leading-order solution. The 2.5PN and 3.5PN radiative pieces of the equations of motion then act as a perturbation that causes the constants of the motion in Eqs.~\eqref{eq:quasikeeqns} to vary with time. Specifically, the mean motion and time eccentricity now vary with time, $n=n(t)$, $e_t=e_t(t)$, and the angles $l$ and $\lambda$ are now given by
\begin{subequations}
\label{eq:positionangles}
\begin{align}
l & \equiv \int_{t_0}^{t} n(t') \, dt' + c_{l}(t)\;,
\label{eq:kepl2} \\
\label{eq:keplambda2}
\lambda & \equiv \int_{t_0}^{t} [1+k(t')] n(t') \, dt' + c_{\lambda}(t) \;,
\end{align}
\end{subequations}
where the positional constants $c_l=c_l(t)$ and $c_{\lambda}=c_{\lambda}(t)$ also vary with time.
Instead of solving a first-order system of differential equations involving $r(t)$, $\phi(t)$, $\dot{r}(t)$, and $\dot{\phi}(t)$, a new first-order system is solved in which the dynamical variables are $c_{\alpha}(t) = [c_1(t), \, c_2(t), \, c_l(t), \, c_{\lambda}(t)]$. In this system the ``constants'' $c_1$ and $c_2$ could be the energy and angular momentum of the binary, but---as indicated earlier---they are more conveniently chosen to be the mean motion $n$ and the time eccentricity $e_t$ (which can be related to the energy and angular momentum). 

To arrive at a first-order system for the $c_{\alpha}$, one proceeds as follows (see Ref.~\cite{DGI} for details): Begin with the PN equations of motion in first-order form,
\bs
\label{eq:fulleqnofmotion}
\begin{align}
\dot{\bm x} &= {\bm v} \, , \\
\dot{\bm v} &= {\bm {\mathcal A}}_0({\bm x},{\bm v}) + {\bm {\mathcal A}}'({\bm x},{\bm v}) \, ,
\end{align}
\es
where the motion is planar and ${\bm {\mathcal A}}_0$ and ${\bm {\mathcal A}}'$ represent the conservative and dissipative pieces of the equations of motion (respectively). These equations are first solved neglecting the ${\bm {\mathcal A}}'$ term, resulting in the parametric quasi-Keplerian solution described in Sec.~\ref{quasikep2}, ${\bm x}={\bm x}_0(t;c_{\alpha})$. The solution to the full equations (including the dissipative term  ${\bm {\mathcal A}}'$) is written such that it has the same functional form as the conservative system, but with the constants $c_{\alpha}$ now as functions of time, ${\bm x}={\bm x}_0[t;c_{\alpha}(t)]$. This exact form for the solution, combined with the full equations of motion \eqref{eq:fulleqnofmotion}, yields a new first-order system for the $c_{\alpha}(t)$,
\be
\frac{d c_{\alpha}}{dt} = F_{\alpha}(l,c_{\beta}) \,; \;\;\;\;\; \alpha, \beta = 1, 2, l , \lambda,
\ee
where $F_{\alpha}$ is linear in ${\bm {\mathcal A}}'$. This is then recast in the form
\be
\frac{d c_{\alpha}}{dl} = G_{\alpha}(l,c_a) \,; \;\;\;\;\; \alpha = 1, 2, l , \lambda; \; a = 1,2 \,,
\ee
where $G_{\alpha} \propto {\bm {\mathcal A}}'$ and is periodic in $l$. Since $G_{\alpha}$ contains both fast, periodic oscillations as well as slowly varying pieces [since radiation reaction causes a slow variation of the $c_a(t)$], the solution is split into a slowly varying piece $\bar{c}_{\alpha}(l)$ and a rapidly varying piece $\tilde{c}_{\alpha}(l)$,
\be
c_{\alpha}(l) = \bar{c}_{\alpha}(l) + \tilde{c}_{\alpha}(l) \;.
\label{eq:c-split}
\ee
For sufficiently long times, the rapidly oscillating terms $\tilde{c}_{\alpha}(l)$ will always be smaller than the slowly varying ones $\bar{c}_{\alpha}(l)$.  Using this splitting, Refs.~\cite{DGI,quasikepphasing35PN} then show how to solve for the quantities $\bar{n}, \, \bar{e}_t, \, \bar{c}_l, \, \bar{c}_{\lambda}, \, \tilde{n}, \, \tilde{e}_t, \, \tilde{c}_l, \, \tilde{c}_{\lambda}$, which are expressed as functions of $l$, $u$, or $t$. The differential equations for each piece of the $c_{\alpha}$ have the form
\bs
\label{eq:dcdleqns}
\begin{align}
\label{eq:dcdlsecular}
\frac{d \bar{c}_{\alpha}}{dl} &= \bar{G}_{\alpha}(\bar{c}_{\alpha}) \;, \\
\label{eq:dcdlperiodic}
\frac{d \tilde{c}_{\alpha}}{dl} &= \tilde{G}_{\alpha}(\bar{c}_{\alpha}) \;,
\end{align}
\es
where $\bar{G}_{\alpha}$ and $\tilde{G}_{\alpha}$ are the orbit-averaged and oscillatory pieces of $G_{\alpha}$.
The time evolution of the angles $l(t)$ and $\lambda(t)$ can be similarly split into secular and oscillatory pieces [by virtue of their definition in Eqs.~\eqref{eq:positionangles}]:
\bs
\begin{align}
l(t) &= \bar{l}(t) + \tilde{l}[l; \bar{c}_a(t)] \;, \\
\lambda(t) &= \bar{\lambda}(t) + \tilde{\lambda}[l; \bar{c}_a(t)] \,.
\end{align}
\es
The next subsections discuss separately the solutions to the oscillatory and secular equations in \eqref{eq:dcdleqns}.
\subsection{\label{sec:periodicsolns}Periodic variation of the constants}
As shown in Ref.~\cite{DGI}, Eq.~\eqref{eq:dcdlperiodic} can be integrated analytically, yielding closed-form expressions for $\tilde{n}(u,\bar{n},\bar{e}_t)$, $\tilde{e}_t(u,\bar{n},\bar{e}_t)$, $\tilde{c}_l(u,\bar{n},\bar{e}_t)$, and $\tilde{c}_{\lambda}(u,\bar{n},\bar{e}_t)$. These are then used in constructing expressions for $\tilde{l}(u,\bar{n},\bar{e}_t)$ and $\tilde{\lambda}(u,\bar{n},\bar{e}_t)$. The full expressions are given by Eqs.~(64) and (67) of Ref.~\cite{DGI} (at 2PN order in ADM gauge) and Eqs.~(36) and (40) of Ref.~\cite{quasikepphasing35PN} (at 3PN order in harmonic gauge). Those expressions are seen to have the following form when 3.5PN-order reactive effects are included,
\be
\tilde{C}_{\gamma} = [\tilde{\xi}/\bar{\xi}, \tilde{e}_t, \tilde{c}_l, \tilde{c}_{\lambda}, \tilde{l}, \tilde{\lambda}] = \eta {\bar{\xi}}^{5/3} \left[ f^{(2.5)}_{\gamma}[u(\bar{l});\bar{e}_t] + {\bar{\xi}}^{2/3} f^{(3.5)}_{\gamma}[u(\bar{l}); \bar{e}_t] \right] \;,
\ee
where the $f_{\gamma}$ label the various expressions listed in Refs.~\cite{DGI,quasikepphasing35PN}, and the index $\gamma$ takes on labels corresponding to the six variables listed on the left-hand side of the equation. The $\tilde{C}_{\gamma}$ are periodic functions of $l$ and have the leading-order scaling $\tilde{C}_{\gamma} \sim O(\eta \bar{\xi}^{5/3})$. This indicates that these terms will generally be small, especially in comparison with the secular pieces that we consider in the next section.

In the low-eccentricity limit, we can simplify the expressions given in Ref.~\cite{quasikepphasing35PN} by using Eq.~\eqref{eq:lexpand} to express $u$ in terms of $l$. At leading PN order, the constants $\tilde{C}_{\gamma}$ reduce to the following when we expand about $\bar{e}_t=0$:
\bs
\label{eq:Cperiodic}
\begin{align}
\label{eq:xitilde}
\tilde{\xi}(\bar{l};\bar{\xi},\bar{e}_t) &= \eta \bar{\xi}^{8/3} \bar{e}_t \left[ 120 \sin \bar{l} + \frac{718}{5} \bar{e}_t \sin 2\bar{l} +  \left( \frac{2269}{5} \sin \bar{l} + \frac{3011}{15} \sin 3\bar{l} \right)\bar{e}_t^2 + O(\bar{e}_t^3)\right] + O(\bar{\xi}^{10/3}) \,,
\\
\label{eq:ettilde}
\tilde{e}_t(\bar{l};\bar{\xi},\bar{e}_t) &= - \eta \bar{\xi}^{5/3} \left[ \frac{64}{5} \sin \bar{l} + \frac{352}{15} \bar{e}_t \sin 2\bar{l} +  \left( \frac{1138}{15} \sin \bar{l} + \frac{358}{9} \sin 3\bar{l} \right)\bar{e}_t^2 + O(\bar{e}_t^3)\right] + O(\bar{\xi}^{7/3}) \,,
\\
\label{eq:clambdatilde}
\tilde{c}_{\lambda}(\bar{l};\bar{\xi},\bar{e}_t) &= \eta \bar{\xi}^{5/3} \bar{e}_t \left[ \frac{64}{3} \cos \bar{l} +  \left(\cos 2\bar{l} - \frac{1}{3} \right)32 \bar{e}_t +  \left( \frac{81}{5} \cos \bar{l} + \frac{2131}{45} \cos 3\bar{l} \right)\bar{e}_t^2 + O(\bar{e}_t^3)\right] + O(\bar{\xi}^{7/3}) \,,
\end{align}
\begin{multline}
\label{eq:cltilde}
\tilde{c}_l(\bar{l};\bar{\xi},\bar{e}_t) = \frac{\eta \bar{\xi}^{5/3}}{\bar{e}_t} \left[ - \frac{64}{5} \cos \bar{l} +  \left( \frac{32}{5} - \frac{352}{15} \cos 2\bar{l} \right)\bar{e}_t +  \left( \frac{146}{15} \cos \bar{l} - \frac{358}{9} \cos 3\bar{l} \right)\bar{e}_t^2  \right. \\ \left. +  \left( \frac{91}{15} + \frac{383}{15} \cos 2\bar{l} - \frac{1289}{20} \cos 4\bar{l} \right)e_t^3 + O(\bar{e}_t^4)\right] + O(\bar{\xi}^{7/3}) \;,
\end{multline}
\begin{multline}
\label{eq:ltilde}
\tilde{l}(\bar{l};\bar{\xi},\bar{e}_t) = - \frac{\eta \bar{\xi}^{5/3}}{\bar{e}_t} \left[ \frac{64}{5} \cos \bar{l} +  (107 + 352 \cos 2\bar{l} )\frac{1}{15} \bar{e}_t  +  \left( \frac{1654}{15} \cos \bar{l}  + \frac{358}{9} \cos 3\bar{l} \right)\bar{e}_t^2 \right. \\ \left. +  \left( -\frac{644}{15} + \frac{694}{15} \cos 2\bar{l}  + \frac{1289}{20} \cos 4\bar{l} \right)\bar{e}_t^3  + O(\bar{e}_t^4)\right] + O(\bar{\xi}^{7/3}) \;,
\end{multline}
\be
\label{eq:lambdatilde}
\tilde{\lambda}(\bar{l};\bar{\xi},\bar{e}_t) = - \eta \bar{\xi}^{5/3} \left[ \frac{203}{15} +   \frac{296}{3} \bar{e}_t  \cos \bar{l}   +  \left( -\frac{131}{5}+ \frac{199}{5} \cos 2\bar{l} \right)\bar{e}_t^2 + O(\bar{e}_t^3)\right] + O(\bar{\xi}^{7/3}) \;.
\ee
\es
To derive the last two equations, we had to evaluate the integrals appearing in Eqs.~(40a) and (40b) of Ref.~\cite{quasikepphasing35PN}. This was done by first expanding the integrands in the small-$e_t$ limit and then computing the indefinite integral, neglecting the constant of integration. Equation \eqref{eq:lexpand} was then substituted, and the result was expanded in the small-$e_t$ limit. Note also that at leading order in $\bar{\xi}$, $\tilde{\lambda}=\tilde{l} - \tilde{c}_l + \tilde{c}_{\lambda}$. The expressions \eqref{eq:Cperiodic} are listed for completeness. As we will discuss in more detail in Sec.~\ref{sec:oscil}, they will be negligible for our purposes.
\subsection{\label{sec:xiphi-define}Secular variation of the constants}
We now focus on computing the secular evolution of the constants of the motion (which will eventually lead to the main results of this paper). The primary expressions that are needed to compute the various PN approximants are the differential equations governing the secular time evolution of $\bar{n} = \bar{\xi}/M$ and $\bar{e}_t$. (The positional constants of the motion are found to have no secular variations, i.e., $\dot{\bar{c}}_l = \dot{\bar{c}}_{\lambda}=0$ \cite{DGI}.) These are given to 2PN order in ADM or harmonic gauge in Refs.~\cite{DGI,quasikepphasing35PN}; the harmonic gauge versions are reproduced in Appendix \ref{app:ndot-edot}. Here, we are more interested in the pair $(\bar{\omega}_{\phi} = \bar{\xi}_{\phi}/M, \bar{e}_t)$ for reasons discussed above. Expressions for $\dot{\bar{\omega}}_{\phi}$ and $\dot{\bar{e}}_{t}$ have been computed to 3PN order in ADM gauge in Ref.~\cite{arun-etal-eccentric-orbitalelements-PRD2009}.\footnote{Reference \cite{arun-etal-eccentric-orbitalelements-PRD2009} includes modified harmonic gauge expressions in an appendix, but Eqs.~(C10)\mbox{--}(C11) there were found to contain an error. This is addressed in a forthcoming erratum. Also note that there are important notational differences between this paper and Ref.~\cite{arun-etal-eccentric-orbitalelements-PRD2009}. Specifically, Ref.~\cite{arun-etal-eccentric-orbitalelements-PRD2009} uses $\zeta \equiv Mn$ in place of our $\xi$. Their $\omega$ is labeled $\omega_{\phi}$ here. They also do not use bars to denote orbit-averaged quantities.} Since we work in modified harmonic (MH) gauge here, we must convert those results from ADM to MH gauge. The explicit relation between the two gauges is given by Eq.~(8.21) of Ref.~\cite{arun-eccentricEflux3PN},
\be
\label{eq:ADMtoMH}
e_t^{\rm ADM} = e_t^{\text{MH}}\left\{1+\left(\frac{1}{4}+\frac{17}{4}\eta\right)\frac{\xi_{\phi}^{4/3}}{1-e_t^2}+\left[\frac{1}{2}+\left(\frac{16739}{1680}-\frac{21}{16} \pi^2 \right)\eta 
-\frac{83}{24} \eta^2+\left(\frac{1}{2}+\frac{249}{16}\eta-\frac{241}{24}\eta^2\right)e_t^2\right]\frac{\xi_{\phi}^2}{\left(1-e_t^2\right)^2}\right\} ,
\ee
where $e_t=e_t^{\rm MH}$ on the right-hand side. Note that the difference between ADM and MH gauges enters at 2PN and higher orders. For reference, we also include the inverse transformation,
\be
\label{eq:MHtoADM}
e_t^{\rm MH} = e^{\rm ADM}\left\{1-\left(\frac{1}{4}+\frac{17}{4}\eta\right)\frac{\xi_{\phi}^{4/3}}{(1-e_t^2)} -\left[\frac{1}{2}+\left(\frac{16739}{1680}-\frac{21}{16} \pi ^2\right)
\eta -\frac{83 }{24}\eta ^2+ \left(\frac{1}{2}+\frac{249 }{16} \eta-\frac{241}{24} \eta ^2\right)e_t^2\right]\frac{\xi_{\phi}^2}{\left(1-e_t^2\right)^2} \right\},
\ee 
where $e_t = e_t^{\rm ADM}$ on the right-hand side. (Everywhere else in this document, $e_t=e_t^{\rm MH}$.) 

Arun \emph{et al}.~\cite{arun-etal-eccentric-orbitalelements-PRD2009} provide expressions for $d\omega_{\phi}/dt$ and $de_t/dt$ for arbitrary eccentricity ($e_t<1$).\footnote{Note that the tail contributions to the expressions for $d\omega_{\phi}/dt$ and $de_t/dt$ in Ref.~\cite{arun-etal-eccentric-orbitalelements-PRD2009} are expressed as infinite series and require careful consideration when used in actual computations. This is discussed further in Sec.~\ref{sec:numerical_comp} below. In the low-eccentricity limit, these tail terms can be expanded as a power series in $e_t$.} They also specialize their results to leading order in $e_t$ and in ADM gauge [see their Eqs.~(7.6c) and (7.6e)]. Using Eq.~\eqref{eq:ADMtoMH} to convert their Eq.~(7.6c) to harmonic gauge gives
\begin{multline}
\label{eq:dxidt-MH-lowe}
\frac{d\xi_{\phi}}{dt} = M \frac{d\omega_{\phi}}{dt} = \frac{96 \eta  \xi_{\phi} ^{11/3}}{5 M} \left(1-\left(\frac{743}{336}+\frac{11}{4}\eta \right) \xi_{\phi} ^{2/3}+4 \pi  \xi_{\phi} +\left(\frac{34103}{18144}+\frac{13661
	 }{2016}\eta+\frac{59 }{18}\eta ^2\right) \xi_{\phi}^{4/3} -\left(\frac{4159}{672}+\frac{189 }{8}\eta\right)\pi
\right. \\ 
 \xi_{\phi}^{5/3}+ \left[\frac{16447322263}{139708800}-\frac{1712
}{105}\gamma_{E}+\frac{16 }{3}\pi ^2  +\left(-\frac{56198689}{217728}+\frac{451}{48} \pi ^2\right) \eta +\frac{541 }{896}\eta ^2-\frac{5605
}{2592}\eta ^3-\frac{856}{105} \ln(16 \xi_{\phi} ^{2/3})\right]\xi_{\phi}^2
\\ 
+e_t^2 \left\{\frac{157}{24}+\left(\frac{713}{112}-\frac{673
	 }{16}\eta\right) \xi_{\phi} ^{2/3}
+\frac{2335 }{48} \pi  \xi_{\phi}+\left(-\frac{479959}{12096}+\frac{80425  }{4032}\eta+\frac{213539 }{1728}\eta ^2\right) \xi_{\phi}
^{4/3}+\left(\frac{7885 }{96}-\frac{27645}{56} \eta\right) \pi \xi_{\phi} ^{5/3}
\right. \\ 
+ \left[\frac{277391496167}{139708800}-\frac{106144 }{315} \gamma_{E}+\frac{992
	}{9}\pi ^2+\left(-\frac{280153957}{120960}+\frac{188231 }{2304}\pi ^2\right) \eta-\frac{73109 }{448}\eta ^2-\frac{6874115 }{31104}\eta ^3+\frac{18832
}{45}\ln2
\right. \\ \left. \left. \left. 
-\frac{234009 }{560}\ln3-\frac{53072}{315} \ln(16 \xi_{\phi} ^{2/3})\right]\xi_{\phi} ^2\right\}\right) \,, 
\end{multline}
where $\gamma_{E}=0.5772156649 \ldots$ is the Euler-Mascheroni constant. 
In the above (and henceforth), we drop overbars where it is clear that we refer to a secular (orbit-averaged) quantity.

To compute $de_t/dt$ in harmonic gauge, one first takes the time derivative of Eq.~\eqref{eq:MHtoADM} and then substitutes Eqs.~(7.6c) and (7.6e) of Ref.~\cite{arun-etal-eccentric-orbitalelements-PRD2009} for $d\xi_{\phi}/dt$ and $de_t^{\rm ADM}/dt$. Next, the gauge transformation in Eq.~\eqref{eq:ADMtoMH} is substituted, and the result is expanded to 3PN order yielding the harmonic gauge expression
\begin{multline}
\label{eq:dedt-MH-lowe}
\frac{de_t}{dt} =  -\frac{304 e_t \eta  \xi_{\phi}^{8/3}}{15 M} \left(1-\left(\frac{2817}{2128}+\frac{1021}{228} \eta \right) \xi_{\phi}^{2/3}+\frac{985}{152}\pi  \xi_{\phi}+\left(-\frac{108197}{38304}+\frac{56407
	 }{4256}\eta+\frac{141 }{19}\eta^2\right) \xi_{\phi}^{4/3}-\left(\frac{55691  }{4256}
\right. \right. \\ \left. 
+\frac{19067}{399}\eta \right) \pi \xi_{\phi} ^{5/3}+ \left[\frac{246060953209}{884822400}-\frac{82283 }{1995}\gamma_{E}+\frac{769 }{57}\pi ^2 +\left(-\frac{613139897}{2298240}+\frac{22345 }{3648}\pi^2\right) \eta -\frac{1046329}{51072}\eta ^2
\right. \\ \left.
-\frac{305005}{49248}\eta^3+\frac{4601 }{105}\ln2-\frac{234009}{5320} \ln3 -\frac{82283
}{3990}\ln(16 \xi_{\phi}^{2/3})\right]\xi_{\phi}
^2+e_t^2 \left\{\frac{881}{304}+\left(\frac{40115}{4256}-\frac{51847
}{1824}\eta \right) \xi_{\phi} ^{2/3}
\right. \\ 
+\frac{21729  }{608}\pi  \xi_{\phi}+\left(-\frac{1368625}{51072}
-\frac{288209  }{17024}\eta+\frac{274515 }{2432}\eta ^2\right)
\xi_{\phi} ^{4/3}+\left(\frac{286789 }{3584}-\frac{7810371 }{17024}\eta \right) \pi \xi_{\phi} ^{5/3}+ \left[\frac{1316189396351}{589881600}
\right. \\ 
-\frac{1500461
}{3990}\gamma_{E}+\frac{14023 }{114}\pi ^2+\left(-\frac{5882746699}{4596480}+\frac{46453 }{1536}\pi ^2\right) \eta -\frac{554719 }{4788}\eta ^2-\frac{100330729
}{393984}\eta ^3
\\ \left. \left. \left.
-\frac{3813587}{3990} \ln2+\frac{6318243 }{21280}\ln3-\frac{1500461}{7980} \ln(16 \xi_{\phi} ^{2/3})\right]\xi_{\phi} ^2\right\}\right). 
\end{multline}

For reference and later use in Sec.~\ref{sec:numerical_comp}, we provide in Appendix \ref{app:ndot-edot}  2PN-order equations for $dn/dt$ and $de_t/dt$ in terms of $n$ and $e_t$ and valid for arbitrary $e_t<1$.
\subsection{\label{sec:e_xi}Analytic eccentricity evolution as a function of frequency}
At Newtonian (0PN) order, the equations for $d\xi_{\phi}/dt$ and $de_t/dt$ can be analytically solved for arbitrary $e_t$ to determine $\xi_{\phi}(e_t)$. Computing $d\xi_{\phi}/de_t = \dot{\xi}_{\phi}/\dot{e}_t$ and integrating using the initial condition that $\xi_{\phi}=\xi_{\phi,0}$ when $e_t=e_0$ gives
\be
\label{eq:xi_et_N}
\frac{\xi_{\phi}(e_t)}{\xi_{\phi,0}} = \left( \frac{e_0}{e_t} \right)^{18/19} \left( \frac{1-e_t^2}{1-e_0^2} \right)^{3/2} \left( \frac{304+121e_0^2}{304+121e_t^2} \right)^{1305/2299}.
\ee
An analogous result in terms of the semimajor axis was first derived by Peters \cite{peters}.

At higher PN orders, the differential equation for $d\xi_{\phi}/de_t$ is not separable, so an exact solution valid for arbitrary eccentricities is not easily found. However, an analytic result can be found if we only include the leading-order eccentricity terms. Expanding the low-eccentricity limit of $de_t/d\xi_{\phi}$ in $\xi_{\phi}$ gives
\begin{multline}
\label{eq:dedxi-lowe}
\frac{de_t}{d\xi_{\phi}} = -\frac{19}{18}\frac{e_t}{\xi_{\phi}}\left\{1+\left(\frac{2833}{3192}-\frac{197  }{114}\eta\right) \xi_{\phi} ^{2/3}+\frac{377  }{152}\pi  \xi_{\phi}+\left(-\frac{1392851}{508032}+\frac{32537 }{6384}\eta-\frac{833
	}{1368}\eta ^2\right) \xi_{\phi} ^{4/3}
\right. \\
+\left(-\frac{253409  }{51072}-\frac{133157 }{12768} \eta\right)\pi \xi_{\phi} ^{5/3}+ \left[\frac{27226918334431}{178380195840}-\frac{3317
}{133}\gamma_{E}-\frac{67 }{38}\pi ^2
\right. \\ \left. \left.
+\left(-\frac{26105879}{3386880}-\frac{3977 }{1216}\pi ^2\right) \eta +\frac{58057 }{153216}\eta ^2-\frac{25
	}{608}\eta ^3+\frac{4601}{105} \ln2-\frac{234009 }{5320}\ln3-\frac{3317}{266} \ln(16 \xi_{\phi} ^{2/3})\right]\xi_{\phi} ^2\right\}.
\end{multline}
Separating variables and integrating gives
\begin{multline}
\label{eq:et_xi_2PN}
e_t = C_1 \xi_{\phi}^{-19/18} \exp\left\{  \left( -\frac{2833}{2016}+ \frac{197}{72}\eta \right)\xi_{\phi}^{2/3} - \frac{377}{144} \pi \xi_{\phi} +  \left(\frac{26464169}{12192768}-\frac{32537}{8064}\eta+\frac{833}{1728}\eta^2 \right) \xi_{\phi}^{4/3}
\right. \\ 
+  \left(\frac{253409}{80640} + \frac{133157}{20160} \eta \right)\pi \xi_{\phi}^{5/3} +  \left[ -\frac{27968380877791}{377983528960}+\frac{3317}{252}\gamma_E+\frac{67}{72}\pi^2+\left( \frac{496011701}{121927680}+\frac{3977}{2304}\pi^2\right) \eta 
\right. \\ \left. \left.
-\frac{58057}{290304}\eta^2+\frac{25}{1152}\eta^3- \frac{87419}{3780}\ln2+\frac{26001}{1120}\ln3 +\frac{3317}{504}\ln(16 \xi_{\phi}^{2/3}) \right]\xi_{\phi}^2\right\}.
\end{multline}
Expanding the above equation in $\xi_{\phi}$ then yields 
\bs
\label{eq:etofxi}
\be
e_t = e_0 \left( \frac{\xi_{\phi,0}}{\xi_{\phi}} \right)^{19/18} \frac{{\mathcal E}(\xi_{\phi})}{{\mathcal E}(\xi_{\phi,0})},
\ee
where
\begin{multline}
\label{eq:E}
{\mathcal E}(\xi_{\phi}) = \left\{1+\left(-\frac{2833}{2016}+\frac{197  }{72}\eta\right) \xi_{\phi} ^{2/3}-\frac{377  }{144}\pi  \xi_{\phi}+\left(\frac{77006005}{24385536}-\frac{1143767
	 }{145152}\eta+\frac{43807 }{10368}\eta ^2\right) \xi_{\phi} ^{4/3}
\right. \\
+\left(\frac{9901567 }{1451520}-\frac{202589   }{362880}\eta\right)\pi \xi_{\phi} ^{5/3}+ \left[-\frac{33320661414619}{386266890240}+\frac{3317 }{252}\gamma_{E}+\frac{180721 }{41472}\pi ^2
+\left(\frac{161339510737}{8778792960}+\frac{3977
}{2304}\pi ^2\right) \eta
\right. \\ \left. \left.
 -\frac{359037739 }{20901888}\eta ^2+\frac{10647791 }{2239488}\eta ^3-\frac{87419 }{3780}\ln2+\frac{26001}{1120} \ln3 +\frac{3317}{504}
\ln(16 \xi_{\phi} ^{2/3})\right]\xi_{\phi}
^2\right\} \,.
\end{multline}
\es
The constant $C_1$ was determined by the initial condition $e_t(\xi_{\phi,0})=e_0$.

To gauge the accuracy of the low-eccentricity approximation, we can compare the 0PN expression \eqref{eq:xi_et_N} with its low-eccentricity version, $\xi_{\phi}/\xi_{\phi,0} = (e_0/e_t)^{18/19}$. For $e_0\lesssim 0.2 \, (0.1)$, these agree to within $7\% \, (2\%)$. The PN corrections have the effect of decreasing the eccentricity more rapidly as the frequency increases.
\subsection{\label{sec:expliciteqns}Explicit evolution equations as a function of frequency or time}
Using the results of the previous subsection, we can determine $d\xi_{\phi}/dt$ explicitly as a function of $\xi_{\phi}$ only (eliminating the frequency dependence in $e_t$) and then solve for  $\xi_{\phi}$ and $e_t$ explicitly as functions of time. This  also allows us to determine the evolution of the phase variables $\lambda$ and $l$ as functions of time or the frequency variable $\xi_{\phi}$.

Substituting Eq.~\eqref{eq:etofxi} for $\bar{e}_t(\xi_{\phi})$ into Eq.~\eqref{eq:dxidt-MH-lowe} and series expanding yields\footnote{When performing the series expansions, we introduce a PN expansion parameter $\varepsilon \sim 1/c$ via $\xi_{\phi}\rightarrow \varepsilon^3 \xi_{\phi}$ and $\xi_{\phi,0} \rightarrow \varepsilon^3 \xi_{\phi,0}$. This parameter $\varepsilon$ is set to $1$ at the end of the calculation.}
\begin{multline}
\label{eq:dxidtexpand2}
\frac{d\xi_{\phi}}{dt} = \frac{96 \eta  \xi_{\phi}^{11/3}}{5M} \left(1-\left(\frac{743}{336}+\frac{11  }{4}\eta\right) \xi_{\phi} ^{2/3}+4 \pi  \xi_{\phi} +\left(\frac{34103}{18144}+\frac{13661
	 }{2016}\eta+\frac{59 }{18}\eta ^2\right) \xi_{\phi} ^{4/3} 
- \left(\frac{4159}{672}+\frac{189 }{8} \eta\right)\pi \xi_{\phi} ^{5/3}
\right. \\
+ \left[\frac{16447322263}{139708800}-\frac{1712
}{105}\gamma_{E}+\frac{16 }{3}\pi ^2
+\left(-\frac{56198689}{217728}  +\frac{451 }{48}\pi ^2\right) \eta +\frac{541 }{896}\eta ^2-\frac{5605
	}{2592}\eta ^3-\frac{856}{105} \ln(16 \xi_{\phi} ^{2/3})\right]\xi_{\phi} ^2
\\  
+\frac{157}{24} e_0^2 \left(\frac{\xi_{\phi,0}}{\xi_{\phi} }\right)^{19/9}
\left\{1- \left(\frac{41539}{22608}+\frac{5413  }{5652}\eta\right) \xi_{\phi} ^{2/3}+\left(\frac{2833}{1008}-\frac{197  }{36}\eta\right)
\xi_{\phi,0}^{2/3}+\frac{24871  }{11304}\pi  \xi_{\phi}+\frac{377 }{72}\pi  \xi_{\phi,0} 
\right.\\  
+\left(-\frac{122085949}{239283072}+\frac{2133953
	 }{712152}\eta-\frac{36497 }{101736}\eta ^2\right) \xi_{\phi} ^{4/3}+\left(-\frac{117679987}{22788864}+\frac{10486813  }{1424304}\eta+\frac{1066361 }{203472}\eta
	 ^2\right) \xi_{\phi} ^{2/3} \xi_{\phi,0}^{2/3} 
\\
+\left(-\frac{1193251}{3048192}-\frac{66317  }{9072}\eta+\frac{18155 }{1296}\eta ^2\right) \xi_{\phi,0}^{4/3}
+\left(\frac{215395661  }{28486080}-\frac{8416733  }{508680}  \eta\right)\pi \xi_{\phi} ^{5/3}
+\left(\frac{10065649
	 }{1627776}
\right. \\ \left.
-\frac{4899587}{406944}\eta\right) \pi \xi_{\phi}  \xi_{\phi,0}^{2/3} -\left(\frac{15660203  }{1627776}+\frac{2040701}{406944} \eta\right)
\pi \xi_{\phi} ^{2/3} \xi_{\phi,0} 
+\left(\frac{764881  }{90720}-\frac{949457    }{22680}\eta\right)\pi \xi_{\phi,0}^{5/3}
\\
+\left(-\frac{345869493517}{241197336576}+\frac{96596798141
	 }{8614190592}\eta-\frac{9444185 }{542592}\eta ^2
+\frac{7189909 }{3662496}\eta ^3\right) \xi_{\phi} ^{4/3} \xi_{\phi,0}^{2/3}
+\frac{9376367 }{813888}\pi ^2 \xi_{\phi}
\xi_{\phi,0}
\\
+\left(\frac{49566453289}{68913524736}+\frac{237857384155  }{17228381184}\eta
-\frac{1281029377 }{68366592}\eta ^2-\frac{98273015
	}{7324992}\eta ^3\right) \xi_{\phi} ^{2/3} \xi_{\phi,0}^{4/3}
\\
+ \left[\frac{9765600648106487}{66329267558400}-\frac{2491067 }{98910}\gamma_{E}
-\frac{10610699
	}{1627776}\pi ^2+\left(-\frac{3409129936301}{8614190592}+\frac{2883161 }{180864}\pi ^2\right) \eta 
\right. \\ \left.
+\frac{1598264033 }{102549888}\eta ^2-\frac{2773315
	}{10987488}\eta ^3 
+\frac{5257873 }{296730}\ln2-\frac{1534059 }{87920}\ln3-\frac{2491067
}{197820}\ln(16 \xi_{\phi} ^{2/3})\right]\xi_{\phi} ^2
\\ 
+\left[\frac{26531900578691}{168991764480}-\frac{3317 }{126}\gamma_{E}+\frac{122833
	}{10368}\pi ^2+\left(\frac{9155185261}{548674560}-\frac{3977 }{1152}\pi ^2\right) \eta  
\right. \\ \left. \left. \left.
-\frac{5732473 }{1306368}\eta ^2-\frac{3090307 }{139968}\eta ^3+\frac{87419
}{1890}\ln2-\frac{26001 }{560}\ln3-\frac{3317}{252} \ln(16 \xi_{\phi,0}^{2/3})\right]\xi_{\phi,0}^2 \right\}\right) .
\end{multline}
This is the key equation that allows us to determine explicit functions for the frequency and phase evolution in the small-eccentricity limit. We illustrate how these results follow from this equation in the remainder of this section, deferring some of the full 3PN expressions to Sec.~\ref{sec:approximants} where they are derived via an equivalent approach that generalizes the quasicircular PN approximants.

The time to coalescence is computed by integrating $dt = d\xi_{\phi}/(d\xi_{\phi}/dt)$. To compute this we first invert Eq.~\eqref{eq:dxidtexpand2}, expand the terms in $\{ \}$-brackets to leading order in $e_0$, and then expand the entire expression to 3PN order [relative $O(\varepsilon^6)$]. Integrating the result with respect to $\xi_{\phi}$ yields
\bs
\label{eq:xi_t}
\be
\label{eq:xi_t_a}
t_c-t \equiv \frac{5M}{\eta} \tau = \frac{5 M}{256 \eta  \xi_{\phi} ^{8/3}} {\mathcal T}(\xi_{\phi}, \xi_{\phi,0},e_0), 
\ee
\begin{multline}
\label{eq:T_xi_t}
 {\mathcal T}(\xi_{\phi}, \xi_{\phi,0},e_0) = \left\{1+\left(\frac{743}{252}+\frac{11  }{3}\eta\right) \xi_{\phi} ^{2/3} + \cdots + O(\xi_{\phi}^2)  \right. \\ \left.
 -\frac{157}{43} e_0^2 \left(\frac{\xi_{\phi,0}}{\xi_{\phi} }\right)^{19/9}
\left[1+\left(\frac{17592719}{5855472}+\frac{1103939  }{209124}\eta\right) \xi_{\phi}^{2/3}
+\left(\frac{2833}{1008}-\frac{197 }{36}\eta\right) \xi_{\phi,0}^{2/3} + \cdots + O(\xi_{\phi}^2) \right] \right\},
\end{multline}
\es
where $t_c$ is the coalescence time and the full 3PN expression can be inferred from Eq.~\eqref{eq:T_PNapprox} via the substitutions ${\mathcal T}(v=\xi_{\phi}^{1/3},v_0=\xi_{\phi,0}^{1/3},e_0)$. 

Note the general structure of the series expansion in Eq.~\eqref{eq:T_xi_t}. The first line shows the quasicircular result.\footnote{The quasicircular results are all known to relative 3.5PN order, but as our starting expressions for eccentric orbits are only known to 3PN order, we restrict to purely 3PN results in this section. The full expressions, listed in Sec.~\ref{sec:approximants}, include the 3.5PN quasicircular terms.} The second line shows the leading order in eccentricity corrections. Since $\xi_{\phi,0}/\xi_{\phi} \sim O(1)$, the first term on the second line is equivalent to a 0PN-order effect. The remainder of the second line schematically shows corrections which we have computed to relative 3PN order [see Eq.~\eqref{eq:T_PNapprox}]. Note that starting at 2PN order, there are terms with the structure $\xi_{\phi}^a \xi_{\phi,0}^b$ where $a+b=2n/3$ for terms at $n$PN order. The other PN series expansions in this section have a similar structure. Note also that the difference between the time of coalescence and the reference time $t_0$ when $e_t(t_0)=e_0$ is given by
\be
\label{eq:tau0}
t_c-t_0 \equiv \frac{5M}{\eta} \tau_0 = \frac{5 M}{256 \eta  \xi_{\phi,0} ^{8/3}} {\mathcal T}(\xi_{\phi,0}, \xi_{\phi,0},e_0). 
\ee

The time evolution of the frequency variable $\xi_{\phi}$ can now be obtained by performing a series reversion on Eq.~\eqref{eq:xi_t}. This is done by expanding $\xi_{\phi}$ as a PN series in $\tau$ and $\tau_0$ with the coefficients undetermined.\footnote{In other words, one can use PN power counting to infer the form of the series shown in the solution \eqref{eq:xi-tau}. The series expansions are more easily performed by introducing a dimensionless parameter $\epsilon \sim O(1/c)$ and substituting $\tau\rightarrow \tau/\epsilon^8$ and $\tau_0\rightarrow \tau_0/\epsilon^8$. The $\epsilon$ parameter is then set to $1$ at the end of the calculation.} This series is then substituted into Eq.~\eqref{eq:xi_t} and expanded in $\epsilon$ and to $O(e_0^2)$. The unknown coefficients are determined by requiring that each term vanish at the appropriate orders in $\epsilon$ and $e_0$. The result is
\bs
\label{eq:xi-tau}
\be
\label{eq:xi-tau_a}
\xi_{\phi}(t) = \frac{1}{8 \tau^{3/8}} \Xi(\tau,\tau_0,e_0) \,,
\ee
\begin{multline}
\label{eq:Xi}
\Xi(\tau,\tau_0,e_0) = \left\{1+\left(\frac{743}{2688}+\frac{11  }{32}\eta\right)\tau^{-1/4} + \cdots + O(\tau^{-3/4}) \right.
\\ \left.
-\frac{471}{344} e_0^2 \left(\frac{\tau }{\tau_0}\right)^{19/24} \left[1+\left(-\frac{7647061}{70265664}+\frac{209353
			}{836496}\eta \right)\tau ^{-1/4}+\left(\frac{4445}{3456}-\frac{185  }{288}\eta\right)\tau_0^{-1/4} + \cdots + O(\tau^{-3/4}) \right]  \right\} \,,
\end{multline}
\es
where the full 3PN-order expression can be inferred from Eq.~\eqref{Xi_PNapprox} (replacing $\theta = \tau^{-1/8}$ and $\theta_0=\tau_0^{-1/8}$ as appropriate). Similar to Eq.~\eqref{eq:xi_t}, there are also cross-terms of the form $\tau^c \tau_0^d$ appearing at 2PN and higher orders. The value of $\xi_{\phi}$ at the reference time [where $e_t(t_0)=e_0$] is determined from  
\be
\label{eq:xi0-tau0}
\xi_{\phi,0}(t_0) = \frac{1}{8 \tau_0^{3/8}} \Xi(\tau_0,\tau_0,e_0).
\ee

Although we do not use this later in our analysis, the evolution of the orbital eccentricity with time can be computed by  substituting Eqs.~\eqref{eq:xi-tau} and \eqref{eq:xi0-tau0} into Eq.~\eqref{eq:etofxi} and expanding to 3PN order. The result has the form
\be
\label{eq:et_tau}
e_t(t)= e_0 \left(\frac{\tau }{\tau_0}\right)^{19/48} \left[1+ \left(-\frac{4445}{6912}+\frac{185  }{576}\eta\right)\left(\tau ^{-1/4}-\tau_0^{-1/4}\right) + \cdots + O(\tau^{-3/4}) \right],
\ee
where the full 3PN expression is given in Appendix \ref{app:longeqns}.

We can also compute the secular evolution of the angular variables $\lambda$ and $l$. (We consider the oscillatory piece of $\lambda$ in the next section.) The secular evolution of $\lambda$ is governed by 
\be
\label{eq:dlambdabardt}
\frac{d\lambda}{dt} = \frac{\xi (1+k)}{M} = \frac{\xi_{\phi}}{M},
\ee
which can be integrated to give
\be
\lambda - c_{\lambda}  =\frac{1}{M} \int \frac{\xi_{\phi}}{(d\xi_{\phi}/dt)} d\xi_{\phi}.
\ee
Series expanding the integrand, evaluating the integral, and simplifying yields
\bs
\label{eq:lambda_xi}
\be
\label{eq:lambda_xi_a}
\lambda(\xi_{\phi}) - c_{\lambda} = -\frac{1}{32 \eta  \xi_{\phi} ^{5/3}} \Lambda_f(\xi_{\phi},\xi_{\phi,0},e_0) \,,
\ee
\begin{multline}
\label{eq:Lambda_f}
\Lambda_f(\xi_{\phi},\xi_{\phi,0},e_0) = \left\{1+\left(\frac{3715}{1008}+\frac{55  }{12}\eta\right) \xi_{\phi} ^{2/3} + \cdots + O(\xi_{\phi}^2) \right. \\
\left. 
-\frac{785}{272} e_0^2 \left(\frac{\xi_{\phi,0}}{\xi_{\phi} }\right)^{19/9} \left[1+\left(\frac{6955261}{2215584}+\frac{436441  }{79128}\eta\right) \xi_{\phi} ^{2/3} +\left(\frac{2833}{1008}-\frac{197
	 }{36}\eta\right) \xi_{\phi,0}^{2/3} + \cdots +O(\xi_{\phi}^2) \right] \right\},
\end{multline}
\es
where the additional terms to 3PN order can be read off of Eq.~\eqref{eq:Lambdaf_PNapprox}. 
Plugging Eqs.~\eqref{eq:xi-tau} and \eqref{eq:xi0-tau0} into the above equation and expanding to 3PN order allows us to determine the time-dependent function $\lambda(t)$,
\bs
\label{eq:lambda_tau}
\be
\label{eq:lambda_tau_a}
\lambda(\tau) - \bar{c}_{\lambda} =- \frac{1}{\eta}\tau^{5/8} \Lambda_t(\tau,\tau_0,e_0) \, ,
\ee
\begin{multline}
\label{eq:Lambda_t}
\Lambda_t(\tau,\tau_0,e_0) =  \left\{ 1+\left(\frac{3715}{8064}+\frac{55  }{96}\eta\right)\tau^{-1/4} + \cdots + O(\tau^{-3/4}) \right. \\
\left. 
-\frac{7065}{11696} e_0^2 \left(\frac{\tau }{\tau_0}\right)^{19/24} \left[1+\left(-\frac{130000037}{983719296}+\frac{3559001 }{11710944}\eta\right)\tau ^{-1/4} +\left(\frac{4445}{3456}-\frac{185 }{288}\eta\right) \tau_0^{-1/4} + \cdots + O(\tau^{-3/4}) \right] \right\} \,,
\end{multline}
\es
with the full 3PN expression given in Eq.~\eqref{eq:Lambda_t_PNapprox}.

Lastly, the secular evolution of $l$ is determined by
\be
\frac{dl}{dt} = \frac{\xi}{M} = \frac{\xi(\xi_{\phi})}{M} \,,
\ee
where $\xi(\xi_{\phi})$ is given by Eq.~\eqref{eq:xi-xiphi}. As was the case for $\lambda$, this is straightforwardly integrated via
\be
l - c_l = \frac{1}{M} \int \frac{\xi(\xi_{\phi})}{d\xi_{\phi}/dt} d\xi_{\phi}.
\ee
Evaluating the integral using the same techniques as above gives
\begin{multline}
 \label{eq:l_xiphi}
l(\xi_\phi)-c_l=-\frac{1}{32 \eta  \xi^{5/3}_\phi }\left\{1+\left(-\frac{1325}{1008}+\frac{55  }{12}\eta\right) \xi^{2/3}_\phi + \cdots + O(\xi_{\phi}^2) \right. \\
\left. 
-\frac{785}{272}e^2_0\bigg(\frac{ \xi_{\phi,0}}{ \xi_\phi }\bigg)^{19/9} \left[1+ \left(\frac{117997}{2215584}+\frac{436441 }{79128}\eta \right) \xi_{\phi}^{2/3} + \left(\frac{2833}{1008}-\frac{197  }{36}\eta\right) \xi^{2/3}_{\phi,0} + \cdots + O(\xi_{\phi}^2) \right] \right\},
\end{multline}
where the full 3PN expression is in Eq.~\eqref{eq:l_xiphi3PN}. A function of time $l(t)$ analogous to Eq.~\eqref{eq:lambda_tau} can also be derived and is given in Eq.~\eqref{eq:l_tau3PN}.

In this section, we have expressed the secular evolution of the intrinsic constants $\xi_{\phi}$ and $e_t$ and the phase functions $\lambda$ and $l$. Results were expressed as functions of time or the frequency variable $\xi_{\phi}$. These results could also have been derived directly in terms of the radial frequency variable $\xi$. They can be converted to functions of $\xi$ via substitution of Eq.~\eqref{eq:xiphi-xi}.
\section{\label{sec:oscil}Effect of oscillatory terms in the phasing}
While the main goal of this work is to analytically compute the secular corrections to the waveform phasing, it is important to also consider the relative sizes of the oscillatory contributions to the phasing. The Newtonian-order GW polarizations in the low-$e_t$ limit depend on the orbital phase $\phi$ via $h_{+,\times} \propto (\cos2\phi, \sin2\phi)$. Recall that the complete orbital phasing is the sum of three terms,
\be
\label{eq:phioscill}
\phi={\lambda}+\tilde{\lambda}+W(e_t,\xi,l),
\ee
where $\lambda = \bar{\lambda}$ is the secularly growing part of the phase [Eq.~\eqref{eq:lambda_xi}], $\tilde{\lambda}$ is the radiation-reaction induced oscillatory contribution to $\lambda$ [Eq.~\eqref{eq:lambdatilde}], and $W(e_t,\xi,l)$ is the eccentricity-induced oscillatory piece of the phase [cf., Eq.~\eqref{eq:phi2pnlowet3}; recall that we have dropped overbars on secularly varying quantities]. Notice that $\tilde{\lambda}$ and $W$ both oscillate at multiples of the radial orbital period, with amplitudes that vary on the radiation-reaction time scale. Since $\tilde{\lambda} \sim O(\xi^{5/3}) \sim O(\xi_{\phi}^{5/3})$ while ${\lambda} \sim O(1/\xi_{\phi}^{5/3})$, this implies that $\tilde{\lambda}$ represents a 5PN relative correction [$\sim O(\xi_{\phi}^{10/3}) \sim O(v^{10})$] to $\phi$. Since our phasing is only accurate to 3PN order, we can safely ignore $\tilde{\lambda}$. The $W$ contribution to the phasing has a leading-order term $W\sim 2 e_t \sin l$; this is potentially of order unity (for large $e_t$) and is not obviously ignorable. However, as we argue below, this order unity contribution is oscillatory, decays with time, and is vastly dominated by the secularly increasing contribution ${\lambda}$ to the total phase.  

To better understand its contribution to the phasing, we can explicitly evaluate the $W$ term as a function of frequency. Restricting for simplicity to 2PN order, $W$ is expressed in Appendix \ref{app:polarization} as a function of the mean anomaly $l$ and a series expansion in $e_t$: 
\begin{multline}
\label{eq:W_l2PN}
 W(l) \approx e_t \sin l \Bigg[ 2 +  (10-\eta)\xi^{2/3} +  \bigg( 72- \frac{259}{12} \eta + \frac{1}{12} \eta^2 \bigg)\xi^{4/3} \Bigg] + e_t^2 \sin 2l \Bigg[ \frac{5}{4} +  \bigg( \frac{31}{4} - \eta \bigg)\xi^{2/3} +  \bigg( \frac{447}{8} - \frac{187}{12} \eta + \frac{1}{12} \eta^2 \bigg)\xi^{4/3} \Bigg] \\ + O(e_t^3) +O(\xi^2),
\end{multline}
where $l = \bar{l}$ and $e_t= \bar{e}_t$ in the above expression. [As with $\tilde{\lambda}$, $\tilde{l}$ and $\tilde{e}_t$ represent relative 5PN corrections and can be ignored.] An explicit expression in terms of the frequency variable $\xi_{\phi}$ can be obtained by substituting Eqs.~\eqref{eq:xi-xiphi}, \eqref{eq:etofxi}, and \eqref{eq:l_xiphi} to 2PN order into \eqref{eq:W_l2PN}; PN expanding then yields 
\begin{multline}
 \label{eq:w_xiphi}
 W(\xi_\phi)=e_0\bigg(\frac{\xi_{\phi,0}}{\xi_\phi}\bigg)^{19/18} \sin l \left[2 +\left(\frac{7247}{1008}+\frac{161
  	 }{36}\eta\right) \xi_{\phi}^{2/3} +\left(\frac{2833}{1008}-\frac{197  }{36}\eta\right) \xi_{\phi,0}^{2/3} + \frac{377 }{72}\pi \left( \xi_{\phi,0}-\xi_\phi\right)
\right. \\
  +\left(\frac{539690101}{12192768}-\frac{477299  }{72576}\eta+\frac{30055 }{5184}\eta^2\right)
  \xi^{4/3}_\phi
   +\left(\frac{20530751}{2032128}-\frac{485773  }{36288}\eta-\frac{31717 }{2592}\eta ^2\right) \xi^{2/3}_{\phi,0}
   \xi^{2/3}_\phi 
  \\ \left.
  	+\left(-\frac{28850671}{12192768}+\frac{27565}{72576}\eta+\frac{33811 }{5184}\eta ^2\right) \xi_{\phi,0}^{4/3} \right] 
  	+e_0^2\bigg(\frac{ \xi_{\phi,0} }{\xi_\phi}\bigg)^{19/9} \sin 2l \left[\frac{5}{4} +\left(\frac{17083}{4032}+\frac{841  }{144}\eta\right) \xi^{2/3}_\phi \right. \\
  	+ \left(\frac{14165}{4032}-\frac{985
 	 }{144}\eta\right) \xi^{2/3}_{\phi,0}  +\frac{1885    }{288}\pi\left( \xi_{\phi,0}-\xi_\phi \right)
+\left(\frac{176530423}{6096384}+\frac{168745  }{72576}\eta+\frac{37667 }{2592}\eta ^2\right) \xi^{4/3}_\phi 
 \\ \left.
 +\left(\frac{48396139}{4064256}-\frac{491399
  	 }{72576}\eta-\frac{165677 }{5184}\eta ^2\right) \xi^{2/3}_{\phi,0} \xi^{2/3}_\phi
 +\left(-\frac{5966255}{12192768}-\frac{331585  }{36288}\eta+\frac{90775
 	}{5184}\eta ^2\right) \xi^{4/3}_{\phi,0}   \right] \,.
\end{multline}

\begin{figure*}[t]
	\includegraphics[width=0.45\textwidth]{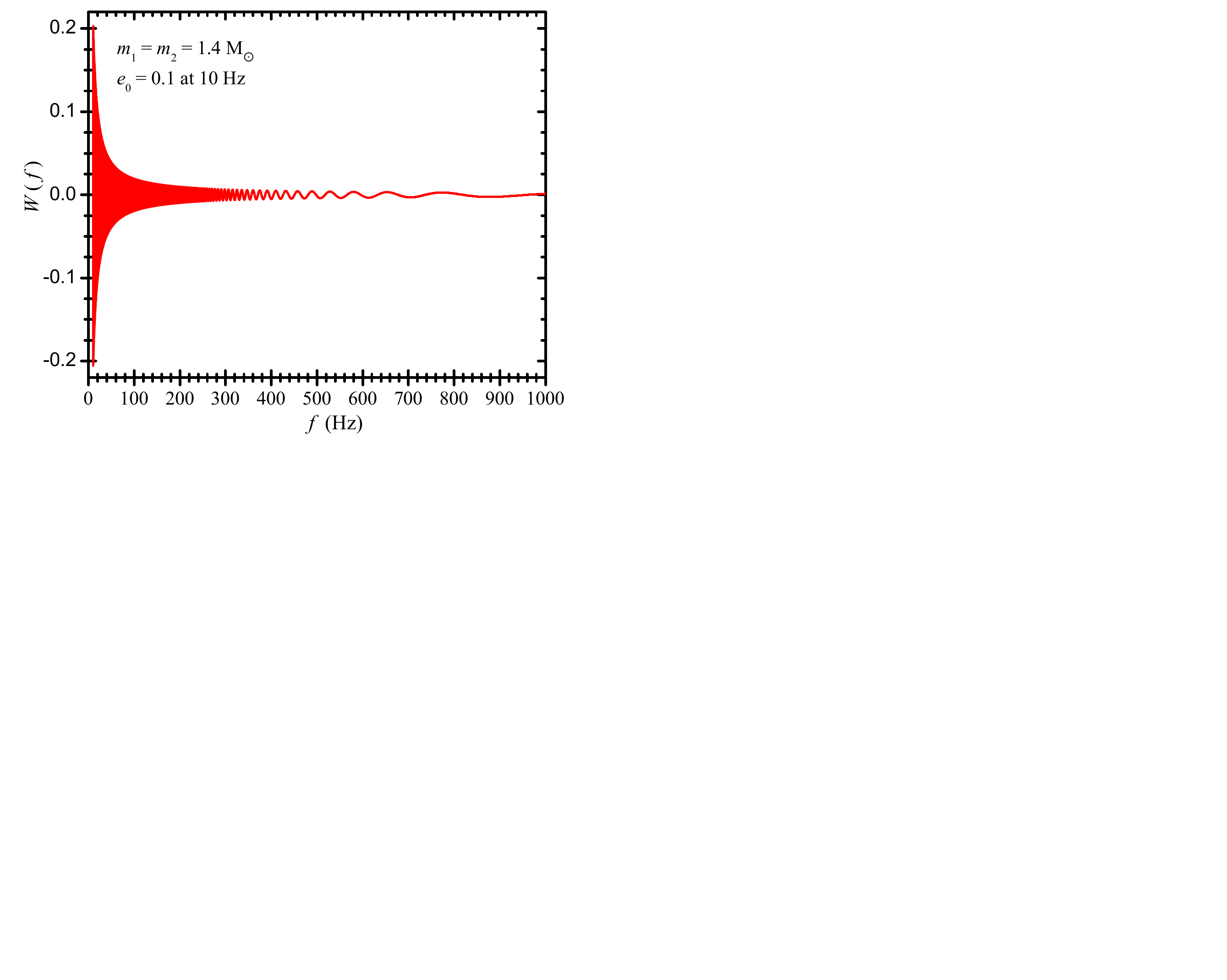}
	\caption{\label{fig:W_f_NSNS} Oscillatory contribution to the orbital phasing $W(f)$ for a NS/NS binary with $e_0=0.1$ at $10$ Hz. The function has a maximum amplitude of $\sim 0.2$ (with little dependence on the binary masses) which decays along with the binary eccentricity. The resulting contribution to the GW phasing is small for low eccentricity.}
\end{figure*}

To estimate the magnitude of $W$, we can inspect the $O(e_0)$ Newtonian-order result, $W_{\rm N}(f)=2e_0 (f_0/f)^{19/18} \sin l$, where we have used $\xi_{\phi} = \pi M f$ and $\xi_{\phi,0} = \pi M f_0$. Clearly, this expression has a maximum value when $f=f_0$ and scales linearly with $e_0$. Section \ref{sec:numerical_comp} below demonstrates that $e_0 \approx 0.1$ is the maximum eccentricity to which our low-eccentricity expressions are valid. We can then estimate that the maximum error $\delta \phi$ in the orbital phasing due to the oscillatory terms is  $\delta \phi \approx W_{\rm N}(e_0=0.1,f=f_0) \approx 0.2$ rad. Figure \ref{fig:W_f_NSNS} shows an evaluation of the full 2PN expression [Eq.~\eqref{eq:w_xiphi}] for a NS/NS binary that evolves through the LIGO band with $e_0=0.1$ at $f_0=10$ Hz (and choosing $c_l=0$). We can see that the amplitude of the periodic term decreases sharply with increasing frequency. Changing the binary masses of the system (which does not affect the Newtonian-order expression $W_{\rm N}$) leads to a slight change in the amplitude; however, the qualitative behavior of the curve remains unchanged. The maximum values of Eq.~\eqref{eq:w_xiphi} for NS/NS, BH/BH, and NS/BH binaries with $e_0=0.1$ are 0.207, 0.224, and 0.217 rad, respectively. This corresponds to a correction to the number of GW cycles of $\delta N_{{\rm cyc}, W} \sim 0.2/\pi \lesssim 0.07$. Since this effect is small and decays rapidly, we ignore it when computing corrections to the PN approximants in the next section. However, we note that this oscillatory contribution is comparable (for some systems) to the 2.5PN and 3PN secular eccentric corrections that we compute (see Sec.~\ref{sec:ncyc} and the tables presented there). We also note that when eccentricities are large (in which case our formalism is not valid), these oscillatory terms will contribute phase errors $\sim O(1)$ and should not be ignored. 
\section{\label{sec:approximants}Post-Newtonian approximants}
For quasicircular inspiralling binaries, the GW signal at leading PN order takes the form given in Eq.~\eqref{eq:hNcirc}. Neglecting corrections to the waveform amplitude that scale as $O(e_t)$ and higher [cf. Eq.~\eqref{eq:hNl}], our low-eccentricity waveforms take the same (quasicircular) form. What remains is a determination of the orbital phase $\phi(t)$. In the quasicircular limit (and in the \emph{adiabatic approximation}---i.e., the assumption that the orbital time scale is much shorter than the radiation reaction time scale), the phase evolution is governed by the following differential equations,
\bs
\label{eq:approx_timedom}
\begin{align}
\label{eq:approx_timedom_phi}
 \frac{d\phi}{dt} &=\frac{\xi_{\phi}}{M}=\frac{v^3}{M},\\
 \label{eq:approx_timedom_v}
 \frac{dv}{dt} &=-\frac{{\mathcal F}(v)}{dE(v)/dv},
 \end{align}
 \es
where in this section we express quantities in terms of the relative orbital velocity parameter $v= \xi_{\phi}^{1/3} = (\pi M f)^{1/3}$. In the above, ${\mathcal F}(v)$ is the gravitational-wave luminosity (often referred to as the energy flux), and $E(v)$ is the orbit energy. Different approaches for solving these differential equations are referred to as different PN {\it approximants} (see e.g., Ref.~\cite{buonanno-iyer-oshsner-pan-sathya-templatecomparison-PRD2009} and references therein, which we follow in this section).\footnote{Note that our notation differs slightly from Ref.~\cite{buonanno-iyer-oshsner-pan-sathya-templatecomparison-PRD2009} in that we take $E$ to be the orbital energy; Ref.~\cite{buonanno-iyer-oshsner-pan-sathya-templatecomparison-PRD2009} uses that symbol to denote the orbital energy divided by $M$. This leads to different factors of $M$ appearing in our Eqs.~\eqref{eq:approx_timedom} and \eqref{eq:approx_freqdom} as compared with the equations in Ref.~\cite{buonanno-iyer-oshsner-pan-sathya-templatecomparison-PRD2009}.} 

For eccentric orbits, the orbital phase includes the oscillatory terms in Eq.~\eqref{eq:phioscill} above. However, as we have argued in Sec.~\ref{sec:oscil}, the oscillatory terms contribute $\lesssim 0.2$ rad for $e_0 \lesssim 0.1$. Assuming this is an acceptably small error, we can ignore these oscillatory terms. In this limit, Eqs.~\eqref{eq:approx_timedom} carry over unchanged to the case of small-eccentricity binaries if we take $\phi \rightarrow \langle \phi \rangle = \lambda$ (recall that we are dropping overbars on secularly varying quantities). The oscillatory effects encapsulated in $W$ could be incorporated into the PN approximants by adding terms equal to $dW/dt$ and $\frac{M}{3v^2} \frac{d^2W}{dt^2}$ to the right-hand sides of Eqs.~\eqref{eq:approx_timedom_phi} and \eqref{eq:approx_timedom_v}, respectively. This will be considered in future work.  

We note that the above equations imply that two initial conditions must be supplied: $\phi(t_0)$ and $v(t_0)$ [or equivalently $f(t_0)$], along with the binary masses and the eccentricity $e_0$ at a reference frequency $f_0$. However, for arbitrarily elliptical orbits, one must specify an additional parameter---equivalent to the argument of pericenter $\varpi$---which determines the orientation of the ellipse that is momentarily tangent to the orbit. Toward the end of Appendix \ref{app:polarization} we discuss how a parameter like $\varpi$ enters the waveform and how it relates to the constants $c_l$ and $c_{\lambda}$. This parameter does not enter our approximants for two reasons: (i) since we ignore $O(e_t)$ and higher corrections to the polarization amplitudes, we need only to evolve the phase variable $\phi(t)$ [and can ignore the other phase variable $l(t)$]; (ii) furthermore, because we ignore the oscillatory corrections to $\phi(t)$ that arise from $W(l)$,  the dependence on the initial orientation of the ellipse (which enters via $c_l$) drops out of our waveforms completely. 

In the remainder of this section, we derive the small-eccentricity extensions to the standard PN approximants to 3PN order, starting with appropriate expressions for the orbital energy and GW luminosity. Most of these approximants can also be derived following the procedure outlined in Sec.~\ref{sec:expliciteqns}. The approximants presented here were derived via both approaches and cross-checked by at least two of the authors. In this section, our goal is to provide a derivation that does not require understanding a significant amount of the ``context'' provided by the quasi-Keplerian formalism. The 3.5PN-order circular terms were not derived here nor in Sec.~\ref{sec:expliciteqns} but can be found in Ref.~\cite{buonanno-iyer-oshsner-pan-sathya-templatecomparison-PRD2009}; for completeness, we added those terms to our expressions below.
\subsection{3PN energy, energy flux, and TaylorT1}
The TaylorT1 approximant is obtained by numerically solving Eqs.~\eqref{eq:approx_timedom} without expanding the ratio in Eq.~\eqref{eq:approx_timedom_v}. To compute this (and the other approximants), we require expressions for the orbital energy and GW luminosity (energy flux). These are given in ADM gauge in Eqs.~(6.5a) and (7.4a) of Ref.~\cite{arun-etal-eccentric-orbitalelements-PRD2009}. Taking those expressions to $O(e_t^2)$, expressing them in MH gauge [via Eq.~\eqref{eq:ADMtoMH}], substituting $e_{t}(v)$ [Eq.~\eqref{eq:etofxi}], and simultaneously expanding in $v$ and $v_0$ yields the low-eccentricity limit of the orbital energy and flux functions (expressed explicitly as functions of $v$):
\begin{multline}
\label{eq:energy_v}
E(v)=-\frac{1}{2}\eta M v^2 \left(1-\left(\frac{3}{4}+\frac{1 }{12}\eta\right)v^2 + \left(-\frac{27}{8}+\frac{19}{8} \eta -\frac{1}{24}\eta ^2\right)v^4 +
\left[-\frac{675}{64}+\left(\frac{34445}{576}-\frac{205 }{96}\pi ^2\right) \eta -\frac{155 }{96}\eta ^2
\right. \right. \\ \left.
-\frac{35}{5184} \eta ^3\right]v^6
+e_0^2\left(\frac{
	v_0}{v}\right)^{19/3} \left\{-2 v^2 + \left(-\frac{6743}{504}+\frac{37
	}{18}\eta \right)v^4+ \left(-\frac{2833}{504}+\frac{197 }{18}\eta \right)v^2 v_0^2+\frac{377 }{36} \pi v^5-\frac{377}{36} \pi  v^2 v_0^3
\right.	\\ 
 +\left(\frac{1193251}{1524096}+\frac{66317
	}{4536}\eta -\frac{18155 }{648}\eta ^2\right)v^2 v_0^4+ \left(-\frac{19102919}{508032}+\frac{179149  }{2268}\eta-\frac{7289 }{648}\eta ^2\right)v^4 v_0^2
\\ \left. \left.
+
\left[-\frac{68531831}{762048}+\left(\frac{1093055}{4536}-\frac{123 }{16}\pi ^2\right) \eta +\frac{1751 }{324}\eta ^2\right]v^6\right\} + O(v^8) \right) \,,
\end{multline}
\begin{multline}
\label{eq:energyflux_v}
{\cal F}(v)=\frac{32}{5} \eta^2v^{10}\left(1- \left(\frac{1247}{336}+\frac{35
}{12}\eta\right)v^2+4 \pi  v^3+ \left(-\frac{44711}{9072}+\frac{9271  }{504}\eta+\frac{65 }{18}\eta ^2\right)v^4- \left(\frac{8191}{672}+\frac{583}{24}\eta\right)\pi  v^5
\right. \\
+ \left[\frac{6643739519}{69854400}-\frac{1712
}{105}\gamma_{E}+\frac{16 }{3}\pi ^2+\left(-\frac{134543}{7776}+\frac{41 }{48}\pi ^2\right) \eta -\frac{94403 }{3024}\eta ^2-\frac{775 }{324}\eta ^3-\frac{856}{105}
\ln(16 v^2)\right]v^6
\\
- \left(\frac{16285}{504}-\frac{214745}{1728}\eta-\frac{193385}{3024}\eta^2\right)
\pi v^7+\frac{157}{24} e_{0}^2 \left(\frac{v_0}{v}\right)^{19/3} \left\{1- \left(\frac{67387}{22608}+\frac{6355
}{5652}\eta\right)v^2+  \left(\frac{2833}{1008}-\frac{197 }{36}\eta \right)v_0^2 
\right. \\
+\frac{24871}{11304} \pi  v^3
+\frac{377 }{72}\pi
v_0^3+ \left(-\frac{1992524449}{239283072}+\frac{30853331  }{2848608}\eta-\frac{32975}{101736} \eta ^2\right)v^4+
\left(-\frac{190907371}{22788864}+\frac{37461479 }{2848608}\eta 
\right. \\ \left.
+\frac{1251935 }{203472}\eta ^2\right) v^2 v_0^2
+  \left(-\frac{1193251}{3048192}-\frac{66317
}{9072}\eta +\frac{18155 }{1296}\eta ^2\right)v_0^4+ \left(\frac{177226481 }{28486080} -\frac{17206531
}{1017360}   \eta \right)\pi v^5+  \left(\frac{10065649 }{1627776} 
\right. \\ \left.
-\frac{4899587 }{406944}  \eta \right)\pi v^3 v_0^2
+ \left(-\frac{25404899 }{1627776} -\frac{2395835
}{406944} \eta \right)\pi v^2  v_0^3+ \left(\frac{764881}{90720}  -\frac{949457  }{22680} \eta \right)\pi v_0^5 +  \left[\frac{7486629590558687}{66329267558400}
\right. \\ \left.
-\frac{2491067
}{98910}\gamma_{E}-\frac{10610699 }{1627776}\pi ^2
+\left(-\frac{1137653498195}{8614190592}+\frac{1161161 }{180864}\pi ^2\right)
\eta +\frac{33498427}{7324992}\eta ^2-\frac{274435 }{2746872}\eta ^3+\frac{5257873 }{296730}\ln2
\right. \\ \left. 
-\frac{1534059 }{87920}\ln3-\frac{2491067
}{197820}	\ln(16 v^2)\right]v^6
+ \left(-\frac{5644821764017}{241197336576}+\frac{327374888311
}{4307095296}\eta -\frac{1028587397 }{17091648}\eta ^2+\frac{6496075}{3662496} \eta ^3\right)v^4 v_0^2 
 \\ 
+\frac{9376367 }{813888}\pi ^2 v^3 v_0^3+ \left(\frac{80409605137}{68913524736}
+\frac{382971019141
}{17228381184}\eta -\frac{764233195 }{22788864}\eta ^2-\frac{115375025 }{7324992}\eta ^3\right)v^2  v_0^4+  \left[\frac{26531900578691}{168991764480}
\right.  \\ 
-\frac{3317
}{126}\gamma_{E}+\frac{122833}{10368} \pi ^2+\left(\frac{9155185261}{548674560}
-\frac{3977}{1152}\pi ^2\right) \eta -\frac{5732473 }{1306368}\eta ^2-\frac{3090307
}{139968}	\eta ^3
\\ \left. \left. \left.
+\frac{87419}{1890} \ln2-\frac{26001 }{560}\ln3-\frac{3317}{252} \ln(16
v_0^2)\right]v_0^6 \right\}\right).
\end{multline}
As it is explicitly needed to compute the TaylorT1 approximant, we also list here the derivative $dE/dv$:
\begin{multline}
\label{eq:denergy_dv}
\frac{dE}{dv}= -\eta M v\left(1- \left(\frac{3}{2}+\frac{1}{6}\eta \right)v^2+ \left[-\frac{81}{8}+\frac{1}{8} (57-\eta ) \eta \right]v^4- \left[\frac{675}{16}+ \left(-62001+2214 \pi ^2+1674 \eta
\right. \right. \right.\\ \left. \left.
+7 \eta^2 \right)\frac{5}{1296}
\eta \right]v^6
+e_{0}^2 \left(\frac{v_0}{v}\right)^{19/3} \left\{\frac{7}{3} v^2 + \left(\frac{2833}{432}-\frac{1379  }{108}\eta\right)v^2
v_0^2+ \left(\frac{6743}{3024}-\frac{37  }{108}\eta\right)v^4+\frac{377}{108} \pi  v^5+\frac{2639}{216} \pi  v^2 v_0^3
\right. \\
+\left(-\frac{1193251}{1306368}-\frac{66317 
}{3888}\eta+\frac{127085 }{3888}\eta ^2\right)v^2 v_0^4
 + \left[-\frac{342659155}{4572288}+\left(\frac{5465275
	 }{27216}-\frac{205  }{32}\pi ^2\right) \eta+\frac{8755 }{1944}\eta ^2\right]v^6
\\ \left. \left.
 + \left(\frac{19102919}{3048192}-\frac{179149  }{13608}\eta+\frac{7289 }{3888}\eta ^2\right)v^4 v_0^2  \right\}\right) \,.
\end{multline}
\subsection{TaylorT2}
The TaylorT2 approximant is obtained by series expanding the ratio in Eq.~\eqref{eq:approx_timedom_v} to the appropriate PN order. One then analytically obtains a parametric solution for the phase $[\langle\phi\rangle(v), t(v)]$ by integrating 
\bs
\label{eq:approx_freqdom}
\begin{align}
\label{eq:approx_freqdom_phi}
\frac{d\langle\phi\rangle}{dv} &= \frac{d\langle\phi\rangle}{dt}\frac{dt}{dv}=-\frac{v^3}{M} \frac{dE(v)/dv}{{\mathcal F}(v)} \,, \\
\label{eq:appox_freqdom_t}
\frac{dt}{dv} &= -\frac{dE(v)/dv}{{\mathcal F}(v)}.
\end{align}
\es
Expanding to 3PN order and $O(e_0^2)$, the resulting solutions are
\bs
\label{eq:taylort2_phi}
\be
\label{eq:taylort2_phi_a}
\langle\phi\rangle-\phi_{c}=-\frac{1}{32 v^5 \eta } \Lambda_f(v,v_0,e_0) \,,
\ee
\begin{multline}
\label{eq:Lambdaf_PNapprox}
\Lambda_f(v,v_0,e_0) = \left(1+ \left(\frac{3715}{1008}+\frac{55 }{12} \eta\right)v^2-10 \pi  v^3+ \left(\frac{15293365}{1016064}+\frac{27145 }{1008} \eta+\frac{3085
}{144}\eta ^2\right)v^4+\left(\frac{38645
}{2016}
\right. \right. \\ \left.
-\frac{65}{24}  \eta  \right)\pi \ln(v^3) v^5 + \left[\frac{12348611926451}{18776862720}-\frac{1712}{21} \gamma_{E}-\frac{160 }{3}\pi^2 +\left(-\frac{15737765635}{12192768}+\frac{2255
}{48}\pi^2 \right) \eta +\frac{76055}{6912} \eta^2-\frac{127825 }{5184}\eta^3
\right.	\\ \left.
-\frac{856}{21} \ln(16 v^2)\right]v^6+ \left(\frac{77096675}{2032128}+\frac{378515}{12096}\eta-\frac{74045}{6048}\eta^2\right)\pi v^7-\frac{785}{272} e_0^2 \left(\frac{v_0}{v}\right)^{19/3}
\left\{1+ \left(\frac{6955261}{2215584}+\frac{436441  }{79128}\eta\right)v^2
\right. \\
+ \left(\frac{2833}{1008}-\frac{197 }{36}\eta \right)v_0^2 
-\frac{1114537 }{141300}\pi  v^3
+\frac{377 }{72}\pi  v_0^3+ \left(\frac{377620541}{107433216}+\frac{561233971
}{31334688}	\eta+\frac{36339727 }{2238192}\eta ^2\right)v^4+ \left(\frac{19704254413}{2233308672}
\right. \\ \left.
-\frac{16718633 }{9970128} \eta-\frac{85978877 }{2848608}\eta
^2\right)v^2 v_0^2 +  \left(-\frac{1193251}{3048192}-\frac{66317  }{9072}\eta+\frac{18155 }{1296}\eta ^2\right)v_0^4 - 
\left(\frac{131697334 }{8456805} + \frac{268652717  }{9664920}  \eta\right)\pi v^5
\\
+ \left(-\frac{3157483321 }{142430400} +\frac{219563789 }{5086800}	\eta \right)\pi v^3 v_0^2 
+  \left(\frac{2622133397 }{159522048} +\frac{164538257  }{5697216}  \eta\right)\pi v^2 v_0^3 +  \left(\frac{764881
}{90720} -\frac{949457 }{22680}  \eta \right) \pi v_0^5
\\
+ \left[-\frac{204814565759250649}{1061268280934400}+\frac{12483797
}{791280}	\gamma_{E}
+\frac{365639621}{13022208}\pi^2+\left(\frac{34787542048195}{137827049472}
-\frac{8764775 }{1446912}\pi ^2\right) \eta
+\frac{80353703837 }{1640798208}\eta^2 
\right. \\ \left.
+\frac{5885194385 }{175799808}\eta ^3+\frac{89383841}{2373840} \ln2-\frac{26079003 }{703360}\ln3
+\frac{12483797}{1582560}\ln(16 v^2)\right]v^6+ \left(\frac{1069798992653}{108292681728}+\frac{5894683956785
}{189512193024}\eta
\right. \\ \left.
-\frac{39391912661}{752032512} \eta ^2-\frac{7158926219 }{80574912}\eta ^3\right)v^4  v_0^2
-\frac{420180449 }{10173600}\pi ^2 v^3 v_0^3+
\left(-\frac{8299372143511}{6753525424128}-\frac{6055808184535  }{241197336576}\eta
\right. \\ \left.
+\frac{3499644089 }{957132288}\eta ^2
+\frac{7923586355 }{102549888}\eta ^3\right)v^2 v_0^4
+ \left[\frac{26531900578691}{168991764480}-\frac{3317 }{126}\gamma_{E}+\frac{122833 }{10368}\pi ^2+\left(\frac{9155185261}{548674560}-\frac{3977
}{1152}\pi ^2\right) \eta 
\right. \\ \left. \left. \left.
-\frac{5732473 }{1306368}\eta ^2
-\frac{3090307 }{139968}\eta ^3
+\frac{87419 }{1890}\ln2-\frac{26001 }{560}\ln3-\frac{3317}{252} \ln(16 v_0^2)\right]v_0^6\right\}\right),
\end{multline}
\es 
\bs
\label{eq:taylort2_t}
\be
\label{eq:taylort2_t_a}
t_{c}-t=\frac{5}{256 } \frac{M}{\eta} \frac{1}{v^8} {\mathcal T}(v,v_0,e_0) \,,
\ee
\begin{multline}
\label{eq:T_PNapprox}
{\mathcal T}(v,v_0,e_0) = \left(1+ \left(\frac{743}{252}+\frac{11 }{3} \eta\right)v^2-\frac{32 }{5}\pi  v^3+ \left(\frac{3058673}{508032}+\frac{5429 }{504}\eta +\frac{617 }{72}\eta ^2\right)v^4+ \left(-\frac{7729 }{252} +\frac{13 }{3}  \eta\right)\pi v^5
\right. \\ 
+ \left[-\frac{10052469856691}{23471078400}+\frac{6848}{105}\gamma_{E}+\frac{128 }{3}\pi ^2+\left(\frac{3147553127}{3048192}-\frac{451 }{12}\pi ^2\right) \eta -\frac{15211 }{1728}\eta ^2+\frac{25565
}{1296}	\eta ^3+\frac{3424}{105} \ln(16 v^2)\right]v^6
\\+ \left(-\frac{15419335}{127008}-\frac{75703}{756}\eta+\frac{14809}{378}\eta^2\right)\pi v^7
-\frac{157}{43} e_0^2 \left(\frac{v_0}{v}\right)^{19/3} \left\{1+  \left(\frac{17592719}{5855472}+\frac{1103939  }{209124}\eta\right)v^2+\left(\frac{2833}{1008}-\frac{197}{36} \eta \right) v_0^2
\right. \\
-\frac{2819123 }{384336}\pi  v^3+\frac{377 }{72}\pi  v_0^3
+ \left(\frac{955157839}{302766336}+\frac{1419591809
}{88306848}	\eta+\frac{91918133 }{6307632}\eta ^2\right)v^4+ \left(\frac{49840172927}{5902315776}-\frac{42288307 }{26349624} \eta
\right. \\ \left.
-\frac{217475983
}{7528464}\eta ^2\right) v^2 v_0^2
+ \left(-\frac{1193251}{3048192}-\frac{66317}{9072} \eta +\frac{18155 }{1296}\eta ^2\right)v_0^4 - \left(\frac{166558393  }{12462660}+\frac{679533343  }{28486080}  \eta\right)\pi v^5+ \left(-\frac{7986575459}{387410688} 
\right. \\ \left.
+\frac{555367231}{13836096}  \eta\right)\pi v^3 v_0^2 + \left(\frac{6632455063 }{421593984} +\frac{416185003  }{15056928}  \eta\right)\pi v^2 v_0^3 + \left(\frac{764881
}{90720}-\frac{949457 }{22680} \eta \right) \pi v_0^5
\\
+ \left[-\frac{2604595243207055311}{16582316889600000}+\frac{31576663
}{2472750}\gamma_{E}+\frac{924853159 }{40694400}\pi ^2+\left(\frac{17598403624381}{86141905920}-\frac{886789}{180864}\pi ^2\right) \eta +\frac{203247603823
}{5127494400}\eta ^2
\right. \\ \left.
+\frac{2977215983 }{109874880}\eta ^3+\frac{226088539 }{7418250}\ln2-\frac{65964537 }{2198000}\ln3+\frac{31576663
}{4945500}\ln(16 v^2)\right]v^6+ \left(\frac{2705962157887}{305188466688}
\right. \\ \left.
+\frac{14910082949515
}{534079816704}\eta-\frac{99638367319 }{2119364352}\eta ^2-\frac{18107872201 }{227074752}\eta ^3\right)v^4 v_0^2 -\frac{1062809371 }{27672192}\pi ^2 v^3 v_0^3+
\left(-\frac{20992529539469}{17848602906624}
\right. \\ \left.
-\frac{15317632466765 }{637450103808} \eta+\frac{8852040931}{2529563904} \eta ^2+\frac{20042012545 }{271024704}\eta
^3\right)v^2v_0^4 +  \left[\frac{26531900578691}{168991764480}-\frac{3317 }{126}\gamma_{E}+\frac{122833 }{10368}\pi ^2
\right. \\ \left. \left. \left.
+\left(\frac{9155185261}{548674560}-\frac{3977
}{1152}\pi ^2\right) \eta -\frac{5732473 }{1306368}\eta ^2-\frac{3090307 }{139968}\eta ^3+\frac{87419 }{1890}\ln2-\frac{26001 }{560}\ln3-\frac{3317}{252} \ln(16 v_0^2)\right]v_0^6\right\}\right).
\end{multline}
\es
\subsection{TaylorT3}
The TaylorT3 approximant is obtained by performing a series reversion of $t(v)$ to obtain $v(t)$. This is then used to compute the phase as an explicit function of time, $\langle\phi\rangle(t)=\langle\phi\rangle[v=v(t)]$. This procedure is equivalent to the derivation of Eqs.~\eqref{eq:xi-tau} and \eqref{eq:lambda_tau}.  Here, we express the result in terms of the notation used in Ref.~\cite{buonanno-iyer-oshsner-pan-sathya-templatecomparison-PRD2009}. They introduce a parameter $\theta$ which is related to our $\tau$ by $\theta=\tau^{-1/8}=[\eta(t_c-t)/(5M)]^{-1/8}$. Our low-eccentricity equations for the time evolution of the fundamental gravitational-wave frequency $F\equiv \xi_{\phi}/(\pi M) = f$ and the secular piece of the orbital phase $\langle\phi\rangle$ then become 
\bs
 \label{eq:taylort3_F}
 \be
\label{eq:taylort3_F_a}
F=\frac{1}{8\pi M} \theta^3 \, \Xi(\theta,\theta_0,e_0) \,,
\ee 
\begin{multline}
\label{Xi_PNapprox}
\Xi(\theta,\theta_0,e_0) = \left(1+\left(\frac{743}{2688}+\frac{11  }{32}\eta\right) \theta ^2-\frac{3   }{10}\pi\theta ^3+\left(\frac{1855099}{14450688}+\frac{56975
	 }{258048}\eta+\frac{371 }{2048}\eta ^2\right) \theta ^4+\left(-\frac{7729  }{21504}+\frac{13    }{256}\eta\right) \pi \theta ^5
\right. \\ 
+ \left[-\frac{720817631400877}{288412611379200}+\frac{107
	}{280}\gamma_{E}+\frac{53 }{200}\pi ^2+\left(\frac{25302017977}{4161798144}-\frac{451 }{2048}\pi ^2\right) \eta -\frac{30913 }{1835008}\eta ^2+\frac{235925
	}{1769472}\eta ^3+\frac{107 }{280}\ln(2\theta)\right]\theta ^6
 \\ 
+\left(-\frac{188516689}{433520640}-\frac{97765}{258048}\eta+\frac{141769}{1290240}\eta^2\right)\pi \theta^7-\frac{471 }{344 }e_0^2 \left(\frac{\theta_0}{\theta}\right)^{19/3} \left\{1+\left(-\frac{7647061}{70265664}+\frac{209353
		 }{836496}\eta\right) \theta ^2+\left(\frac{4445}{3456}
	\right. \right.	\\ \left.
	-\frac{185  }{288}\eta\right) \theta_0^2
	-\frac{256883   }{3074688}\pi\theta ^3+\frac{61 }{2880}\pi\theta_0^3
	+\left(-\frac{1172375466061}{6586984406016}+\frac{439689491  }{8477457408}\eta+\frac{91590343
		}{1867059072}\eta ^2\right) \theta ^4+\left(\frac{1024948915}{1170505728}
\right.	\\  \left.
	-\frac{1026871
	}{6967296}\eta+\frac{14971 }{165888}\eta ^2\right) \theta_0^4
	+\left(-\frac{4855883735}{34691162112}+\frac{495545305  }{1264781952}\eta-\frac{1046765
		}{6511104}\eta ^2\right) \theta ^2 \theta_0^2+\left(\frac{49653615200993  }{96325793464320}
\right.	\\ \left.
	-\frac{449103385817    }{1146735636480}\eta\right) \pi \theta^5
	+\left(-\frac{12410299
		 }{17418240}+\frac{576391    }{1451520}\eta\right) \pi \theta_0^5+\left(-\frac{1141844935  }{10626121728}+\frac{47523355    }{885510144}\eta\right) \pi \theta
	^3 \theta_0^2
	\\
	+\left(-\frac{466470721  }{202365112320}+\frac{12770533
	}{2409108480}\eta\right)\pi  \theta ^2 \theta_0^3
	+\left(-\frac{744458420948735}{3252088301027328}+\frac{12376149277895  }{68362216538112}\eta+\frac{896664159665 }{30111928713216}\eta ^2
\right.	\\ \left.
	-\frac{457951715
	}{14532784128}\eta ^3\right) \theta^4 \theta_0^2-\frac{15669863  }{8855101440}\pi^2\theta ^3 \theta_0^3
	+\left(-\frac{7837846874888815}{82246362193723392}+\frac{76760406851419 
	}{326374453149696}\eta-\frac{60493466573 }{1295136718848}\eta ^2
\right.	\\ \left.
	+\frac{3134223763 }{138764648448}\eta^3\right) \theta ^2 \theta_0^4 
	+ \left[\frac{71857244107315089475141}{32866417392257433600000}-\frac{75936937
	}{158256000}\gamma_{E}-\frac{9863961577 }{44275507200}\pi ^2+\left(-\frac{672050112032567}{87421728522240}
\right. \right. \\ \left. \left.
+\frac{1214953 }{3858432}\pi ^2\right)
\eta 
+\frac{380720733285643 }{1129197326745600}\eta ^2-\frac{231474834959 }{8065695191040}\eta ^3-\frac{53821 }{14836500}\ln2-\frac{65964537
	}{140672000}\ln3
\right.	\\ \left.
	-\frac{75936937 }{158256000} \ln \theta \right]\theta ^6
	+ \left[-\frac{55579234653596057}{23361421521715200}+\frac{15943
	}{40320}\gamma_{E}+\frac{3968617 }{16588800}\pi ^2+\left(\frac{21736949245913}{1685528248320}-\frac{12751 }{24576}\pi ^2\right) \eta 
\right. \\ \left. \left. \left. 
-\frac{1742350567
}{4013162496}\eta ^2
+\frac{4790953 }{143327232}\eta ^3+\frac{8453 }{7560}\ln2-\frac{26001}{35840}\ln3+\frac{15943 }{40320} \ln \theta_0 \right]\theta _0^6\right\}\right),
\end{multline}
\es 
\bs
\label{eq:taylort3_phi}
\be
\label{eq:taylort3_phi_a}
\langle\phi\rangle-\phi_c=-\frac{1}{\eta  \theta ^5} \Lambda_t(\theta,\theta_0,e_0) \,,
\ee
\begin{multline}
\label{eq:Lambda_t_PNapprox}
\Lambda_t(\theta,\theta_0,e_0) = \left(1+\left(\frac{3715}{8064}+\frac{55  }{96}\eta\right) \theta ^2-\frac{3  }{4}\pi\theta ^3+\left(\frac{9275495}{14450688}+\frac{284875 
	}{258048}\eta+\frac{1855 }{2048}\eta ^2\right) \theta ^4+  \left(\frac{38645}{21504}
	 \right. \right.	\\ \left.
	 -\frac{65}{256}  \eta\right)\pi \theta^5 \ln(\theta)
	+ \left[\frac{831032450749357}{57682522275840}-\frac{107 }{56}\gamma_{E}-\frac{53
	}{40}\pi ^2+\left(-\frac{126510089885}{4161798144}+\frac{2255 }{2048}\pi ^2\right) \eta +\frac{154565 }{1835008}\eta ^2
\right. \\ \left.
 -\frac{1179625 }{1769472}\eta ^3-\frac{107}{56}\ln(2\theta )\right]\theta ^6
+\left(\frac{188516689}{173408256}+\frac{488825}{516096}\eta-\frac{141769}{516096}\eta^2\right)\pi\theta^7-\frac{7065 }{11696 } e_0^2 \left(\frac{\theta_0}{\theta}\right)^{19/3} \left\{1+\left(-\frac{130000037}{983719296}
\right.	\right. \\ \left.
+\frac{3559001
}{11710944}\eta\right) \theta ^2
	+\left(\frac{4445}{3456}-\frac{185  }{288}\eta\right) \theta_0^2-\frac{256883
		  }{2260800}\pi\theta ^3+\frac{61   }{2880}\pi\theta_0^3+\left(-\frac{19930382923037}{72456828466176}+\frac{7474721347
		 }{93252031488}\eta
\right.	\\  \left.
	+\frac{1557035831 }{20537649792}\eta ^2\right) \theta ^4
	+\left(-\frac{82550023495}{485676269568}+\frac{8424270185  }{17706947328}\eta-\frac{17795005
		}{91155456}\eta ^2\right) \theta ^2 \theta_0^2+\left(\frac{1024948915}{1170505728}-\frac{1026871  }{6967296}\eta
\right.	\\ \left.
	+\frac{14971 }{165888}\eta ^2\right)
	\theta_0^4
	+\left(\frac{49653615200993  }{53829119877120}-\frac{449103385817   }{640822855680}\eta\right)
	\pi \theta ^5+\left(-\frac{228368987  }{1562664960}+\frac{9504671    }{130222080}\eta\right)\pi \theta ^3 \theta_0^2
	\\  
+\left(-\frac{7930002257
}{2833111572480}
	+\frac{217099061   }{33727518720}\eta\right) \pi \theta ^2 \theta_0^3+\left(-\frac{12410299  }{17418240}+\frac{576391
		  }{1451520} \eta\right) \pi \theta_0^5+\left(-\frac{12655793156128495}{35772971311300608}
\right.	\\ \left. 
+\frac{210394537724215
}{751984381919232}\eta
	+\frac{15243290714305 }{331231215845376}\eta ^2-\frac{7785179155 }{159860625408}\eta ^3\right) \theta ^4 \theta_0^2-\frac{15669863
		 }{6511104000}\pi ^2\theta ^3 \theta_0^3
\\  
+\left(-\frac{133243396873109855}{1151449070712127488}+\frac{1304926916474123  }{4569242344095744}\eta
	-\frac{1028388931741
		}{18131914063872}\eta ^2+\frac{53281803971 }{1942705078272}\eta ^3\right) \theta ^2 \theta _0^4
	\\ 
+ \left[\frac{1171304704862747874258197}{262931339138059468800000}-\frac{1290927929
}{1266048000}\gamma_{E}-\frac{9863961577 }{20835532800}\pi ^2
	+\left(-\frac{11424851904553639}{699373828177920}+\frac{20654201 }{30867456}\pi ^2\right)
	\eta 
\right.	\\ \left. 
+\frac{6472252465855931 }{9033578613964800}\eta ^2-\frac{3935072194303 }{64525561528320}\eta ^3-\frac{914957 }{118692000}\ln2
	-\frac{1121397129
		}{1125376000}\ln3-\frac{1290927929  }{1266048000} \ln \theta  \right]\theta ^6
\\ 
+ \left[-\frac{55579234653596057}{23361421521715200}+\frac{15943
}{40320}\gamma_{E}+\frac{3968617 }{16588800}\pi ^2+\left(\frac{21736949245913}{1685528248320}
	-\frac{12751 }{24576}\pi ^2\right) \eta -\frac{1742350567
		}{4013162496}\eta ^2
\right. \\		\left. \left. \left.
		+\frac{4790953 }{143327232}\eta ^3+\frac{8453 }{7560}\ln2-\frac{26001 }{35840}\ln3+\frac{15943}{40320} \ln \theta_0 \right]\theta_0^6\right\}\right) \,.
\end{multline}
\es
In the above, $\theta_0$ is the dimensionless reference time defined by $e_t(\theta_0)=e_0$ (analogous to $\tau_0$ in Sec.~\ref{sec:expliciteqns}).

The TaylorT3 approximant has been shown to behave very differently from the other approximants. It displays a nonmonotonic frequency evolution (the orbital frequency starts to decrease before the ISCO is reached; see Fig.~1 of Ref.~\cite{buonanno-iyer-oshsner-pan-sathya-templatecomparison-PRD2009}). It also has worse overlaps with an EOBNR model in comparison to the other approximants \cite{buonanno-iyer-oshsner-pan-sathya-templatecomparison-PRD2009}. We have performed our own comparison by examining the frequency evolution as a function of time for TaylorT1, TaylorT3, and TaylorT4 (defined in the next section). While all three agree well at early times (large orbital separations), as the ISCO is approached, TaylorT3 deviates substantially from the other two approximants (which remain close to each other). In fact, TaylorT3 diverges before the Schwarzschild ISCO frequency is reached.  This behavior remains when the comparison is performed by truncating the series at different PN orders; it is present for small and large mass ratios. In contrast, TaylorT1 and TaylorT4 remain relatively close to each other for all PN orders and are finite at the ISCO. We also observe the nonmonotonic behavior reported in Ref.~\cite{buonanno-iyer-oshsner-pan-sathya-templatecomparison-PRD2009}: for some PN orders, TaylorT3 reaches a peak frequency and decreases before the ISCO. TaylorT1 and TaylorT4 display a monotonically increasing frequency. For these reasons we recommend that the TaylorT3 approximant should {\emph not} be used in practical applications. We include it here only for completeness. 
\subsection{TaylorT4}
In the TaylorT4 approximant, the right-hand side of Eq.~\eqref{eq:approx_timedom_v} is series expanded; the resulting system \eqref{eq:approx_timedom} is then solved numerically. Our low-eccentricity equation for $dv/dt$ is equivalent to substituting $\xi_{\phi} \rightarrow v^3$ and $\xi_{\phi,0} \rightarrow v_0^3$ in Eq.~\eqref{eq:dxidtexpand2}:
\begin{multline}
\label{eq:taylort4_dvdt}
\frac{dv}{dt}=\frac{32}{5} \frac{\eta}{M} v^9\left(1- \left(\frac{743}{336}+\frac{11 }{4}\eta \right)v^2+4 \pi  v^3+ \left(\frac{34103}{18144}+\frac{13661 }{2016}\eta+\frac{59 }{18}\eta ^2\right)v^4- \left(\frac{4159}{672}+\frac{189
}{8}  \eta \right)\pi v^5
\right. \\ \left.
+ \left[\frac{16447322263}{139708800}-\frac{1712
}{105}\gamma_{E}+\frac{16 }{3}\pi ^2+\left(-\frac{56198689}{217728}+\frac{451 }{48}\pi ^2\right) \eta +\frac{541}{896} \eta ^2-\frac{5605
}{2592}\eta ^3-\frac{856}{105} \ln(16 v^2)\right]v^6
\right.	\\ -\left(\frac{4415}{4032}-\frac{358675}{6048}\eta-\frac{91495}{1512}\eta^2\right) \pi v^7
+\frac{157}{24} e_{0}^2 \left(\frac{v_0}{v}\right)^{19/3} \left\{1- \left(\frac{41539}{22608}+\frac{5413 }{5652} \eta\right)v^2+   \left(\frac{2833}{1008}-\frac{197 }{36}\eta \right)v_0^2+\frac{24871 }{11304}\pi  v^3
\right. \\
+\frac{377 }{72}\pi  v_0^3
+ \left(-\frac{122085949}{239283072}+\frac{2133953 
}{712152}\eta-\frac{36497 }{101736}\eta ^2\right)v^4+ \left(-\frac{117679987}{22788864}+\frac{10486813 }{1424304} \eta+\frac{1066361 }{203472}\eta ^2\right)v^2 v_0^2
\\
+  \left(-\frac{1193251}{3048192}-\frac{66317 }{9072}\eta +\frac{18155 }{1296}\eta ^2\right)v_0^4 + \left(\frac{215395661
}{28486080}-\frac{8416733 }{508680}  \eta \right)\pi v^5+  \left(\frac{10065649  }{1627776}-\frac{4899587 }{406944}  \eta \right) \pi v^3 v_0^2
\\
- \left(\frac{15660203  }{1627776}+\frac{2040701 }{406944}   \eta\right)\pi v^2 v_0^3 + \left(\frac{764881  }{90720}-\frac{949457  }{22680}  \eta\right)\pi  v_0^5
+ \left[\frac{9765600648106487}{66329267558400}-\frac{2491067 }{98910}\gamma_{E}
\right.	\\
-\frac{10610699 }{1627776}\pi ^2+\left(-\frac{3409129936301}{8614190592}+\frac{2883161
}{180864}\pi ^2\right) \eta +\frac{1598264033 }{102549888}\eta ^2-\frac{2773315 }{10987488}\eta ^3+\frac{5257873 }{296730}\ln2-\frac{1534059
}{87920}\ln3
\\ \left.
-\frac{2491067 }{197820}\ln(16 v^2)\right]v^6+ \left(-\frac{345869493517}{241197336576}+\frac{96596798141
}{8614190592}\eta -\frac{9444185 }{542592}\eta ^2+\frac{7189909 }{3662496}\eta ^3\right)v^4 v_0^2+\frac{9376367 }{813888}\pi ^2 v^3 v_0^3
\\
+ \left(\frac{49566453289}{68913524736}+\frac{237857384155
}{17228381184}\eta -\frac{1281029377 }{68366592}\eta ^2-\frac{98273015 }{7324992}\eta ^3\right)v^2 v_0^4 + \left[\frac{26531900578691}{168991764480}-\frac{3317
}{126}\gamma_{E}+\frac{122833}{10368} \pi ^2
\right. \\ \left. \left. \left.
+\left(\frac{9155185261}{548674560}-\frac{3977 }{1152}\pi ^2\right) \eta -\frac{5732473 }{1306368}\eta ^2-\frac{3090307
}{139968}\eta ^3+\frac{87419 }{1890}\ln2-\frac{26001 }{560}\ln3-\frac{3317}{252} \ln(16
v_0^2)\right]v_0^6 \right\}\right).
\end{multline}

\subsection{TaylorF2 (SPA)}
Unlike the previous \emph{time-domain} approximants, TaylorF2 is a \emph{frequency-domain} approximant consisting of the phase of the Fourier transform of the GW signal evaluated in the stationary-phase approximation (SPA). Working at 0PN order and $O(e_t^0)$ in the waveform amplitude significantly simplifies the calculation as one can begin with Eqs.~\eqref{eq:hNcirc} (which contain no harmonics beyond the fundamental GW frequency). Accounting for the antenna patterns $F_{+,\times}$ of the detector allows us to write the GW signal in the form
\bs
\begin{align}
h & = F_+ h_+ + F_{\times} h_{\times}, \\
  & = A(t) \cos\Phi(t) = A(t) \cos [2\phi(t) - 2\Phi_0],
\end{align}
\es
where
\be
A(t) = -\frac{2 \eta M}{D} [v(t)]^2 \left[(1+C^2)^2 F_{+}^2 + 4 C^2 F_{\times}^2 \right]^{1/2} \, \;\;\;\;\text{and}
\ee
\be
\Phi_0 = \frac{1}{2} \arctan\left[ \frac{2 F_{\times} C}{F_+ (1+C^2)} \right] \,.
\ee
(Recall that $C=\cos \iota$ with $\iota$ the inclination angle.)

To compute the Fourier transform
\be
\tilde{h}(f) \equiv \int_{-\infty}^{\infty} h(t) e^{2\pi i f t} \, dt 
\label{eq:fourier2}
\ee
via the SPA, we first use $\cos \Phi = (e^{i\Phi} + e^{-i\Phi})/2$ to split $\tilde{h}(f)$ into two integrals,
\be
\label{eq:hfsplit}
\tilde{h}(f) = \frac{1}{2} \int_{-\infty}^{+\infty} A(t) [e^{i f \varphi_{+}(t)} + e^{i f \varphi_{-}(t)}] \, dt\;,
\ee
where $\varphi_{\pm}=2\pi t \pm \Phi(t)/f$. Here, $f$ represents the observed GW frequency (which for general orbits is {\it not} the same as twice the azimuthal orbital frequency [i.e., in this section, we initially assume $f \neq  \xi_{\phi}/(\pi M)$].
The method of stationary phase can then be used to compute integrals of the form \cite{haberman}
\be
I(f) = \int_a^b A(t) e^{i f \varphi(t)} \,dt \;.
\ee
For large values of the parameter $f$, the integrand oscillates rapidly and causes large cancellations when integrated. [Integration by parts shows that this integral scales like $I(f)=O(1/f)$ if $\dot{\varphi}\neq 0$.]

The SPA relies on the fact that the integral \eqref{eq:hfsplit} is dominated by contributions near times when the $\varphi_{\pm}$  are stationary ($\dot{\varphi}_{\pm}=0$). This is equivalent to the statement
\be
\dot{\lambda} + \dot{W} = \mp \pi f \,.
\ee 
Because $\dot{\lambda}>0$, in the quasicircular limit ($\dot{W}=0$), it can be shown that $\varphi_+$ has no real stationary points for positive $f$, while $\varphi_{-}$ has a single stationary point at $t=t_0$ given by $2\pi f=d\Phi/dt(t_0)$.  
When $\dot{W}\neq 0$, the condition $\dot{\varphi}_{+}=0$ is satisfied for $f>0$ provided that $\dot{W} < - \dot{\lambda}$. A numerical investigation indicates that this is safely satisfied for low-eccentricity waveforms ($e_t \lesssim 0.3$) but not for large eccentricity ones. Because the $\varphi_{+}$ contribution to the integral has no stationary points in the low-eccentricity limit, the rapidly varying phase results in a near perfect cancellation when that part of the integral is evaluated. For this reason, the first integral in Eq.~\eqref{eq:hfsplit} can be ignored.  However, if $\varphi(t)$ is roughly constant over some interval (as is the case for $\varphi_{-}$), then there will be less cancellation, and most of the value of the integral will come from the region near $t=t_0$ where $\dot{\varphi}(t_0)=0$. Using this fact, the second integral in \eqref{eq:hfsplit} can be evaluated by Taylor expanding the phase function $\varphi_{-}$ about its stationary point: $\varphi_{-}(t) \approx \varphi_{-}(t_0) + \frac{1}{2} \ddot{\varphi}_{-}(t_0) (t-t_0)^2 + \cdots$.

Note that, while $\varphi_{-}$ has a single stationary point $t_0$ in the quasicircular case, when eccentricity oscillations are included $\varphi_{-}$ acquires multiple stationary points (even for low values of $e_t$). Because of this, eccentric waveforms cannot be strictly treated in the SPA approximation. However, when eccentricity is small, the effect of the oscillatory terms encapsulated in $W$ is to add a rapid but small oscillation to $\varphi_{-}$. These oscillations are on a much shorter time scale than the evolution of $\varphi_{-}$ near the stationary point, so they are essentially averaged away when the integral is performed. The quadratic order Taylor expansion of $\varphi_{-}$ given above remains an accurate approximation for the small eccentricities we consider here. We will explore these issues in more detail in a future work; at the level of accuracy suitable for our present purposes, we set $W \rightarrow 0$ and proceed as in the quasicircular case.

Substituting the above expansion for $\varphi_{-}$, the Fourier transform then becomes\footnote{Note also that the function $h(t)$ must additionally satisfy the following conditions \cite{flanagancutler}:  i) its amplitude must vary slowly compared to the phase, $\dot{A}/A \ll \dot{\Phi}$, so that the amplitude can be approximated as nearly constant near the stationary point, and ii) the phase must satisfy $\ddot{\Phi} \ll \dot{\Phi}^2$, which guarantees that the phase does not vary too quickly and that a real stationary point exists. These conditions are satisfied provided the eccentricity is small and the binary is not too close to its last stable orbit. Detailed discussion of the validity of the SPA and corrections to it can be found in  Refs.~\cite{damour-iyer-sathya-usefulcyc,poissonowen-stationaryphase}.} 
\be
\tilde{h}(f) \approx \frac{1}{2} A(t_0) e^{if\varphi_{-}(t_0)} \int_{-\infty}^{+\infty} e^{\frac{if}{2} \ddot{\varphi}_{-}(t_0) (t-t_0)^2} \, dt \,,
\ee
where $f \varphi_{-}(t_0) = 2\pi f t_0-\Phi(t_0)$ and $f \ddot{\varphi}_{-}(t_0)=-\ddot{\Phi}(t_0)$. Since the orbital frequency is monotonically increasing (when ignoring orbital time scale oscillations due to $W$), $\ddot{\Phi}>0$, and the integral is convergent. The integral can be evaluated using
\be
\int_{-\infty}^{+\infty} e^{-au^2} \, du = \sqrt{\pi/a} \;\; \text{for  } a>0\;.
\ee
The Fourier transform then becomes
\be
\label{eq:hSPA}
\tilde{h}(f) \approx \frac{A(t_0(f))}{2} \sqrt{\frac{2\pi}{\ddot{\Phi}(t_0(f))}} e^{i[2\pi f t_0(f) - \Phi(t_0(f)) - \pi/4]} \;,
\ee
where we have used $i^{-1/2} = e^{-i\pi/4}$. The function $t_0(f)$ is determined by the stationarity condition $2\pi f = \dot{\Phi}(t_0)$.\footnote{If the phase has more than one stationary point, then this procedure can be repeated at each point and the resulting integrals summed.}
Using 
\begin{align}
\ddot{\Phi}[t_0(f)] &=2 \ddot{\phi} \approx 2 \ddot{\lambda} = 2 \frac{\dot{\xi}_{\phi}}{M} \approx \frac{192}{5} \frac{\eta}{M^2} (\pi M f)^{11/3}, \\
\label{eq:A_f}
A[t_0(f)] &= -2 \frac{\eta M}{D} (\pi M f)^{2/3} \left[(1+C^2)^2 F_{+}^2 + 4 C^2 F_{\times}^2 \right]^{1/2},
\end{align}
we can write the SPA as 
\be
\tilde{h}(f) = {\mathcal A} e^{i \Psi},
\ee
where
\begin{align}
{\mathcal A} &= - M \sqrt{\frac{5\pi}{96}} \left(\frac{M}{D}\right) \sqrt{\eta} (\pi M f)^{-7/6} \left[(1+C^2)^2 F_{+}^2 + 4 C^2 F_{\times}^2 \right]^{1/2}, \\
\label{eq:PsiFT}
\Psi &= 2 \pi f t_0(f) - 2 \phi[t(f)] + 2\Phi_0 - \frac{\pi}{4},
\end{align}
and $\phi \approx \lambda$ in the above equation (i.e., we are ignoring all oscillatory terms). Using Eqs.~\eqref{eq:taylort2_t} and \eqref{eq:taylort2_phi} for $t_0$ and $\phi$ and simplifying, we arrive at our final result for the SPA phase,
\begin{multline}
\label{eq:PsiFTecc}
\Psi = \psi_0 + 2 \pi f t_c +\frac{3}{128} \frac{1}{\eta v^5} \left(1+\left(\frac{3715}{756}+\frac{55  }{9} \eta\right)v^2-16 \pi  v^3+\left(\frac{15293365}{508032}+\frac{27145
}{504}\eta+\frac{3085 }{72}\eta^2\right)v^4
\right. \\
+ \left\{\left[1+\ln(v^3)\right]\left(\frac{38645}{756}-\frac{65}{9}\eta\right)\right\}\pi v^5+ \left[\frac{11583231236531}{4694215680}-\frac{6848 }{21}\gamma_{E}-\frac{640 }{3}\pi^2+\left(-\frac{15737765635}{3048192}
\right. \right. \\ \left. \left.
+\frac{2255
}{12}\pi ^2\right) \eta +\frac{76055 }{1728}\eta ^2-\frac{127825 }{1296}\eta ^3-\frac{3424}{21} \ln(16 v^2)\right]v^6+\left(\frac{77096675}{254016}+\frac{378515}{1512}\eta-\frac{74045}{756}\eta^2\right)\pi v^7
\\ 
-\frac{2355}{1462}
e_0^2 \left(\frac{v_0}{v}\right)^{19/3} \left\{1+ \left(\frac{299076223}{81976608}+\frac{18766963}{2927736} \eta\right)v^2+ \left(\frac{2833}{1008}-\frac{197
}{36}\eta\right)v_0^2 -\frac{2819123 }{282600}\pi  v^3+\frac{377 }{72}\pi  v_0^3
\right. \\
+ \left(\frac{16237683263}{3330429696}+\frac{24133060753
}{971375328}\eta+\frac{1562608261 }{69383952}\eta^2\right)v^4+ \left(\frac{847282939759}{82632420864}-\frac{718901219  }{368894736}\eta-\frac{3697091711
}{105398496}\eta^2\right) v^2 v_0^2
\\
+  \left(-\frac{1193251}{3048192}-\frac{66317 }{9072}\eta +\frac{18155 }{1296} \eta^2\right)v_0^4- \left(\frac{2831492681  }{118395270}+\frac{11552066831  }{270617760}  \eta\right)\pi v^5+ \left(-\frac{7986575459  }{284860800}
\right. \\ \left.
+\frac{555367231
}{10173600}  \eta\right) \pi v^3 v_0^2
+ \left(\frac{112751736071 }{5902315776} +\frac{7075145051  }{210796992}  \eta\right)\pi v^2 v_0^3 + \left(\frac{764881
}{90720} -\frac{949457  }{22680}  \eta\right)\pi v_0^5 
\\
+ \left[-\frac{43603153867072577087}{132658535116800000}+\frac{536803271
}{19782000}\gamma_E
+\frac{15722503703}{325555200} \pi^2+\left(\frac{299172861614477}{689135247360}-\frac{15075413 }{1446912}\pi ^2\right)
\eta 
\right. \\ \left.
+\frac{3455209264991 }{41019955200}\eta^2+\frac{50612671711 }{878999040}\eta^3+\frac{3843505163 }{59346000}\ln2
-\frac{1121397129
}{17584000}\ln3+\frac{536803271 }{39564000}\ln(16 v^2)\right]v^6
\\ 
+ \left(\frac{46001356684079}{3357073133568}+\frac{253471410141755
}{5874877983744}\eta -\frac{1693852244423 }{23313007872}\eta^2
-\frac{307833827417 }{2497822272}\eta^3\right)v^4 v_0^2 -\frac{1062809371 }{20347200}\pi^2 v^3
v_0^3
 \\
+ \left(-\frac{356873002170973}{249880440692736}-\frac{260399751935005  }{8924301453312}\eta+\frac{150484695827 }{35413894656}\eta^2
+\frac{340714213265
}{3794345856}\eta^3\right)v^2 v_0^4 + \left[\frac{26531900578691}{168991764480}
\right. \\ 
-\frac{3317}{126} \gamma_E+\frac{122833 }{10368}\pi^2+\left(\frac{9155185261}{548674560}-\frac{3977
}{1152}\pi^2\right) \eta -\frac{5732473 }{1306368}\eta^2
-\frac{3090307 }{139968}\eta^3
\\ \left. \left. \left.
+\frac{87419 }{1890}\ln2-\frac{26001 }{560}\ln3-\frac{3317}{252} \ln(16 v_0^2)\right]v_0^6 \right\}\right),
\end{multline}
where the constant $\psi_0 = 2 \Phi_0 - 2 \phi_c - \pi/4$. Recall that $t_c$ and $\phi_c$ represent the time and phase of coalescence. Along with the other PN approximants in this section, Eq.~\eqref{eq:PsiFTecc} represents one of our main results.
\section{\label{sec:ncyc}post-Newtonian contributions to the number of wave cycles}
With the above results in hand, we can now examine the significance of the various PN correction terms to the secular phasing. There are various ways of quantifying this, and here we choose to examine three easily computable quantities, applying them to fiducial sources for LIGO and eLISA.

The first quantity we compute is the number of GW cycles in the time-domain waveform as the signal sweeps from an initial frequency $f_1$ to a final frequency $f_2$,
\be
\label{eq:Ncycorb}
\Delta N_{\rm cyc} = \frac{1}{\pi} \left[ \lambda(f_2)- \lambda(f_1) \right],
\ee
where $\lambda = \langle \phi \rangle$ is given by Eqs.~\eqref{eq:taylort2_phi}.
Notice that this is simply twice the number of (azimuthal) orbits. The initial frequency $f_1$ is taken to be $10$ Hz for LIGO (an estimate of the seismic cutoff for the final design sensitivity) and one year or one month before the ISCO frequency for eLISA sources. The latter is computed via 
\be
\label{eq:finit1yr}
f_1(\delta t) = f_{\rm isco} (1+ \delta t/\tau_{\rm rr, isco})^{-3/8},
\ee
where $\delta t$ is the time before the ISCO frequency and $\tau_{\rm rr, isco} = (5/256)\eta^{-1} M^{-5/3} (\pi f_{\rm isco})^{-8/3}$. (This 0PN-order formula for the initial frequency assumes circular orbits and makes a few percent error in the initial frequency.) The reference frequency $f_0$ entering all the PN approximants is set equal to $f_1$. The termination frequency $f_2$ is taken to be $1000$ Hz for NS/NS systems (in between the tidal disruption and last-stable-orbit frequencies) and the Schwarzschild ISCO $f_{\rm isco} = (6^{3/2} \pi M)^{-1}$ for all other sources. These choices are made for simplicity and to conform with standard practice for comparing the sizes of different PN terms (see, e.g., Refs.~\cite{kidder-spineffects,faye-buonanno-luc-higherorderspinII,*faye-buonanno-luc-higherorderspinIIerratum,*faye-buonanno-luc-higherorderspinIIerratum2}). The results are shown in Tables \ref{tab:Ncyc-LIGO}, \ref{tab:Ncyc-SMBHs}, and \ref{tab:Ncyc-EMRI} for LIGO and eLISA sources, where each row shows the contribution to $\Delta N_{\rm cyc}$ from each PN order. The eccentric contributions are for an initial eccentricity $e_0=0.1$; those numbers can be scaled to other values by multiplying by $(e_0/0.1)^2$.  In computing these numbers, we made use of the following constants:\footnote{The kilometer-second relation comes from the definition $c=1=299\,792\,458$ \text{m/s} \cite{nistconstants}. The year is taken as the Julian year which is precisely $365.25$ days, with one day $=86400$ s \cite{astroconstants}. The relations involving the solar mass come from the value $G M_{\odot} = (1.32712442099 \times 10^{20} \pm 1 \times 10^{10}) \, \text{m}^3/\text{s}^2$ in Barycentric Coordinate Time (TCB) \cite{astroconstants}. This is the value used by the LIGO Algorithms Library (LAL). Using Barycentric Dynamical Time (TDB) would cause a difference in the seventh decimal place of Eq.~\eqref{eq:Msun}.}
\begin{align}
1\, \text{km} &= 3.335640952 \times 10^{-6} \, \text{s} \, ,\\
1\, \text{yr} &= 3.1557600 \times 10^{7} \, \text{s} \, , \\
\label{eq:Msun}
1\, M_{\odot} &= 4.925491025 \times 10^{-6} \, \text{s} \, , \\
& = 1.4766225061 \, \rm{km} \, .
\end{align}

The quantity $\Delta N_{\rm cyc}$ above represents the simplest approach for assessing the importance of PN phase corrections. Alternative (but more computationally difficult) methods include the computation of waveform overlaps, fitting factors, or systematic parameter estimation errors (the latter being of greatest observational significance). We leave analysis via these methods to future work. Instead, we investigate two additional methods that are comparable in simplicity to Eq.~\eqref{eq:Ncycorb}.

The first is the difference $\Delta N_{\rm cyc, \Psi}$ in the number of accumulated GW cycles computed via the SPA phase $\Psi$ of the waveform's Fourier transform $\tilde{h}(f)$. This quantity provides a better representation of the importance of PN corrections than $\Delta N_{\rm cyc}$ because it is $\Psi$ which directly enters the inner product between a GW signal and a template waveform. For example, consider a ``true'' GW signal $h_{\rm T}$ with Fourier transform of the form $\tilde{h}_{\rm T} = {\mathcal A}_0 e^{i (\Psi_0+ \delta \Psi)}$. Suppose an approximate search template $h_{\rm AP}$ has the Fourier transform  $\tilde{h}_{\rm AP} = {\mathcal A}_0 e^{i \Psi_0}$. The template differs from the true signal by an unmodeled PN phase correction $\delta \Psi$. A straightforward calculation shows that the standard noise-weighted inner product between $h_{\rm T}$ and $h_{\rm AP}$ is
\be
\label{eq:innerprod1}
\left( h_{\rm AP} | h_{\rm T} \right)  = 2 \int_0^{\infty} \frac{\tilde{h}_{\rm AP} \tilde{h}_{\rm T}^{\ast} + \tilde{h}_{\rm AP}^{\ast} \tilde{h}_{\rm T}}{S_n(f)} df =  4 \int_0^{\infty} \frac{|{\mathcal A}_0|^2}{S_n(f)} \cos \delta \Psi \, df \,, 
\ee   
where $S_n(f)$ is the detector noise spectral density.
The above expression indicates that a phase error $\delta \Psi \sim \pi$ rad will cause a near cancellation in the inner-product integral, leading to a significant loss in template overlap. It is thus errors in the SPA phase (not twice the orbital phase $2 \phi$) that more directly impact GW data analysis. One approach to quantifying these SPA phase errors that is analogous to Eq.~\eqref{eq:Ncycorb} is to define the quantity
\be
\label{eq:NcycFT}
\Delta N_{{\rm cyc}, \Psi} = \frac{1}{2\pi} \left[ \Psi(f_2)- \Psi(f_1)+(f_1-f_2)\frac{d\Psi}{df_1} \right] \,,
\ee
where $\Psi$ is given by Eq.~\eqref{eq:PsiFTecc}. The above equation is constructed by requiring that constants $\psi_0$ and $t_c$ are chosen such that $\Delta N_{{\rm cyc}, \Psi}$ and  $\frac{d\Delta N_{{\rm cyc}, \Psi}}{df}$ vanish at $f=f_1$.\footnote{The use of this particular metric for waveform errors is less common in the literature. A version of it was brought to our attention by \'{E}. Flanagan \cite{DFH,favata-phd}.} Any phase contributions from PN corrections $\Delta N_{{\rm cyc}, \Psi} \gtrsim \frac{1}{2}$ cycle will negatively impact data analysis. These phase errors are listed in the tables along with $\Delta N_{\rm cyc}$.

In Ref.~\cite{damour-iyer-sathya-usefulcyc}, an alternative measure called the number of \emph{useful} cycles $\Delta N_{\rm useful}$ was introduced.\footnote{Shortly before submitting this paper, we learned of an additional measure called the \emph{effective cycles of phase} \cite{sampson-yunes-etal-NSNSscalar-PRD2014}. This is equivalent to the waveform distinguishability criterion of Ref.~\cite{lindblom-owen-brown-waveformaccuracy-PRD2008} and is closely related to $\Delta N_{\rm useful}$.} This effectively \emph{weights} the number of cycles according to their contributions to the signal-to-noise ratio (SNR). To define this quantity, we first note that Eq.~\eqref{eq:Ncycorb} can be expressed in the form
\be
\label{eq:Ncyc-alt}
\Delta N_{\rm cyc} = \int_{f_1}^{f_2} \frac{dN_{\rm cyc}}{df} \, df \,,
\ee
where 
\be
\frac{dN_{\rm cyc}}{df} = \frac{1}{\pi}\frac{d\lambda}{df} = \frac{1}{\pi}\frac{d\lambda/dt}{df/dt} = \frac{f}{df/dt}.
\ee
The SNR $\rho$ is defined via
\be
\label{eq:SNR}
\rho^2 = \left(h | h \right) = 4 \int_0^{\infty} \frac{|\tilde{h}(f)|^2}{S_n(f)} df.
\ee
This can be expressed in terms of $dN_{\rm cyc}/df$ using the fact that $|\tilde{h}(f)| = {\mathcal A} = A[t_0(f)]/(2 \sqrt{df/dt})$ [see Eq.~\eqref{eq:hSPA} and the associated discussion]:
\be
\label{eq:SNR2}
\rho^2 = \int_0^{\infty} \left( \frac{A^2}{f S_n} \right) \frac{dN_{\rm cyc}}{df} df \,.
\ee
If we define the weight function
\be
\label{eq:weight}
{\mathcal W}(f) = \frac{A(f)^2}{f S_n(f)} \,,
\ee 
we see that the SNR $\rho$ is simply $\sqrt{\Delta N_{\rm cyc}}$ appropriately weighted by ${\mathcal W}(f)$. This leads us to define the number of \emph{useful} cycles according to
\be
\label{eq:Nuseful}
\Delta N_{\rm useful} = \left( \int_{f_1}^{f_2} {\mathcal W} \frac{dN_{\rm cyc}}{df} df  \right) \left( \int_{f_1}^{f_2} {\mathcal W} \frac{df}{f} \right)^{-1}.
\ee  
While our notation and method of presentation differ from Sec.~IIB of Ref.~\cite{damour-iyer-sathya-usefulcyc}, our Eq.~\eqref{eq:Nuseful} for $\Delta N_{\rm useful}$ is completely equivalent to their Eq.~(2.24). Here, we tried to make clear the connection between $\Delta N_{\rm cyc}$ and $\Delta N_{\rm useful}$. Note that if the weight function ${\mathcal W} \rightarrow 1$, then $\Delta N_{\rm useful} \rightarrow  \Delta N_{\rm cyc}/[\ln(f_2/{\rm Hz}) - \ln(f_1/{\rm Hz})]$. However, the definition in Eq.~\eqref{eq:Nuseful} is problematic in that it allows the number of useful cycles to vastly exceed the number of (actual) cycles. This is especially apparent for EMRI sources. Considering this, we suggest that a more appropriate definition of the useful cycles incorporates the following normalization:
\be
\label{eq:Nuseful-norm}
\Delta N_{\rm useful}^{\rm norm} = \left( \int_{f_1}^{f_2} {\mathcal W} \frac{dN_{\rm cyc}}{df} df  \right) \left( \int_{f_1}^{f_2} {\mathcal W} \frac{df}{f} \right)^{-1} \left[\ln(f_2/{\rm Hz}) - \ln(f_1/{\rm Hz})\right].
\ee  
This \emph{normalized useful cycles} definition then precisely reduces to $\Delta N_{\rm cyc}$ when the weight function ${\mathcal W} \rightarrow 1$. Note that $\Delta N_{\rm useful}^{\rm norm}$ is a factor $\sim 3 \mbox{--} 6$ larger than $\Delta N_{\rm useful}$ (for LIGO band binaries).  

Along with the previous measures, we also show $\Delta N_{\rm useful}^{\rm norm}$ in Tables \ref{tab:Ncyc-LIGO}, \ref{tab:Ncyc-SMBHs}, and \ref{tab:Ncyc-EMRI}. 
In evaluating Eq.~\eqref{eq:Nuseful-norm}, $f_{1,2}$ are chosen as before, $A = \beta f^{2/3}$ [the value of $\beta$ can be inferred from Eq.~\eqref{eq:A_f} but does not enter $\Delta N_{\rm useful}^{\rm norm}$], and the detector spectral density curves come from Eq.~(4.7) of Ref.~\cite{ajith-spin-PRD2011} for LIGO and Eq.~(1) of Ref.~\cite{klein-etal-2015} for eLISA.\footnote{Additional information on the design sensitivity curves for Advanced LIGO can be found in Ref.~\cite{aLIGOnoise-DCC}. For eLISA, we consider only the instrumental noise and do not include the galactic white dwarf foreground. We choose the configuration in Ref.~\cite{klein-etal-2015} corresponding to noise model N2A1 (optimistic acceleration noise and 1 million km arm-lengths). This corresponds to the blue-dashed curve in the right panel of their Fig.~1. Our choices of $1$ month and $1$ year for the SMBH and EMRI sources (respectively) is motivated by the time scale for which the characteristic amplitude of those sources will be above the eLISA noise (see, e.g., Fig.~13 of Ref.~\cite{eLISA-whitepaper2013}).}
From examining these tables, we can make the following observations: 
\begin{enumerate}[(a)]
\item Following standard expectations, the 0PN eccentric correction is always negative. Eccentricity therefore reduces the overall number of cycles and causes binaries to merge faster.
\item For all cases examined, the circular PN corrections are non-negligible (even at the highest PN orders known). This suggests the possibility of systematic parameter biases if only 3.5PN-order circular templates are used. This issue was examined in more detail in Ref.~\cite{favata-PRL2014}.\footnote{See also Ref.~\cite{kapadia-nathan-ajith2016PhRvD} for a recent attempt to address the issue of unknown PN corrections by fitting higher-order terms in the orbital energy and energy flux to EOBNR calculations.}
\item For LIGO band binaries, the importance of the various eccentric PN corrections depends on which measure one uses. $\Delta N_{\rm cyc}$ becomes $\ll 1$ at the level of the 2.5PN eccentric terms, while $\Delta N_{\rm cyc, \Psi}$ is not similarly small until 3PN order is reached. The same conclusion holds for SMBH binaries. However, for LIGO band binaries $\Delta N_{\rm useful}^{\rm norm}$ becomes clearly negligible for 2PN eccentric terms, while for SMBH binaries, that measure is clearly very small even for the 0PN-order eccentric term. This does not suggest that small initial eccentricities in SMBH binaries are undetectable. Rather, this is an artifact of the several-orders-of-magnitude difference between the minimum of the eLISA sensitivity curve and the value of the sensitivity at $f_1$. In other words, most of the ``useful'' cycles are accumulated near the minimum of the noise curve; the binary eccentricity has been significantly reduced before those frequencies are reached.  But this does not mean that a strong SNR source would not have observable eccentric effects. Because it does not account for the overall strength of the signal, the number of cycles (even when weighted by the detector response as in $\Delta N_{\rm useful}^{\rm norm}$) is an imperfect measure of the detectability of a particular PN effect. The various $\delta N$ provide an indication that an effect is important when $\delta N \gtrsim O(1)$, but ambiguity can be introduced when the different $\delta N$ differ significantly. Parameter estimation errors from neglecting a PN effect provide a more effective measure (although with some increase in computational complexity) \cite{favata-PRL2014}. 
\item For comparable-mass binaries and for eccentricities $e_0 \lesssim 0.1$, the eccentric PN terms indicate convergence in the sense that higher-order eccentric PN corrections are smaller than lower-order ones. This suggests that 3PN-order eccentric terms are sufficient to treat low-eccentricity, comparable-mass systems.
\item For EMRI systems, it is clear that the PN series has not converged (as is well known). This holds for both the circular and eccentric terms for all the measures indicated. We also note that the tail terms are especially large for EMRI systems.
\item The normalized useful cycles for EMRIs is often comparable to (and in a few cases larger than) the number of cycles for a given PN term. This is likely due to the fact that EMRI systems accumulate many cycles even in regions of frequency space far from the minimum of the detector noise curve. (In effect, \emph{all} the cycles are useful for these systems). We note that if we had used the standard definition [Eq.~\eqref{eq:Nuseful}], $\Delta N_{\rm useful}$ would far exceed $\Delta N_{\rm cyc}$ for the two EMRI systems involving a $10^6$ $M_{\odot}$ SMBH. 
\end{enumerate}
\begin{table}
\caption{\label{tab:Ncyc-LIGO}Post-Newtonian contributions to the number of gravitational-wave cycles $\Delta N_{\rm cyc}$, $\Delta N_{\rm cyc, \Psi}$, and $\Delta N_{\rm useful}^{\rm norm}$  for compact-object binaries in the LIGO band. The first column lists the post-Newtonian order and the type of term: ``(circ)'' refers to the quasicircular contributions at that PN order, and ``(ecc)'' refers to the leading-order eccentric PN terms that are computed here. The initial frequency $f_1$ is the seismic cutoff for Advanced LIGO (10 Hz). The final frequency $f_2$ is the ISCO frequency for systems involving BHs and $1000$ Hz for NS/NS binaries. The numbers in the ``(ecc)'' rows assume an initial eccentricity $e_0=0.1$ at $f_0=f_1$. These values can be scaled to other values of $e_0$ by multiplying by $(e_0/0.1)^2$. All numbers are rounded to at least three significant digits.}
\begin{tabular}{|l|rrr|rrr|rrr|}
	\hline 
	& \multicolumn{3}{|c|}{$1.4 M_{\odot} + 1.4 M_{\odot}$, $f_2=1000$ Hz} & \multicolumn{3}{|c|}{$1.4 M_{\odot} + 10 M_{\odot}$, $f_2=386$ Hz} & \multicolumn{3}{|c|}{$10 M_{\odot} + 10 M_{\odot}$, $f_2=220$ Hz} \\
	\hline
	PN order & $\Delta N_{\rm cyc}$ & $\Delta N_{\rm cyc,\Psi}$ & $\Delta N_{\rm useful}^{\rm norm}$ & $\Delta N_{\rm cyc}$ & $\Delta N_{\rm cyc,\Psi}$ &  $\Delta N_{\rm useful}^{\rm norm}$ & $\Delta N_{\rm cyc}$ & $\Delta N_{\rm cyc,\Psi}$ & $\Delta N_{\rm useful}^{\rm norm}$\\
	\hline
	0PN(circ) & $16031$ & $986372$ & $1821$& $3577$ & $82853$ & $492$ & $602$ & $7715$  & $101$ \\
	0PN(ecc)  &  $-463$ & $-36137$  &$-6.37$& $-103$ & $-3052$  &$-1.77$& $-17.5$ & $-286$  &$-0.385$ \\
	1PN(circ) & $439$ & $21743$  &$125$& $213$ & $4003$  &$69.0$& $59.3$ & $622$ &$21.8$ \\
	1PN(ecc) & $-15.8$ & $-1193$  &$-0.332$& $-9.00$ & $-258$  &$-0.221$& $-2.21$ & $-35.0$ &$-0.0743$ \\
	1.5PN(circ) & $-208$ & $-8520$  &$-94.8$& $-181$ & $-2877$  &$-89.3$& $-51.4$ & $-463$ &$-27.4$\\
	1.5PN(ecc) & $1.67$ & $103$  &$0.113$& $1.52$ & $35.2$  &$0.128$& $0.450$ & $5.75$ &$0.0482$ \\
	2PN(circ) & $9.54$ & $294$  &$6.70$& $9.79$ & $123$  &$7.08$& $4.06$ & $30.1$ &$3.04$ \\
	2PN(ecc) & $-0.215$ & $-15.4$  &$-0.00817$& $-0.285$ & $-7.77$  &$-0.0118$& $-0.112$ & $-1.67$ &$-0.00669$ \\
	2.5PN(circ) & $-10.6$ & $-218$  &$-10.6$& $-20.0$ & $-186$  &$-20.0$& $-7.14$ & $-41.4$ &$-7.14$ \\
	2.5PN(ecc) & $0.0443$ & $2.61$  &$0.00473$& $0.106$ & $2.34$  &$0.0130$& $0.0442$ & $0.539$ & $0.00725$ \\
	3PN(circ) & $2.02$ & $18.2$  &$2.80$& $2.30$ & $9.14$  &$3.55$& $2.18$ & $8.29$ &$2.98$ \\
	3PN(ecc) & $0.00200$ & $0.119$  &$-0.000238$& $0.0173$  & $0.412$ &$-0.0000672$& $0.00508$ & $0.0719$ & $-0.000724$ \\
	3.5PN(circ) & $-0.662$ & $-4.39$ &$-0.977$& $-1.82$ & $-7.65$  &$-2.79$& $-0.818$ & $-2.56$ &$-1.24$ \\ \hline
	Total & $15785$ & $962445$  & $1843$ & $3488$ & $80637$ & $458$ & $589$ & $7552$  & $92.6$ \\
	\hline 
\end{tabular}
\end{table}
\begin{table}
\caption{\label{tab:Ncyc-SMBHs}Same format as Table \ref{tab:Ncyc-LIGO}, except listing SMBH sources for eLISA. Here, $f_2$ is the Schwarzschild ISCO GW frequency, and $f_1$ is the GW frequency of the binary $1$ month before the ISCO.}
	\begin{tabular}{|l|rrr|rrr|rrr|}
		\hline 
		& \multicolumn{3}{|c|}{$10^6 M_{\odot} + 10^6 M_{\odot}$} & \multicolumn{3}{|c|}{$10^5 M_{\odot} + 10^6 M_{\odot}$} & \multicolumn{3}{|c|}{$10^5 M_{\odot} + 10^5 M_{\odot}$} \\
		& \multicolumn{3}{|c|}{$f_1=0.000115 \, {\rm Hz}, f_2=0.00220 \, {\rm Hz}$} & \multicolumn{3}{|c|}{$f_1=0.000252 \, {\rm Hz},f_2=0.00400 \, {\rm Hz}$} & \multicolumn{3}{|c|}{$f_1=0.000483\, {\rm Hz}, f_2=0.0220 \,  {\rm Hz}$} \\
		\hline		
	PN order & $\Delta N_{\rm cyc}$ & $\Delta N_{\rm cyc,\Psi}$ &$\Delta N_{\rm useful}^{\rm norm}$ &  $\Delta N_{\rm cyc}$ & $\Delta N_{\rm cyc,\Psi}$  & $\Delta N_{\rm useful}^{\rm norm}$ & $\Delta N_{\rm cyc}$ & $\Delta N_{\rm cyc,\Psi}$  &$\Delta N_{\rm useful}^{\rm norm}$ \\
		\hline
		0PN(circ) & $479$ & $5303$  &$27.9$& $1051$& $9461$  &$82.9$& $2030$ & $55788$ & $120$ \\
		0PN(ecc)  &  $-13.9$ & $-197$  &$-0.0128$& $-30.6$ & $-353$  &$-0.0632$& $-58.7$ & $-2052$  & $-0.104$ \\
		1PN(circ) & $51.4$ & $467$  &$10.8$& $107$ & $793$  &$26.5$& $126$ & $2796$ & $26.5$ \\
		1PN(ecc) & $-1.93$ & $-26.4$  &$-0.00498$& $-4.82$ & $-54.0$ &$-0.0212$& $-4.58$ & $-155$ &$-0.0177$ \\
		1.5PN(circ) & $-46.2$ & $-363$  &$-17.4$& $-120$ & $-778$  &$-50.2$& $-88.1$ & $-1648$ &$-33.6$\\
		1.5PN(ecc) & $0.411$ & $4.53$  &$0.00662$& $1.09$ & $9.78$  &$0.0300$& $0.731$ & $20.1$ &$0.0144$ \\
		2PN(circ) & $3.78$ & $24.6$ &$2.40$& $7.77$ & $42.1$  &$5.17$& $5.79$ & $85.1$ &$3.74$ \\
		2PN(ecc) & $-0.107$ & $-1.38$  &$-0.000948$& $-0.271$ & $-2.89$  &$-0.00313$& $-0.142$ & $-4.54$ & $-0.00153$ \\
		2.5PN(circ) & $-6.83$ & $-35.2$ &$-6.83$& $-19.8$ & $-86.6$  &$-19.8$& $-8.82$ & $-94.0$ &$-8.82$ \\
		2.5PN(ecc) & $0.0441$ & $0.465$  &$0.00190$& $0.138$ & $1.18$  &$0.00768$& $0.0443$ & $1.16$ & $0.00202$ \\
		3PN(circ) & $2.14$ & $7.44$  &$3.54$& $2.18$ & $5.27$  &$4.14$& $2.37$ & $14.0$ &$3.68$ \\
		3PN(ecc) & $0.00542$ & $0.0665$  &$-0.000713$& $0.0348$  & $0.328$  &$-0.00150$& $0.00360$ & $0.106$ &$-0.00468$ \\
		3.5PN(circ) & $-0.807$ & $-2.34$ &$-1.60$& $-2.14$ &  $-5.56$  &$-4.01$&  $-0.864$ &  $-3.95$ &$-1.52$ \\ \hline
		Total & $467$ & $5181$ & $18.8$ & $991$ & $9032$  & $44.7$ & $2004$ & $54747$  & $110$\\	
		\hline 
	\end{tabular}
\end{table}
\begin{table}
\caption{\label{tab:Ncyc-EMRI}Same as Table \ref{tab:Ncyc-SMBHs}, except listing EMRI sources for eLISA and with the initial frequency 1 yr before ISCO.}
	\begin{tabular}{|l|rrr|rrr|rrr|}
		\hline 
		& \multicolumn{3}{|c|}{$1 M_{\odot} + 10^6 M_{\odot}$} & \multicolumn{3}{|c|}{$10 M_{\odot} + 10^5 M_{\odot}$} & \multicolumn{3}{|c|}{$10 M_{\odot} + 10^6 M_{\odot}$} \\
		& \multicolumn{3}{|c|}{$f_1=0.00404 \, {\rm Hz}, f_2=0.00440 \, {\rm Hz}$} & \multicolumn{3}{|c|}{$f_1=0.00551 \, {\rm Hz},f_2=0.0440 \, {\rm Hz}$} & \multicolumn{3}{|c|}{$f_1=0.00274 \, {\rm Hz}, f_2=0.00440 \, {\rm Hz}$} \\
		\hline	
		PN order & $\Delta N_{\rm cyc}$ & $\Delta N_{\rm cyc,\Psi}$  & $\Delta N_{\rm useful}^{\rm norm}$ &  $\Delta N_{\rm cyc}$ & $\Delta N_{\rm cyc,\Psi}$ &  $\Delta N_{\rm useful}^{\rm norm}$ & $\Delta N_{\rm cyc}$ & $\Delta N_{\rm cyc,\Psi}$  & $\Delta N_{\rm useful}^{\rm norm}$ \\
		\hline
		0PN(circ) & $132844$ & $5920$ &$132474$& $270628$ & $1116876$  &$226651$& $105270$ & $33493$  & $94979$ \\
		0PN(ecc)  &  $-7975$ & $-366$ &$-7925$& $-8060$ & $-42797$  &$-5766$& $-4637$ & $-1694$ & $-3669$ \\
		1PN(circ) & $47573$ & $2100$  & $47494$ &$37604$ & $131181$  & $33805$ & $32596$ & $9860$ & $30660$\\
		1PN(ecc) & $-6867$ & $-314$  &$-6827$& $-2001$ & $-10344$  &$-1461$& $-3223$ & $-1154$ &$-2604$ \\
		1.5PN(circ) & $-108805$ & $-4781$  &$-108684$& $-56117$ & $-175279$ & $-52344$& $-69528$ & $-20481$ &  $-66760$\\
		1.5PN(ecc) & $1257$ & $51.8$ &$1265$& $568$ & $2289$ & $484$&$1045$ & $289$ & $1042$ \\
		2PN(circ) & $10491$ & $459$  &$10485$& $3662$ & $10040$  &$3541$& $6265$ & $1796$ &$6139$ \\
		2PN(ecc) & $-1875$ & $-85.0$  &$-1865$& $-167$  & $-826$  &$-126$& $-731$ & $-253$ &$-611$ \\
		2.5PN(circ) & $-50686$ & $-2206$  &$-50686$& $-12442$ & $-29361$ &$-12442$& $-28341$ & $-7899$ & $-28341$ \\
		2.5PN(ecc) & $974$ & $41.1$  &$977$& $114$ & $443$ & $98.6$& $574$ & $162$ & $566$ \\
		3PN(circ) & $4082$ & $175$  &$4090$& $251$ & $-78.3$ & $306$& $1839$ & $460$ & $1973$ \\
		3PN(ecc) & $1131$ & $47.7$  &$1136$& $46.6$ & $198$ & $38.8$& $534$ & $154$  &$517$\\
		3.5PN(circ) & $-10838$ &  $-467$  &$-10850$&  $-1481$ &  $-2464$  &$-1565$&  $-5345$ &  $-1406$ &  $-5561$ \\ \hline		
		Total & $11308$ & $576$ & $11083$ & $232605$ & $999878$  & $191219$ & $36317$& $13326$ &  $28332$\\
		\hline 
	\end{tabular}
\end{table}
\section{\label{sec:numerical_comp}Comparison with large-eccentricity formulas}
Since the primary results of this work are analytical expressions for the PN approximants in the low-eccentricity limit, it is helpful to determine the range of eccentricities for which these expressions are valid. To do this, we compare our $O(e_0^2)$ analytic expression for the orbital phase $\langle \phi(f) \rangle = \lambda(f)$ [Eq.~\eqref{eq:lambda_xi} or \eqref{eq:taylort2_phi}] against a numerical solution that is valid for arbitrary $e_0<1$. (We neglect oscillatory pieces of the phasing which were investigated in Sec.~\ref{sec:periodicsolns}.) For simplicity (and since we only wish to test the low-eccentricity approximation), we perform the comparison using only 2PN-order accurate expressions.

The numerical phase evolution for the secular phasing as a function of frequency is determined by the following system of ODEs,
\bs
\label{eq:2pnODEsys}
\be
\label{eq:2pnlambda-full-e}
\frac{d\lambda}{d\xi_{\phi}}=\frac{d\lambda/dt}{d\xi_{\phi}/dt}=\frac{1}{M}\frac{\xi_{\phi}}{d\xi_{\phi}/dt}=\frac{5 }{\eta}\xi_{\phi}^{-8/3}\left( {\Lambda}^{\rm N}+ {\Lambda}^{\rm 1PN}+ {\Lambda}^{\rm 1.5PN}+{\Lambda}^{\rm 2PN}\right) \,,
\ee
\be
\label{eq:2pnet-full-e}
\frac{de_{t}}{d\xi_{\phi}}=\frac{de_{t}/dt}{d\xi_{\phi}/dt}=\frac{e_{t}}{3\xi_{\phi}}\left( {E_{t}}^{\rm N}+{E_{t}}^{\rm 1PN}+{E_{t}}^{\rm 1.5PN}+{E_{t}}^{\rm 2PN}\right) \,,
\ee
\es
where the detailed expressions on the right-hand side are given in Appendix \ref{app:numericaleqns}. 
To derive Eqs.~\eqref{eq:2pnODEsys}, first Eq.~\eqref{eq:xiphi-xi} was differentiated with respect to time to obtain $d\xi_{\phi}/dt$ as a function of $\xi$, $e_{t}$, $\dot{\xi}$, and $\dot{e}_t$. Equations \eqref{eq:DGI-n-et} (see also Ref.~\cite{quasikepphasing35PN}) were then substituted for $\dot{\xi}$ and $\dot{e}_t$.  Lastly, $\xi$ was replaced with $\xi_{\phi}$ via Eq.~\eqref{eq:xi-xiphi} and the result was PN expanded in $\xi_{\phi}$. 

\begin{figure}[t]
\includegraphics[angle=0, width=0.45\textwidth]{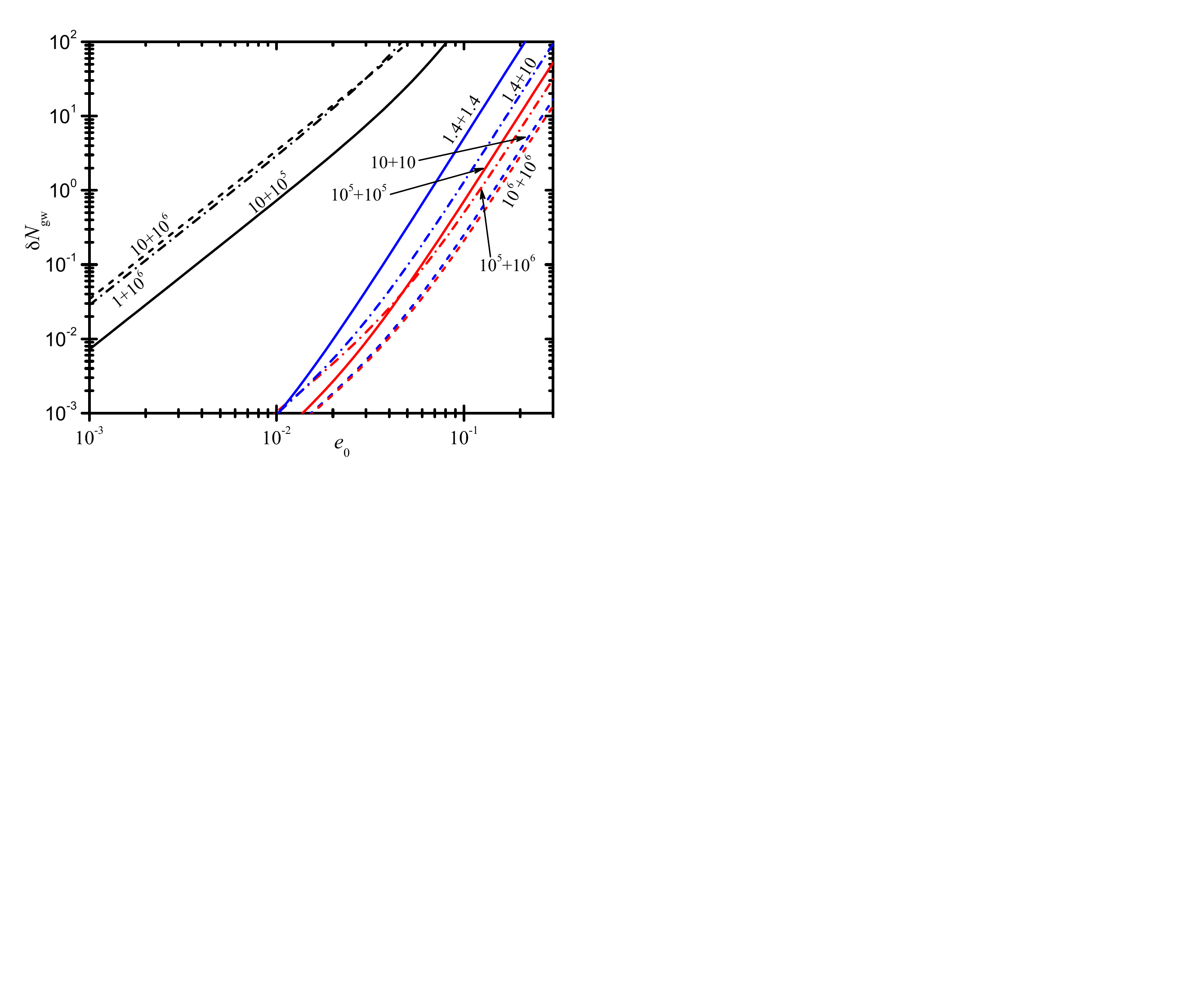}
\caption{\label{fig:large-et} Difference in the number of gravitational-wave cycles between a numerical evolution of $\lambda$ (valid for any $e_t$) and the analytical formula (valid for small $e_t$); see Eq.~\eqref{eq:deltaNgw}. Different binaries are labeled according to their masses (in solar mass units). We show three groups of sources that correspond to those studied in Tables \ref{tab:Ncyc-LIGO}, \ref{tab:Ncyc-SMBHs}, and \ref{tab:Ncyc-EMRI} (using the same frequency ranges discussed there). LIGO-band sources are shown in blue, SMBH binaries in red, and EMRIs in black. $\delta N_{\rm gw} \sim 1$ roughly represents where the phase error becomes significant and our approximation breaks down.}
\end{figure}

Using the substitutions $\xi_{\phi}\rightarrow (\pi M f)$ and $\xi_{\phi,0}\rightarrow (\pi M f_{0})$, we numerically integrated Eqs.~\eqref{eq:2pnODEsys} over a frequency interval $f\in(f_1,f_2)$ for various fiducial LIGO and eLISA sources, using $\lambda(f_{1})=0$ as our initial condition and setting $f_0=f_1$. (Appendix \ref{app:ndot-edot} discusses how we approximated the tail terms.) We then calculated the difference in the number of in-band GW cycles between our numerical solution $\lambda_{\rm num}$ and the analytical formula  $\lambda_{\rm anl}$ in Eq.~\eqref{eq:taylort2_phi},\footnote{The constant $\phi_c=c_{\lambda}$ was chosen to enforce $\lambda(f_1)=0$.}
\be
\label{eq:deltaNgw}
\delta N_{\rm gw} = \frac{1}{\pi} [\lambda_{\rm num}(f_2)-\lambda_{\rm anl}(f_2)] \,.
\ee
The results are plotted in Fig.~\ref{fig:large-et} for the same binary systems and frequency ranges considered in the previous section. Using the criterion that phase errors should satisfy $\delta N_{\rm gw} \lesssim 1$, the plot indicates that the small-eccentricity approximation is valid if $e_0 \lesssim 0.06 \mbox{--} 0.15$ for comparable-mass systems and $e_0 \lesssim 0.005 \mbox{--} 0.01$ for EMRI systems. 

In addition to the 2PN-order comparison, we have also performed the comparison using only 0PN, 1PN, and 1.5PN accurate expressions (in both the analytical and numerical parts). The resulting phase errors $\delta N_{\rm gw}$ show the expected convergence behavior for PN series as the PN order is increased. For the systems investigated above, the fractional error between the 2PN and 1.5PN calculation of $\delta N_{\rm gw}$ at the value where the 2PN $\delta N_{\rm gw}(e_0) \approx 1$ varies from $\sim 1\%$ (comparable-mass systems) to $\sim 10\%$ (EMRI systems). This gives us confidence that including 3PN corrections (for which the evaluation of the tail terms becomes difficult) will not significantly change our assessment of the range of validity in $e_0$ for our low-eccentricity formulas.

\section{Conclusions and Discussion}
Our main objective has been to reduce the complexity of fully eccentric waveforms to a level comparable to that of the standard circular PN approximants. This is achieved by working at $O(e_0^2)$ in the secular GW phasing. The resulting low-eccentricity PN approximants are found in Sec.~\ref{sec:approximants}.\footnote{The formulas in Sec.~\ref{sec:approximants} are available as a Mathematica notebook \cite{suppMathematicaNotebook}.} We have clearly illustrated how these results follow from a more general treatment, starting with fully eccentric formulas for the GW amplitude and phase (Sec.~\ref{sec:quasikep} and Appendix \ref{app:polarization}). We provided enough detail for the reader to generate arbitrarily eccentric waveforms including all effects in the quasi-Keplerian formalism \cite{DGI,quasikepphasing35PN}. We have also reexamined previous arguments \cite{DGI} concerning the realm of validity of the quasi-Keplerian formalism and argue that it should be extendable to frequencies comparable to (but less than) circular waveforms, provided the eccentricity is small (Sec.~\ref{subsec:domain}). While all of our results are expressed in terms of the eccentricity variable $e_t$ and the radial or azimuthal dimensionless frequencies $\xi$ or $\xi_{\phi}$, we show in Appendix \ref{app:gaugeinvar} how to replace $e_t$ in favor of the gauge-invariant periastron advance constant $k$. 

Although we have focused on computing the secular contributions to the GW phasing via the various PN approximants, our work also contains the necessary formulas to fully describe both secular and oscillatory contributions to the orbit and waveform in terms of analytic functions of time or frequency [working to $O(e_0^2)$; Sec.~\ref{sec:expliciteqns} and Appendix \ref{app:longeqns}]. Using these results, we have evaluated the relative importance of periodic contributions to the phasing arising from both (i) Newtonian orbital effects and (ii) perturbations from the radiation reaction force, with the latter being negligible (see Secs.~\ref{sec:periodicsolns} and \ref{sec:oscil}). The oscillatory terms in (i) do not exceed $0.07$ GW cycles for $e_0=0.1$ at $10$ Hz. While these oscillatory terms are small, they are comparable to the 2.5PN secular eccentric corrections to the phasing that we compute here. It is not entirely clear that these oscillatory terms are completely negligible, and their effect on parameter estimation will be investigated in future work. [For large eccentricities $e_0\sim O(1)$, oscillatory terms will contribute $\sim O(1)$ GW cycles.] We have also briefly discussed how these oscillatory terms can be included in the time-domain low-eccentricity PN approximants (Sec.~\ref{sec:approximants}). 

In addition to the secular approximation, our PN approximants ignore $O(e_0)$ corrections to the waveform amplitude (i.e., we treat the amplitude as circular). Our low-eccentricity approximation has been tested by comparing the orbital phase to a numerical calculation valid for any eccentricity; our analytic expressions are valid up to $e_0 \lesssim 0.1$ (the precise upper limit depends on the specific system and one's accuracy threshold).

The relative importance of the low-eccentricity PN corrections to the secular phasing is quantified in Sec.~\ref{sec:ncyc} and Tables \ref{tab:Ncyc-LIGO}, \ref{tab:Ncyc-SMBHs}, and \ref{tab:Ncyc-EMRI} for various compact binary sources. There, we investigated three measures for gauging the importance of each PN term, including an improved version of the \emph{useful cycles} contribution introduced in Ref.~\cite{damour-iyer-sathya-usefulcyc}.  Except for EMRI systems (for which the PN series is known to converge very slowly), our 3PN-order phase corrections are likely to be sufficient for any low-eccentricity binary with comparable masses. 

While compact-object binaries observable by ground-based detectors are likely to be nearly circular, binaries with a small eccentricity are more likely than those with a large one. The low-eccentricity case---despite its obvious limitations---is thus of compelling interest. Our expressions also have the advantage of being relatively simple (only slightly more complex than for circular binaries); they will thus be useful for situations where computational speed is a priority (e.g., parameter estimation). While we have focused on compact binaries relevant to LIGO and eLISA, these results may be applicable to modeling very low-frequency SMBH binaries that could be sources for pulsar timing arrays (PTAs). Eccentricity effects will also be important for third-generation ground-based detectors (such as the Einstein Telescope \cite{etweb}), where systematic errors from ignoring eccentricity can easily swamp statistical errors \cite{favata-PRL2014}. 

Additionally, many of the explicit expressions that we derive could be useful for numerical relativity (NR) or gravitational self-force (GSF) calculations. In both of those areas, it has been fruitful to compare numerical calculations with PN results (e.g., Refs.~\cite{boyle-etal-Efluxcomparison,favata-iscostudy,letiec-2014IJMPD,akcay-etal-GSFeccentricPRD2015,bini-damour-eccentricGSF-a,Akcay-vandeMeent-eccentricEOBpotential}). We provide the needed formulas to enable these comparisons (including orbital variables and waveform phasing as explicit functions of frequency or time). The conversion of our results to a set of gauge-invariant variables (Appendix \ref{app:gaugeinvar}) might be helpful in this regard.  

In the case of NR, our results could also be useful for reducing the initial eccentricity in binary merger simulations \cite{pfeiffer-etal-e-reductionCQG2007,husa-etal-e-reductionPRD2008,walther-brugmann-muller-e-reductionPRD2009,abdul-etal-eBHBH2010PRD,buonanno-etal-e-reductionPRD2011,tichy-marronetti-e-reductionPRD2011,purrer-husa-hannam-e-reductionPRD2012,zhang-szilagyi-e-reductionPRD2013,kyutoku-shibata-taniguchi-BNSe-reductionPRD2014,moldenhauer-etal-eBNS2014PRD}. For example, eccentricity removal typically involves fitting NR data to Newtonian-order expressions that depend on $r(t)$, $\phi(t)$, or their derivatives. Small eccentricities in NR simulations might be reduced more efficiently by instead fitting to some of the explicit analytic expressions provided here.

Lastly, our work could be useful for attempts to extend recent phenomenological inspiral-merger-ringdown (IMR) waveform templates \cite{ajith-etal-IMR-2007CQG,ajith-etal-BBHtemplates-PRD,ajith-etal-IMR-2011prl,santamaria-etal-phenom,Hannam:2013oca,Husa:2015iqa,Khan:2015jqa} to eccentric binaries. Our extension to the SPA/TaylorF2 approximant could be easily added on to the existing framework used to treat the inspiral portion of the analytic IMR models. This may be useful for parameter estimation of compact binaries that have both their inspiral and merger in the detector's frequency band. 
\begin{acknowledgments}
M.~F.~and B.~M.~were supported by NSF Grant No.~PHY-1308527 and the Montclair State Science Honors Innovation Program (SHIP). K.~G.~A.~was partially funded by a grant from Infosys Foundation. C.~K.~M.~acknowledges support from the Max Planck Society and India's Department of Science and Technology through a Max Planck Partner Group at the International Centre for Theoretical Sciences. For helpful discussions or comments on this manuscript, we gratefully acknowledge Luc Blanchet, Alejandro
Boh\'{e}, Bala Iyer, A. Gopakumar, and Harald Pfeiffer, and Nicol\'{a}s Yunes. We especially thank Chunglee Kim, Jeongcho Kim, and Hyung Won Lee for helping code the TaylorF2 approximant in LAL. M.~F.~also thanks \'{E}anna Flanagan for suggesting the initial calculation that led to an earlier version of this work. This manuscript is assigned LIGO DCC number P1500268.
\end{acknowledgments}
\appendix
\section{\label{app:polarization}Waveform polarizations for eccentric binaries}
Here, we provide more explicit formulas for the GW polarizations. We first compute the Newtonian-order polarizations in terms of the true anomaly for general elliptical orbits. We then express the polarizations in terms of the mean anomaly as an eccentricity expansion to $O(e_t^3)$. The latter expressions more clearly illustrate the time-domain nature of the waveform and clarify the approximations made in this work.

Using the Newtonian-order expressions in Eqs.~\eqref{eq:hN}, the $+$ and $\times$ polarizations can be written in terms of the eccentric anomaly (for arbitrary $e_t$) using the Newtonian-order equations for $(r, \dot{r}, \phi, \dot{\phi})$ given in Sec.~\ref{subsec:quasikep-Newt}. Plugging in and simplifying gives
\bs
\label{eq:hNu}
\begin{multline}
\label{eq:hplusu}
h_{+} = - \frac{\eta M}{D} \frac{\xi^{2/3}}{(1-e_t \cos u)^2} \bigg( (1+C^2) \Big\{ \big[ 2(1-e_t^2) - e_t \cos u (1-e_t \cos u) \big] \cos 2\phi
\\ + 2 e_t \sqrt{1-e_t^2} \sin u \sin 2\phi \Big\} - S^2 e_t \cos u (1-e_t \cos u) \bigg) \;,
\end{multline}
\be
\label{eq:hcrossu}
h_{\times} = - 2\frac{\eta M C}{D} \frac{\xi^{2/3}}{(1-e_t \cos u)^2} \Big\{ \big[ 2(1-e_t^2) - e_t \cos u (1-e_t \cos u) \big] \sin 2\phi + 2 e_t \sqrt{1-e_t^2} \sin u \cos 2\phi \Big\} \;.
\ee
\es
To evaluate these expressions as a function of time (neglecting radiation-reaction effects), one must numerically invert Kepler's equation to determine $u[l(t)]$. While the amplitude here is at Newtonian order, the PN order of the waveform phasing depends on the expressions used for $\phi$ and Kepler's equation [e.g., Eqs.~\eqref{eq:phi2pn1}, \eqref{eq:phi2pn3}, and \eqref{eq:keplereqn}].

In the limit of small eccentricity, Kepler's equation [Eq.~\eqref{eq:keplereqn}] can be solved as a series expansion in both $e_t$ and the PN expansion parameter $\xi$. This is done by first expanding the 3PN-order Kepler equation (Eq.~\eqref{eq:keplereqn} or Eq.~(27) of Ref.~\cite{quasikepphasing35PN}) for small $e_t$, then writing a similar expansion for the inverted series, $u=l + c_1(l) e_t + c_2(l) e_t^2 + \ldots$. The unknown coefficients $c_i(l)$ are determined by substituting one series into the other, requiring that the resulting coefficients of the appropriate powers of $e_t$ vanish. After expanding in the PN variable $\xi$, the $O(e_t^3)$ solution to the 3PN Kepler equation is 
\begin{multline}
\label{eq:lexpand}
u = l +  \Bigg[ 1 +  \Bigg( -\frac{15}{2} +\frac{9}{8} \eta + \frac{1}{8}\eta^2 \Bigg)\xi^{4/3} + \Bigg( -85 + \frac{112153}{1680} \eta + \frac{5}{4} \eta^2 + \frac{1}{24} \eta^3 \Bigg) \xi^2  \Bigg]e_t \sin l \\
+  \Bigg\{ 1 + \Bigg( - \frac{75}{4} + \frac{15}{8} \eta + \frac{3}{8} \eta^2 \Bigg) \xi^{4/3} +  \Bigg[ -\frac{465}{2} + \left(\frac{241819}{1680}  + \frac{41 }{128}\pi^2 \right) \eta + \frac{59}{4} \eta^2 - \frac{3}{8} \eta^3 \Bigg]\xi^2 \Bigg\}\frac{1}{2} e_t^2 \sin 2l\\
+  \Bigg( (3 \sin 3l - \sin l) +  \Bigg[ -(15 \sin l + 95 \sin 3l) + \frac{1}{8} (93 \sin l + 49 \sin 3l) \eta + \frac{1}{8} (-3 \sin l + 17 \sin 3l) \eta^2 \Bigg]\xi^{4/3}
\\ +  \Bigg\{ -(770 \sin l + 1250 \sin 3l) + \Bigg[ \Bigg( \frac{310967}{336}-\frac{205 }{64}\pi^2 \Bigg) \sin l + \Bigg(\frac{3106493}{5040} + \frac{533}{192} \pi^2 \Bigg) \sin 3l \Bigg] \eta
\\ + \Bigg( - \frac{775}{8} \sin l + \frac{3143}{24} \sin 3l \Bigg) \eta^2 + \Bigg( \frac{1}{2} \sin l - \frac{29}{6} \sin 3l \Bigg) \eta^3 \Bigg\}\xi^2 \Bigg)\frac{1}{8} e_t^3.
\end{multline}
Using this expansion, one can then replace $u=u(l)$ in the equations for $r$, $\dot{r}$, $\phi$, and $\dot{\phi}$ [Eqs.~\eqref{eq:quasikep2pn}] and expand to the appropriate order in $e_t$. The resulting equations to 2PN order and to $O(e_t^2)$ are
\bs
\label{eq:quasikepsmallet}
\begin{multline}
\label{eq:r2pnlowet}
r \approx M \xi^{-2/3} \Bigg( \bigg[ 1 +  \bigg( -3 + \frac{1}{3}\eta \bigg)\xi^{2/3} +  \bigg( - 8 + \frac{113}{12} \eta + \frac{1}{9} \eta^2 \bigg)\xi^{4/3} \bigg] - \bigg[ 1 +  \bigg( 1 - \frac{7}{6} \eta \bigg)\xi^{2/3} +  \bigg( \frac{7}{2} - \frac{101}{24} \eta + \frac{35}{72} \eta^2 \bigg)\xi^{4/3} \bigg]e_t \cos l 
\\
+  \Bigg\{ \frac{1}{2} (1-\cos 2l) +  (1-\cos 2l) \bigg( 1- \frac{7}{6} \eta \bigg)\frac{\xi^{2/3}}{2} +  \bigg[ -\frac{17}{2} + \frac{155}{24} \eta + \frac{11}{36} \eta^2 + \bigg( 2 + \frac{37}{24} \eta - \frac{11}{36} \eta^2 \bigg) \cos 2l \bigg]\xi^{4/3} \Bigg\}e_t^2 +O(e_t^3) + O(\xi^2) \Bigg) ,
\end{multline}
\begin{multline}
\label{eq:rdot2pnlowet}
\dot{r} \approx \xi^{1/3} e_t \sin l \Bigg\{ 1+  \bigg( 1 - \frac{7}{6}\eta \bigg)\xi^{2/3} +  \bigg( \frac{7}{2} - \frac{101}{24} \eta + \frac{35}{72} \eta^2 \bigg)\xi^{4/3}
+  \bigg[ 2 +  \bigg( 2 - \frac{7}{3} \eta \bigg)\xi^{2/3} +  \bigg( - 8 - \frac{37}{6} \eta + \frac{11}{9} \eta^2 \bigg)\xi^{4/3} \bigg]e_t \cos l \\ + O(e_t^2) + O(\xi^2) \Bigg\} ,
\end{multline}
\be
\label{eq:phi2pnlowet1}
\phi = \lambda + W(l),
\ee
\begin{multline}
\label{eq:phi2pnlowet3}
W(l) \approx  \Bigg[ 2 +  (10-\eta)\xi^{2/3} +  \bigg( 72- \frac{259}{12} \eta + \frac{1}{12} \eta^2 \bigg)\xi^{4/3} \Bigg]e_t \sin l +  \Bigg[ \frac{5}{4} +  \bigg( \frac{31}{4} - \eta \bigg)\xi^{2/3} +  \bigg( \frac{447}{8} - \frac{187}{12} \eta + \frac{1}{12}\eta^2 \bigg)\xi^{4/3} \Bigg]e_t^2 \sin 2l \\ + O(e_t^3) +O(\xi^2) ,
\end{multline}
\begin{multline}
\label{eq:phidot2pnlowet}
\dot{\phi} \approx \frac{\xi}{M}  \Bigg( 1 + 3 \xi^{2/3} +  \bigg( \frac{39}{2}-7 \eta \bigg)\xi^{4/3} +  \bigg[ 2 +  (10-\eta)\xi^{2/3} +  \bigg( 72 - \frac{259}{12} \eta + \frac{1}{12}\eta^2 \bigg)\xi^{4/3} \bigg]e_t \cos l
\\
+  \Bigg\{ 3 \xi^{2/3} +  \bigg( \frac{207}{4} - \frac{41}{2}\eta \bigg)\xi^{4/3} + \cos 2l \bigg[ \frac{5}{2} +  \bigg( \frac{31}{2} - 2\eta \bigg)\xi^{2/3} +  \bigg( \frac{447}{4} - \frac{187}{6} \eta + \frac{1}{6} \eta^2 \bigg)\xi^{4/3} \bigg]\Bigg\}e_t^2  + O(e_t^3) + O(\xi^2) \Bigg) .
\end{multline}
\es

To express the polarizations in terms of the mean anomaly, one plugs Eq.~\eqref{eq:lexpand} into \eqref{eq:hNu} and expands in $e_t$. The result is
\bs
\label{eq:hNl}
\begin{multline}
\label{eq:hplusl}
h_{+} = - \frac{\eta M \xi^{2/3}}{D} \bigg( (1+C^2) \bigg\{ \bigg[ 2+ 3 e_t \cos l +  (4 \cos 2l -1)e_t^2 +  (43 \cos 3l - 19 \cos l)\frac{e_t^3}{8} \bigg] \cos 2\phi
\\ + \bigg[ 2 e_t \sin l + 3 e_t^2 \sin 2l +  (17 \sin 3l - 7 \sin l)\frac{e_t^3}{4}  \bigg]  \sin 2\phi \bigg\} - S^2 \bigg[ e_t \cos l + e_t^2 \cos 2l +  (9 \cos 3l - \cos l)\frac{e_t^3}{8}   \bigg] + O(e_t^4) \bigg) \;,
\end{multline}
\begin{multline}
\label{eq:hcrossl}
h_{\times} = - 2\frac{\eta M \xi^{2/3} C}{D} \bigg\{ \bigg[ 2+ 3 e_t \cos l +  (4 \cos 2l -1)e_t^2 +  (43 \cos 3l - 19 \cos l)\frac{e_t^3}{8} \bigg] \sin 2\phi
\\ + \bigg[ 2 e_t \sin l + 3 e_t^2 \sin 2l +  (17\sin 3l - 7 \sin l)\frac{e_t^3}{4}  \bigg] \cos 2\phi + O(e_t^4) \bigg\} \;.
\end{multline}
\es
Equations~\eqref{eq:hNl} provide the Newtonian-order waveform as a function of $e_t$, $\xi$, $l$, and $\phi$. (Note that unlike the amplitude, the phases $l$ and $\phi$ are not yet restricted to any PN order or expansion in $e_t$.) 
Expressed in this way, the polarizations contain three types of terms proportional to expressions of the form $A_1(e_t) \cos(2\phi + \alpha_1)$, $A_2(e_t) \cos(j l + \alpha_2)$, and $A_3(e_t) \cos(2\phi \pm j l + \alpha_3)$. Here, the $A_i(e_t)$ represent amplitude terms which depend also on the orbital inclination and satisfy $A_2(e_t=0)=A_3(e_t=0)=0$. The $\alpha_i$ represent phase constants and $j=1,2,3 \ldots$. These three different terms correspond to components of the signal with respective frequencies $2(1+k) n$, $j n$, and $[2(1+k) \pm j]n$. The signal power in these different harmonics is determined by the functions $A_i(e_t)$. 

For nonevolving orbits (no radiation reaction), the explicit phasing as a function of time is determined by $l = n(t-t_0) + c_l$ and $\lambda = (1+k)n(t-t_0) +c_{\lambda}$. Note that in addition to the two \emph{intrinsic} constants of the motion $n$ and $e_t$, one must also specify two \emph{extrinsic} constants corresponding to the phase variables $c_l$ and $c_{\lambda}$. Unlike circular waveforms which contain one phase function (and an associated reference phase constant), eccentric waveforms depend on two evolving phase functions $l(t)$ and $\phi(t)$; the constants $(c_l,c_{\lambda})$ set the  values of those phases at some reference time $t_0$. These two extrinsic constants are equivalent to specifying the  argument of periastron $\varpi$ and the initial orbital phase  $\phi(t_0)$ (see Fig.~\ref{fig:ellipse}). Given the constants $[\varpi, \phi(t_0)]$, the constants $(c_l, c_{\lambda})$ are then determined by numerically solving the equations
\begin{align}
\label{eq:positionalconsts}
\varpi &= c_{\lambda} - (1+k) c_l \,, \\
\phi(t_0) &= c_{\lambda} + W(l=c_l).
\end{align}
The above follow from Eqs.~\eqref{eq:quasikeeqns} along with the fact that the argument of periastron corresponds to the angle $\phi=\varpi$ when $l=v=u=W=0$. In the Newtonian limit, these equations simplify slightly because $k \rightarrow 0$, but they still must be solved numerically or via a series expansion. In the limit of circular (but post-Newtonian) orbits, $W \rightarrow 0$ and one finds that $c_{\lambda}=\phi(t_0)$ and $c_l=[\phi(t_0)-\varpi]/(1+k)$. However, $c_l$ becomes irrelevant as any dependence on $l(t)$ drops out of the expressions \eqref{eq:hNl} in the $e_t \rightarrow 0$ limit. 

In this paper, we consider only $O(e_t^0)$ terms in the amplitude [but $O(e_t^2)$ in the phasing]. In that case, any \emph{explicit} dependence on $l(t)$ (and hence $c_l$) again drops out of the expressions for $h_{+,\times}$ in \eqref{eq:hNl}, yielding Eqs.~\eqref{eq:hNcirc}. However,  since $\phi(t) = \lambda(t) + W[l(t)]$, a dependence on $c_l$ enters the oscillatory piece of the phasing via $W(l)$ (see, e.g., Sec.~\ref{sec:oscil}). If we ignore those oscillatory contributions to the phasing (as we do when deriving the PN approximants in Sec.~\ref{sec:approximants}) then dependence on $c_l$ drops out completely, and only one phase constant (equivalent to $c_{\lambda}$) enters our expressions.

In this Appendix, we have so far neglected radiation reaction. If it is present, then $\xi$ and $e_t$ will evolve secularly according to Eqs.~\eqref{eq:DGI-n-et}, supplemented by (much smaller) periodic correction terms $\tilde{\xi}$ and $\tilde{e}_t$ that vary on multiples of the orbital time scale (Sec.~\ref{sec:periodicsolns}). The evolution of the phase variables $l(t) = \bar{l}(t) + \tilde{l}(t)$ and $\phi(t)=\bar{\lambda}(t) + \tilde{\lambda}(t) + W(t)$ is governed by the ODEs 
\begin{align}
\label{eq:secularODEs}
\frac{d\bar{l}}{dt} &= \frac{\bar{\xi}(t)}{M},\\
\frac{d\bar{\lambda}}{dt} &= \frac{(1+\bar{k}(t)) \bar{\xi}(t)}{M},
\end{align}
which must be solved numerically along with the secular evolution equations \eqref{eq:DGI-n-et} and the initial conditions $\bar{l}(t_0) = \bar{c}_l$ and $\bar{\lambda}(t_0) = \bar{c}_{\lambda}$ [these are related to $\varpi$ and $\phi(t_0)$ as above]. The numeric solutions $\bar{l}(t)$ and $\bar{\lambda}(t)$ are then combined with the semianalytic periodic components $W$, $\tilde{\lambda}$, $\tilde{l}$, $\tilde{c}_l$, and $\tilde{c}_{\lambda}$ (the latter four quantities provide relative 5PN corrections and can be ignored for our purposes). In the main text, we focus on solving Eqs.~\eqref{eq:secularODEs} analytically in terms of $\xi_{\phi}$ rather than $\xi$ and including only terms to order $O(e_t^2)$. The above description clarifies how one could additionally compute waveforms for arbitrary elliptical orbits.
\section{\label{app:ndot-edot}2PN-order secular evolution equations for $\bar{n}$ and $\bar{e}_t$}
Here, we list explicit equations for the secular (orbit-averaged) evolution equations for the mean motion $n$ and time eccentricity $e_t$. Defining $\bar{\xi} \equiv M \bar{n}$, the harmonic gauge evolution equations for $\bar{n}$ and $\bar{e}_t$ to 2PN order are
\begin{subequations}
\label{eq:DGI-n-et}
\begin{align}
\label{eq:dxibardt}
\frac{ d \bar{\xi} }{ dt } & =
\frac{\eta}{M} \bar{\xi}^{11/3}
\left(
\dot{\bar{n}}^{\rm N}
+ \dot{\bar{n}}^{\rm 1PN} + \dot{\bar{n}}^{\rm 1.5PN}
+ \dot{\bar{n}}^{\rm 2PN} 
\right) \,,
\\
\label{eq:detbardt}
\frac{ d \bar{e}_t }{ dt } & =
- \frac{\eta}{M}{\bar{\xi}}^{8/3} \, \bar{e}_t
\left(
\dot{\bar{e}}_t^{\rm N}
+ \dot{\bar{e}}_t^{\rm 1PN} + \dot{\bar{e}}_t^{\rm 1.5PN}
+ \dot{\bar{e}}_t^{\rm 2PN} 
\right) \,,
\end{align}
\end{subequations}
where the bars emphasize that these are the orbit-averaged quantities (the bars are dropped in parts of the main text where their meaning is clear). The  various contributions in \eqref{eq:DGI-n-et} to 2PN order are \cite{quasikepphasing35PN}
\begin{subequations}
\begin{align}
\dot{\bar{n}}^{\rm N} & =
\frac{96 + 292 \bar{e}_t^2 + 37 \bar{e}_t^4}{ 5 (1 - \bar{e}_t^2)^{7/2} } \,,
\\
\dot{\bar{n}}^{\rm 1PN} & =
\frac{ \bar{\xi}^{2/3} }{ 280 (1 - \bar{e}_t^2)^{9/2} }
\left[ 20368 - 14784 \eta + ( 219880 - 159600 \eta ) \bar{e}_t^2  + ( 197022 - 141708 \eta ) \bar{e}_t^4 + ( 11717 - 8288 \eta ) \bar{e}_t^6 \right] \,,
\\
\dot{\bar{n}}^{\rm 1.5PN} & = \frac{384}{5} \pi \kappa_{\rm E} \bar{\xi} \,, \\
\dot{\bar{n}}^{\rm 2PN} & =
\frac{ \bar{\xi}^{4/3} }{ 30240 (1 - \bar{e}_t^2)^{11/2} }
\bigg[
12592864 - 13677408 \eta + 1903104 \eta^2
+ ( 133049696 - 185538528 \eta + 61282032 \eta^2 ) \bar{e}_t^2  \nonumber \\
& \quad + ( 284496744 - 411892776 \eta + 166506060 \eta^2 ) \bar{e}_t^4 + ( 112598442 - 142089066 \eta + 64828848 \eta^2 ) \bar{e}_t^6
\nonumber \\ 
& \quad + ( 3523113 - 3259980 \eta + 1964256 \eta^2 ) \bar{e}_t^8 + 3024
( 96 + 4268 \bar{e}_t^2 + 4386 \bar{e}_t^4  + 175 \bar{e}_t^6 ) ( 5 - 2 \eta ) \sqrt{1 - \bar{e}_t^2}
\bigg] \,,
\\
\dot{\bar{e}}_t^{\rm N} & =
\frac{ 304 + 121 \bar{e}_t^2 }{ 15 (1 - \bar{e}_t^2)^{5/2} } \,,
\\
\dot{\bar{e}}_t^{\rm 1PN} & =
\frac{ \bar{\xi}^{2/3} }{ 2520 (1 - \bar{e}_t^2)^{7/2} }
\left[ 340968 - 228704 \eta + ( 880632 - 651252 \eta ) \bar{e}_t^2  + ( 125361 - 93184 \eta ) \bar{e}_t^4 \right]
\,,
\\
\label{eq:edottail}
\dot{\bar{e}}_t^{\rm 1.5PN} & = \frac{128}{5} \frac{\pi}{\bar{e}_t^2} \bar{\xi} \left[ (1-\bar{e}_t^2) \kappa_{\rm E} - \sqrt{1-\bar{e}_t^2} \kappa_{\rm J}  \right] \,, \\
\dot{\bar{e}}_t^{\rm 2PN} & =
\frac{ \bar{\xi}^{4/3} }{ 30240 (1 - \bar{e}_t^2)^{9/2} }
\bigg[ 20815216 - 25375248 \eta + 4548096 \eta^2 + ( 87568332  - 128909916 \eta + 48711348 \eta^2 ) \bar{e}_t^2
\nonumber \\
& \quad  + ( 69916862 - 93522570 \eta + 42810096 \eta^2 ) \bar{e}_t^4 + ( 3786543 - 4344852 \eta + 2758560 \eta^2 ) \bar{e}_t^6
\nonumber \\
& \quad + 1008 ( 2672 + 6963 \bar{e}_t^2 + 565 \bar{e}_t^4 ) ( 5 - 2 \eta ) \sqrt{1 - \bar{e}_t^2} \bigg] \,.
\end{align}
\end{subequations}

The tail contributions $\dot{\bar{n}}^{\rm 1.5PN}$ and $\dot{\bar{e}}_t^{\rm 1.5PN}$ were derived in Sec.~VI of Ref.~\cite{DGI} using the Keplerian orbital parametrization and the tail corrections to the orbit-averaged expressions for the far-zone energy and angular momentum fluxes derived in Refs.~\cite{blanchetschafertails,riethschafer}. The $\kappa_{\rm E}$ and $\kappa_{\rm J}$ appearing in the tail terms are expressed as infinite sums involving quadratic products of the Bessel function $J_p (p\,\bar{e}_t)$ and its derivative $J'_p (p\,\bar{e}_t)\equiv \frac{dJ_p(p\,\bar{e}_t)}{d(p\,\bar{e}_t)}$:
\bs
\label{eq:kappaEJ}
\begin{align}
\kappa_{\rm E} &=
\lim_{E_{\rm max} \rightarrow +\infty} \sum_{p=1}^{E_{\rm max}} \frac{p^3}{4}
\biggl \{ ( J_p (p\,\bar{e}_t))^2
\biggl [ \frac{1}{\bar{e}_t^4} - \frac{1}{\bar{e}_t^2} + \frac{1}{3}
+ p^2 \biggl ( \frac{1}{\bar{e}_t^4} - \frac{3}{\bar{e}_t^2} + 3 -\bar{e}_t^2
\biggr ) \biggr ]
\no
& \quad
+ p\,
\biggl [ -\frac{4}{\bar{e}_t^3} + \frac{7}{\bar{e}_t} -3\,\bar{e}_t \biggr ]
J_p (p\,\bar{e}_t)\,J'_p (p\,\bar{e}_t)
+ ( J'_p (p\,\bar{e}_t) )^2
\biggl [ \frac{1}{\bar{e}_t^2} - 1
+ p^2 \left (
\frac{1}{\bar{e}_t^2} -2 + \bar{e}_t^2 \right )
\biggr ]
\biggr \}\,,
\\
\kappa_{\rm J} &=
\lim_{J_{\rm max} \rightarrow +\infty} \sum_{p=1}^{J_{\rm max}} \frac{p^2}{2} \sqrt{ 1-\bar{e}_t^2}
\biggl \{
p\, \biggl [ \frac{3}{\bar{e}_t^2} -\frac{2}{\bar{e}_t^4} -1 \biggr ]
\,( J_p (p\,\bar{e}_t))^2
+ \biggl [ \frac{2}{\bar{e}_t^3} -\frac{1}{\bar{e}_t}
\no
&
\quad + 2\,p^2 \biggl ( \frac{1}{\bar{e}_t^3} -\frac{2}{\bar{e}_t} + \bar{e}_t
\biggr ) \biggr ]
 J_p (p\,\bar{e}_t)\,J'_p (p\,\bar{e}_t)
 + 2\,p \biggl ( 1 -\frac{1}{\bar{e}_t^2} \biggr )\,
( J'_p (p\,\bar{e}_t) )^2
\biggr \}\,.
\end{align}
\es
They satisfy $\kappa_{\rm E}(\bar{e}_t=0) = \kappa_{\rm J}(\bar{e}_t=0) = 1$. Expanding these functions for small $\bar{e}_t$ gives\footnote{To consistently expand $\kappa_{\rm E}$ and $\kappa_{\rm J}$ to a given order $O(\bar{e}_t^n)$ requires that the infinite sum be expanded to sufficiently high values of $p=\{E_{\rm max},J_{\rm max}\}$. For the $O(\bar{e}_t^8)$ accurate expressions shown here, $E_{\rm max}=J_{\rm max}=6$ is sufficient.}
\bs
\label{eq:kappaexpand}
\begin{align}
\kappa_{\rm E} &= 1 + \frac{2335}{192} \bar{e}_t^2 + \frac{42955}{768} \bar{e}_t^4 + \frac{6204647}{36864} \bar{e}_t^6 +\frac{352891481}{884736}\bar{e}_t^8+ O(\bar{e}_t^{10}) \;, \\
\kappa_{\rm J} &= 1 + \frac{209}{32} \bar{e}_t^2 + \frac{2415}{128} \bar{e}_t^4 + \frac{730751}{18432} \bar{e}_t^6 +\frac{10355719}{147456}\bar{e}_t^8+ O(\bar{e}_t^{10}) \;.
\end{align}
\es
Note that the expression for $\dot{\bar{e}}_t^{\rm 1.5PN}$ in \eqref{eq:edottail} is convergent in the $e_t \rightarrow 0$ limit. In practical applications, the sums in Eqs.~\eqref{eq:kappaEJ} must be evaluated to some finite number of terms $\{E_{\rm max},J_{\rm max}\}$. To maintain a specified level of accuracy for $\kappa_{\rm E, J}$, the required number of terms in the sum increases dramatically as the eccentricity approaches $1$. In Table \ref{tab:EJmax}, we indicate the values of $\{E_{\rm max},J_{\rm max}\}$ that are needed to keep $\kappa_{\rm E,J}$ accurate to within $0.1\%$. 
\begin{table}[t]
\caption{\label{tab:EJmax}Maximum values $\{E_{\rm max}, J_{\rm max}\}$ of the summation index $p$ needed to evaluate the tail corrections $\kappa_E$ and $\kappa_J$ [Eqs.~\eqref{eq:kappaEJ}] to a fractional accuracy of $0.1\%$ for a given eccentricity $\bar{e}_t$ [i.e., we show the values of $\{E_{\rm max}, J_{\rm max}\}$ that satisfy $\kappa_{\rm E,J}(E_{\rm max}, J_{\rm max})/\kappa_{\rm E,J}(E_{\rm max}=J_{\rm max}=500) -1$) $<0.001$ for a given value of $\bar{e}_t$].  Note that the number of needed terms rapidly increases for large eccentricities.}
\begin{tabular}{|r|r|r|}
\hline
$\bar{e}_t$ & $E_{\rm max}$ & $J_{\rm max}$  \\
\hline
0.1 & 4 & 4 \\
0.3 & 9 & 8 \\
0.5 & 17 & 15 \\
0.7 & 42 & 37 \\
0.9 & 239 & 212 \\
\hline
\end{tabular}
\end{table}

The time evolution of $\bar{n}$ and $\bar{e}_t$ for arbitrary  $\bar{e}_t<1$ is obtained by solving Eqs.~\eqref{eq:DGI-n-et} numerically. This evolution is ``adiabatic'' in the sense that i) the above equations become invalid when the radiation reaction time scale becomes comparable to the orbital time scale, and ii) it neglects rapidly varying contributions to $n$ and $e_t$ that average to zero on an orbital time scale. These oscillatory contributions $\tilde{n}$ and $\tilde{e}_t$ are discussed in Sec.~\ref{sec:periodicsolns}. For simplicity, we have only provided the 2PN-order secular evolution equations. The 2.5PN- and 3PN-order terms (both instantaneous and hereditary contributions) can be found in Ref.~\cite{arun-etal-eccentric-orbitalelements-PRD2009}. 
\section{\label{app:longeqns}Secular evolution for the eccentricity and mean anomaly}
Here, we collect some additional long expressions which are not essential to the main text. The full 3PN expression for the time evolution of the eccentricity $e_t(t)$ [see Eq.~\eqref{eq:et_tau}] is given by 
\begin{multline}
\label{eq:et_tau-3PN}
e_t(t)= e_0 \left(\frac{\tau }{\tau_0}\right)^{19/48} \left\{1+ \left(-\frac{4445}{6912}+\frac{185  }{576}\eta\right)\left(\tau ^{-1/4}-\tau_0^{-1/4}\right)-\frac{61  }{5760 }\pi\left(\tau ^{-3/8}-\tau_0^{-3/8}\right)+\left(\frac{854531845}{4682022912}
\right. \right. \\ \left. 
-\frac{15215083  }{27869184}\eta+\frac{72733 }{663552}\eta ^2\right)\tau^{-1/2}+\left(\frac{1081754605}{4682022912}+\frac{3702533
		 }{27869184}\eta-\frac{4283 }{663552}\eta ^2\right)\tau_0^{-1/2}+\left(-\frac{19758025}{47775744}+\frac{822325  }{1990656}\eta
\right. \\ \left. 
	-\frac{34225 }{331776}\eta
	^2\right)\tau ^{-1/4} \tau_0^{-1/4}+\left(\frac{104976437  }{278691840}-\frac{4848113   }{23224320}\eta\right)\pi \tau ^{-5/8}+\left(-\frac{101180407
		 }{278691840}+\frac{4690123   }{23224320}\eta\right)\pi \tau_0^{-5/8}
	\\ 
	+\pi \left(-\frac{54229  }{7962624}+\frac{2257    }{663552}\eta\right)\left( \tau
	^{-1/4} \tau_0^{-3/8}+\tau^{-3/8}\tau_0^{-1/4}\right)+\left(-\frac{686914174175}{4623163195392}-\frac{10094675555  }{898948399104}\eta+\frac{501067585
		}{10701766656}\eta ^2
\right.	\\ \left. 
	-\frac{792355 }{382205952}\eta ^3\right)\tau ^{-1/4} \tau_0^{-1/2}-\frac{3721 }{33177600 }\pi ^2\tau ^{-3/8}\tau_0^{-3/8}+\left(\frac{542627721575}{4623163195392}-\frac{122769222935
		 }{299649466368}\eta+\frac{2630889335 }{10701766656}\eta ^2
\right.	\\ \left.
	-\frac{13455605 }{382205952}\eta ^3\right)\tau^{-1/2} \tau_0^{-1/4}+\left[\frac{255918223951763603}{186891372173721600}-\frac{15943
		}{80640}\gamma_{E}-\frac{7926071 }{66355200}\pi ^2+\left(-\frac{81120341684927}{13484225986560}
\right. \right.	\\ \left. \left.
+\frac{12751 }{49152}\pi ^2\right) \eta
	 -\frac{3929671247
		}{32105299968}\eta ^2+\frac{25957133 }{1146617856}\eta ^3-\frac{8453}{15120}\ln2+\frac{26001 }{71680}\ln3+\frac{15943 }{645120} \ln \tau \right]\tau ^{-3/4}
	\\ 
+\left[-\frac{250085444105408603}{186891372173721600}
	+\frac{15943 }{80640}\gamma_{E}+\frac{7933513
		}{66355200}\pi ^2+\left(\frac{86796376850327}{13484225986560}-\frac{12751 }{49152}\pi ^2\right) \eta -\frac{5466199513 }{32105299968}\eta ^2
\right. \\	\left. \left.
+\frac{16786747
}{1146617856}\eta ^3
	+\frac{8453 }{15120}\ln2-\frac{26001 }{71680}\ln3-\frac{15943 }{645120} \ln \tau_0 \right]\tau_0^{-3/4}\right\} \,.
\end{multline}

The full 3PN expression for the secular frequency evolution of the phase variable $l$ [Eq.~\eqref{eq:l_xiphi}] is given by
\begin{multline}
 \label{eq:l_xiphi3PN}
 l(\xi_\phi)-c_l=-\frac{1}{32 \eta  \xi_\phi^{5/3}}\left(1+\left(-\frac{1325}{1008}+\frac{55  }{12}\eta\right) \xi_\phi ^{2/3}-10 \pi  \xi_\phi +\left(-\frac{41270555}{1016064}+\frac{20845  }{1008}\eta+\frac{3085
 		}{144}\eta ^2\right) \xi_\phi^{4/3} 	- \left(\frac{1675}{2016} 
\right. 	\right. \\  \left.
+\frac{65}{24}   \eta  \right)\pi \xi_\phi^{5/3} \ln \xi_\phi + \left[\frac{15398147061251}{18776862720}-\frac{1712 }{21}\gamma_E-\frac{160 \pi ^2}{3}+\left(-\frac{22272871555}{12192768}+\frac{6355
 		}{96}\pi ^2\right) \eta +\frac{96935 }{6912}\eta ^2
 \right.	\\ \left. 
 	-\frac{127825 }{5184}\eta ^3-\frac{856}{21} \ln\left(16 \xi_\phi ^{2/3}\right)\right]\xi_\phi ^2-\frac{785}{272}
 		e_0^2 \left(\frac{\xi_{\phi,0}}{\xi_\phi}\right)^{19/9} \left\{1+ \left(\frac{117997}{2215584}+\frac{436441  }{79128}\eta\right) \xi_\phi ^{2/3}+\left(\frac{2833}{1008}-\frac{197
 			 }{36}\eta\right) \xi_{\phi,0}^{2/3}
 \right.		\\ 
 		-\frac{1114537  }{141300}\pi  \xi_\phi+\frac{377 }{72}\pi  \xi_{\phi,0}+\left(-\frac{732350735}{68366592}+\frac{271164331  }{31334688}\eta+\frac{36339727 }{2238192}\eta ^2\right) \xi_\phi ^{4/3}+\left(\frac{334285501}{2233308672}+\frac{151648993
 			 }{9970128}\eta
 \right.	\\ \left. 
 		-\frac{85978877 }{2848608}\eta ^2\right) \xi_\phi ^{2/3} \xi_{\phi,0}^{2/3}+\left(-\frac{1193251}{3048192}-\frac{66317  }{9072}\eta+\frac{18155
 			}{1296}\eta ^2\right) \xi_{\phi,0}^{4/3}+  \left(\frac{270050729}{33827220}-\frac{268652717}{9664920}\eta\right)  \pi
 		\xi_\phi^{5/3}
 		\\ 
 		+\left(-\frac{3157483321  }{142430400}+\frac{219563789    }{5086800}\eta\right)\pi \xi_\phi  \xi_{\phi,0}^{2/3}+\left(\frac{44484869 
 		}{159522048}+\frac{164538257    }{5697216}\eta\right) \pi \xi_\phi ^{2/3} \xi_{\phi,0}+  \left(\frac{764881  }{90720}-\frac{949457}{22680}\eta\right) 
 		\pi \xi_{\phi,0}^{5/3}
 		\\ 
 		+\left(-\frac{2074749632255}{68913524736}+\frac{15718279597553  }{189512193024}\eta-\frac{1296099941
 			}{752032512}\eta ^2-\frac{7158926219 }{80574912}\eta ^3\right) \xi_\phi ^{4/3} \xi_{\phi,0}^{2/3}-\frac{420180449  }{10173600}\pi ^2\xi_\phi  \xi_{\phi,0}
 		\\ 
 		+\left(-\frac{140800038247}{6753525424128}-\frac{614686144279
 			 }{241197336576}\eta-\frac{37877198551 }{957132288}\eta ^2+\frac{7923586355 }{102549888}\eta ^3\right) \xi_\phi ^{2/3} \xi_{\phi,0}^{4/3}+ \left[-\frac{231385908692247049}{1061268280934400}
 	\right.	\\ 
 		+\frac{12483797}{791280}\gamma_{E}+\frac{365639621 }{13022208}\pi ^2+\left(\frac{43054867314787}{137827049472}-\frac{14711579 }{1446912}\pi ^2\right) \eta
 		+\frac{55988213933 }{1640798208}\eta ^2+\frac{5885194385 }{175799808}\eta ^3
 		\\ \left.
 		+\frac{89383841}{2373840}\ln 2 -\frac{26079003 }{703360}\ln 3 +\frac{12483797}{1582560} \ln\left(16 \xi_{\phi}^{2/3}\right)\right]\xi_\phi ^2+ \left[\frac{26531900578691}{168991764480}-\frac{3317
 			}{126}\gamma_{E}+\frac{122833 }{10368}\pi ^2
\right. 		\\	\left. \left. \left.
 			+\left(\frac{9155185261}{548674560}-\frac{3977 }{1152}\pi ^2\right) \eta -\frac{5732473 }{1306368}\eta ^2-\frac{3090307
 			}{139968}\eta ^3+\frac{87419 }{1890}\ln 2 -\frac{26001 }{560}\ln 3-\frac{3317}{252} \ln\left(16\xi_{\phi,0}^{2/3}\right)\right]\xi_{\phi,0}^2\right\}\right) .
 \end{multline}
Using the same techniques as in deriving Eq.~\eqref{eq:lambda_tau}, the time evolution of $l(t)$ to 3PN order is 
\begin{multline}
\label{eq:l_tau3PN}
l(t)-c_l= -\frac{1}{\eta }\tau ^{5/8} \left(1+\left(-\frac{6365}{8064}+\frac{55  }{96}\eta\right)\tau ^{-1/4}-\frac{3  }{4
}\pi\tau ^{-3/8}+\left(-\frac{36010585}{14450688}+\frac{294955  }{258048}\eta+\frac{1855 }{2048}\eta ^2\right)\tau^{-1/2}
 \right. \\ 
+\left(\frac{1675}{172032} +\frac{65}{2048}\eta\right) \pi \tau ^{-5/8} \ln \tau 
+\left[\frac{999649208535757}{57682522275840}-\frac{107}{56}\gamma_{E}-\frac{53
			}{40}\pi ^2-\left(\frac{159394981085  }{4161798144}-\frac{1435   }{1024}\pi ^2\right)\eta
	\right.	\\ \left.
	+\frac{1403375 }{5505024}\eta ^2
		-\frac{1179625 }{1769472}\eta ^3-\frac{107
			}{56} \ln\left(2 \tau^{-1/8} \right) \right]\tau^{-3/4}-\frac{7065}{11696} e_0^2 \left(\frac{\tau }{\tau_0}\right)^{19/24} \left\{1+\left(-\frac{968824373}{983719296}+\frac{3559001 
			}{11710944}\eta\right)\tau ^{-1/4}
		\right.	\\ 
		+\left(\frac{4445}{3456}
			-\frac{185  }{288}\eta\right)\tau_0^{-1/4}-\frac{256883 
		}{2260800}\pi \tau ^{-3/8}+\frac{61  }{2880}\pi \tau_0^{-3/8}+\left(-\frac{3948564655159}{6586984406016}+\frac{1084152188431
			 }{3450325165056}\eta
	\right.	\\ \left.
		+\frac{1557035831 }{20537649792}\eta ^2\right)\tau^{-1/2} 
		+\left(\frac{1024948915}{1170505728}-\frac{1026871  }{6967296}\eta+\frac{14971}{165888}
		\eta ^2\right)\tau_0^{-1/2}+\left(-\frac{615203476855}{485676269568}+\frac{9061588285  }{8853473664}\eta
	\right.	\\ \left.
		-\frac{17795005 }{91155456}\eta ^2\right)\tau
		^{-1/4} \tau_0^{-1/4}
		+\left(\frac{56295003356801
		 }{53829119877120}-\frac{449103385817    }{640822855680}\eta\right)\pi \tau ^{-5/8}+\left(-\frac{12410299 }{17418240}+\frac{576391    }{1451520}\eta\right)\pi \tau_0^{-5/8}
	\\ 
+\left(-\frac{59098286753  }{2833111572480}
	+\frac{217099061   }{33727518720} \eta\right)\pi \tau ^{-1/4} \tau_0^{-3/8}+\left(-\frac{228368987
		 }{1562664960}+\frac{9504671   }{130222080}\eta\right)\pi \tau^{-3/8} \tau_0^{-1/4}
	\\ 
+\left(-\frac{992995489931905295}{1151449070712127488}
	+\frac{1879169839629227  }{4569242344095744}\eta-\frac{2423726613325
		}{18131914063872}\eta ^2+\frac{53281803971 }{1942705078272}\eta ^3\right)\tau ^{-1/4} \tau_0^{-1/2}
	\\ 
	-\frac{15669863 }{6511104000}\pi ^2 \tau ^{-3/8}
	\tau_0^{-3/8}
	+\left(-\frac{2507338556025965}{3252088301027328}+\frac{21958237212008275  }{27823422131011584}\eta-\frac{34557938365535 }{331231215845376}\eta
	^2
\right.	\\ \left.
	-\frac{7785179155 }{159860625408}\eta ^3\right)\tau^{-1/2} \tau_0^{-1/4} 
	+\left[\frac{1414005101903318632008197}{262931339138059468800000}-\frac{1290927929}{1266048000}\gamma_{E}-\frac{9863961577 }{20835532800}\pi^2
\right.	\\ \left.
+\left(-\frac{326342634216461209}{21680588673515520}+\frac{18186821 }{30867456}\pi ^2\right)
\eta 
	+\frac{7463874079487531 }{9033578613964800}\eta ^2-\frac{3935072194303 }{64525561528320}\eta ^3
	\right.	\\  \left.
		-\frac{914957 }{118692000}\ln 2 -\frac{1121397129
		}{1125376000}\ln 3 +\frac{1290927929
	}{10128384000}\ln \tau \right]\tau ^{-3/4}
		+\left[-\frac{55579234653596057}{23361421521715200}
\right.	\\ 
+\frac{15943
}{40320}\gamma_{E}+\frac{3968617 }{16588800}\pi ^2+\left(\frac{21736949245913}{1685528248320}-\frac{12751 }{24576}\pi ^2\right) \eta -\frac{1742350567
}{4013162496}\eta ^2
	+\frac{4790953 }{143327232}\eta ^3
	\\ \left. \left. \left.
	+\frac{8453}{7560}\ln 2 -\frac{26001 }{35840}\ln 3 -\frac{15943 }{322560}\ln \tau_0\right]\tau_0^{-3/4}\right\}\right) \,.
\end{multline}
\section{\label{app:numericaleqns}Frequency evolution of $e_t$ and $\lambda$}
Here, we will list the full expressions for $d\lambda/d\xi_{\phi}$ and $de_t/d\xi_{\phi}$ that were numerically solved in Sec.~\ref{sec:numerical_comp}. The equations for the frequency derivative of $\lambda$ and $e_t$ have the form
\bs
\label{eq:2pnODEsysapp}
\begin{align}
\label{eq:2pnlambda-full-eapp}
\frac{d\lambda}{d\xi_{\phi}} &= \frac{d\lambda/dt}{d\xi_{\phi}/dt}=\frac{1}{M}\frac{\xi_{\phi}}{d\xi_{\phi}/dt}=\frac{5}{\eta}\xi_{\phi}^{-8/3}\left({\Lambda}^{\rm N}+{\Lambda}^{\rm 1PN}+{\Lambda}^{\rm 1.5PN}+{\Lambda}^{\rm 2PN}\right),
\\
\label{eq:2pnet-full-eapp}
\frac{de_{t}}{d\xi_{\phi}} &= \frac{de_{t}/dt}{d\xi_{\phi}/dt}= \frac{e_{t}}{3\xi_{\phi}}\left({E_{t}}^{\rm N}+{E_{t}}^{\rm 1PN}+{E_{t}}^{\rm 1.5PN}+{E_{t}}^{\rm 2PN}\right).
\end{align}
\es
(Recall that we have dropped overbars; the quantities $\lambda$, $\xi_{\phi}$, and $e_t$ here contain only secular pieces.)
Section \ref{sec:numerical_comp} outlines how Eqs.~\eqref{eq:2pnODEsysapp} were derived. The various PN contributions to these equations are
\bs
\label{eq:PNparts_fulle}
\begin{equation}
{\Lambda}^{\rm N}= \frac{(1-e_{t}^2)^{7/2}}{96+292e_{t}^2+37e_{t}^4},
\end{equation}
\begin{multline}
{\Lambda}^{\rm 1PN}=\frac{(1-e_{t}^2)^{5/2}}{56(96+292e_{t}^2+37e_{t}^4)^2} \xi_{\phi}^{2/3}\left[16(743+924\eta)+(-731+1330\eta)120e_{t}^2
\right.	\\ \left.
+(-12217+10122\eta)14e_{t}^4+(-11717+8288\eta)e_{t}^6\right],
\end{multline}
\begin{equation}
{\Lambda}^{\rm 1.5PN}=-\frac{384(1-e_{t}^2)^7 \pi \kappa_{\rm E} \xi_{\phi}}{(96+292e_{t}^2+37e_{t}^4)^2},
\end{equation}
\begin{multline}
{\Lambda}^{\rm 2PN}=\frac{\left(1-e_{t}^2\right)^{3/2}}{ \left(96+292 e_{t}^2+37 e_{t}^4\right)^3} \xi_{\phi}^{4/3} \left\{\frac{22395332}{441}+\frac{282944  }{7}\eta+39488 \eta
^2+\left(\frac{1092911608}{1323}-\frac{241508}{3}\eta+\frac{1320560}{3}\eta^2\right)e_t^2
\right.	\\ 
+ \left(\frac{2482558957}{441}-\frac{20856653  }{3}\eta+\frac{34607570 }{9}\eta
^2\right)e_t^4+ \left(\frac{11491137626}{1323}-\frac{94648088 
}{7}\eta+5059075 \eta ^2\right)e_t^6
\\ 
+
\left(\frac{41598328573}{7056}-\frac{827463337 
}{84}\eta+\frac{220841935 }{72}\eta ^2\right)e_t^8+ \left(\frac{12550222}{21}-\frac{353621581  }{336}\eta+\frac{1545527 }{6}\eta
^2\right)e_t^{10}
\\  
+ \left(\frac{209089541}{9408}-\frac{2351387 
}{56}\eta+\frac{88985 }{9}\eta ^2\right)e_t^{12}+\sqrt{1-e_t^2}\left[-23040+9216 \eta+ ( -1094400+437760 \eta)e_t^2
\right. \\  \left. \left.
+ ( -4177160+1670864 \eta)e_t^4 +( -3638570+1455428
\eta) e_t^6 +( -533455+213382 \eta )e_t^8 + \left( -\frac{32375}{2}+6475 \eta\right)e_t^{10}\right]\right\} ,
\end{multline}
\begin{equation}
{E_{t}}^{\rm N}=-\frac{(1-e_{t}^2)(304+121e_{t}^2)}{96+292e_{t}^2+37e_{t}^4},
\end{equation}
\begin{multline}
{E_{t}}^{\rm 1PN}=-\frac{(1-e_{t}^2)\xi_{\phi}^{2/3}}{84(96+292e_{t}^2+37e_{t}^4)^2} \left[-768\left(-2833+5516\eta \right)+ \left(-1384161+821716\eta \right)8e_{t}^2
\right. \\ \left.
+ \left(4607952-3626672 \eta \right)e_{t}^4+ \left(-192543+219632\eta \right)e_{t}^6 \right],
\end{multline}
\begin{equation}
{E_{t}}^{\rm 1.5PN}=\frac{384 \pi (1-e_{t}^2)^{7/2} \xi_{\phi}}{e_{t}^2(96+292e_{t}^2+37e_{t}^4)^2}\left[-12(1-e_{t}^2)^2 (8+7e_{t}^2) \kappa_{\rm E}+\sqrt{1-e_{t}^2}(96+292e_{t}^2+37e_{t}^4)\kappa_{\rm J} \right],
\end{equation}
\begin{multline}
{E_{t}}^{\rm 2PN}=\frac{\xi_{\phi}^{4/3}}{ \left(e_{t}^2-1\right) \left(96+292 e_{t}^2+37 e_{t}^4\right)^3}\left\{-\frac{27447809152}{441}+\frac{252718080 
}{7}\eta-1705984 \eta ^2+
\left(-\frac{280235917148}{1323}
\right. \right. \\ \left. 
+201957536 \eta-\frac{263680960 }{3}\eta ^2\right)e_t^2+ \left(\frac{311523561058}{1323}-\frac{8973484400  }{21}\eta+\frac{1174947296 }{9}\eta
^2\right)e_t^4+
\left(-\frac{97410004618}{1323}
\right. \\ \left. 
+\frac{1205297000 
}{7}\eta-\frac{737557936 }{9}\eta ^2\right)e_t^6+ \left(\frac{123041288107}{2646}+\frac{2729307556  }{21}\eta-\frac{74924956 }{9}\eta
^2\right)e_t^8+
\left(\frac{183339693991}{2352}
\right. \\ \left.
-\frac{879377631 
}{7}\eta+\frac{482513675 }{9}\eta ^2\right)e_t^{10}+ \left(-\frac{23619720463}{2016}+\frac{519652379  }{42}\eta-\frac{38871571 }{9}\eta
^2\right)e_t^{12}+ \left(-\frac{2428416731}{4704}
\right. \\ \left. 
+\frac{4886701 
}{14}\eta+\frac{290228 }{9}\eta ^2\right)e_t^{14}+\sqrt{1-e_t^2}\left[54558720-21823488 \eta+ (199449600-79779840 \eta )e_t^2+ (203690880
\right. \\ 
-81476352 \eta)e_t^4+ (219866160-87946464 \eta )e_t^6+(-86681310+34672524 \eta )e_t^8 
\\ \left. \left.
+ \left(-\frac{30233775}{2}+6046755 \eta \right)e_t^{10}+ (-24975+9990 \eta )e_t^{12}\right]\right\} .
\end{multline}
\es
\section{\label{app:gaugeinvar}Gauge-invariant parametrization for eccentric waveforms}
In this work we have presented our results in terms of a ``frequencylike'' variable ($\xi$ or $\xi_{\phi}$) and a particular ``eccentricitylike'' parameter $e_t$. The latter reduces to the Newtonian definition of eccentricity in the Newtonian limit and allows us to easily read off the circular 3PN limit of our expressions when $e_t \rightarrow 0$. However, the choice of $e_t$ is somewhat arbitrary: we could have chosen to use the other eccentricity variables $e_r$ or $e_{\phi}$ or a definition based on the angular frequencies at the pericenter and apocenter \cite{mora-will-PRD2002,mora-will-PRD2004,*mora-will-PRD2004-erratum}.  All of these eccentricity parameters are gauge dependent. For example, expressions for $e_t$, $e_r$, or $e_{\phi}$ in terms of the conserved energy per reduced mass $E$ and reduced orbital angular momentum $h$ depend on the choice of gauge [see Eqs.~(345) of Ref.~\cite{blanchet-LRR2014} and the subsequent discussion there]. However, the quantities $n$ and $k$ have expressions in terms of $E$ and $h$ that do not depend on the coordinate system; these quantities are gauge invariant. Likewise, the related quantities $\xi$, $\xi_{\phi}$, $K=k+1$, as well as $E$ and $h$ themselves, are also gauge invariants. 

To obtain a gauge-invariant form for the expressions derived in this paper, it is necessary to eliminate $e_t$ and to express our results in terms of $k$ and either $\xi$ or $\xi_{\phi}$. The relation $e_t=e_t(k,\xi)$ can be derived from Eqs.~(25d) and (28) of Ref.~\cite{quasikep3PN}.\footnote{Note that the $k'$ in Ref.~\cite{quasikep3PN} is related to $k$ via $k'=k/3$. Also, Ref.~\cite{quasikep3PN} uses the notation $x$ to mean $\xi^{2/3}$.} These provide $e_t^2(E,h)$ along with $E(\xi,k/\xi^{2/3})$ and $h(\xi,k/\xi^{2/3})$. Substituting for $E$ and $h$ into $e_t$ (and using the fact that $\xi \sim c^{-3}$ and $k \sim c^{-2}$) gives the 3PN accurate result:
\begin{multline}
\label{eq:et_k_xi}
e_t^2 = 1- \frac{3 \xi^{2/3}}{k}  +  \left[ -\frac{43}{4} +\frac{9}{2}\eta + \left( \frac{51}{4}-\frac{13}{2}\eta \right) \left(\frac{\xi^{2/3}}{k} \right) \right] \xi^{2/3}
+  \left\{ \frac{649}{16} + \left( -\frac{157}{4} + \frac{41 }{128}\pi^2 \right) \eta + 9 \eta^2  + \left( -\frac{39}{2} + \frac{55}{4} \eta
\right. \right. \\ \left. \left.
 - \frac{65}{8} \eta^2 \right) \left( \frac{\xi^{2/3}}{k} \right)  + \sqrt{3} (10-4 \eta) \sqrt{\frac{\xi^{2/3}}{k}} + \sqrt{3} (-5 + 2\eta) \sqrt{\frac{k}{\xi^{2/3}}} +  \left[  -\frac{187}{48} +  \left( \frac{557}{36} -  \frac{205}{384}\pi^2 \right) \eta+ \frac{5}{8} \eta^2 \right] \left( \frac{k}{\xi^{2/3}} \right)  \right\} \xi^{4/3} 
 \\ 
+  \left\{ -\frac{1199}{16} + \left( 90 -\frac{123 }{128}\pi^2 \right) \eta + \left( -\frac{3121}{72} +\frac{205 }{384}\pi^2 \right) \eta^2 + \frac{31}{3} \eta^3  + \left( \frac{35}{4} - \frac{23}{16} \eta + \frac{25}{4} \eta^2 - \frac{1565}{216} \eta^3  \right) \left( \frac{\xi^{2/3}}{k}  \right) 
\right. \\
 + \sqrt{3} \left( -\frac{105}{4} + \frac{73}{3} \eta - \frac{25}{3} \eta^2 \right) \sqrt{\frac{\xi^{2/3}}{k}}  + \sqrt{3} \left[\frac{1055}{24} + \left( -\frac{1394}{27} + \frac{41}{144}\pi^2  \right) \eta  + \frac{55}{6} \eta^2 \right] \sqrt{\frac{k}{\xi^{2/3}}} + \left[ \frac{209}{48} + \left( - \frac{35857}{5040}
\right. \right. \\ \left. \left.
  + \frac{41 }{128}\pi^2  \right) \eta + \left( \frac{3509}{216} - \frac{1025 }{1152}\pi^2  \right) \eta^2 
   + \frac{25}{24}\eta^3 \right] \left( \frac{k}{\xi^{2/3}} \right) + \sqrt{3} \left[ -\frac{35}{8} + \left( \frac{301}{27} - \frac{41}{288}\pi^2 \right) \eta -\frac{5}{6}\eta^2 \right] \left( \frac{k}{\xi^{2/3}} \right)^{3/2}
\\\left.
+ \left[ -\frac{2}{3} + \left( -\frac{14081}{1890} + \frac{41}{72}\pi^2 \right) \eta -\frac{17}{9} \eta^2 \right] \left( \frac{k}{\xi^{2/3}} \right)^2
 \right\}\xi^2.
\end{multline}
Note that terms of order $O(\xi^{2/3}/k)$ are formally Newtonian order (since $k \sim \xi^{2/3}$). 

To derive $e_t = e_t(k,\xi_{\phi})$, we use series reversion on Eq.~\eqref{eq:k3pnxiphi}, assuming a result of the form \eqref{eq:et_k_xi} with $\xi \rightarrow \xi_{\phi}$. However, because the expansion \eqref{eq:k3pnxiphi} contains only relative 2PN corrections beyond its leading-order term, we can only derive relative 2PN corrections to $e_t^2$ if we use $k$ as our starting point.\footnote{Note that the leading-order term in $k$ represents a 1PN effect in the equations of motion. In the derivation of Eq.~\eqref{eq:et_k_xi}, we started from relative 3PN accurate expressions for $e_t(E,h)$, $E(\xi,k)$, and a relative 2PN accurate expression for $h(\xi,k)$ (which is sufficient as it enters the relevant expressions only in terms of the combination $E h^2$).} Specifying a solution with the form
\be
\label{eq:etsq_xiphiexpand}
e_t^2 = 1 - \frac{3 \xi_{\phi}^{2/3}}{k} +  \left[ A + B \left( \frac{\xi_{\phi}^{2/3}}{k} \right) \right]\xi_{\phi}^{2/3} +  \left[ C + D \left( \frac{\xi_{\phi}^{2/3}}{k} \right) +E \sqrt{\frac{\xi_{\phi}^{2/3}}{k} } +F \sqrt{ \frac{k}{\xi_{\phi}^{2/3}} } + G \left( \frac{k}{\xi_{\phi}^{2/3}} \right)  \right]\xi_{\phi}^{4/3} + O(\xi_{\phi}^2), 
\ee
the coefficients $A$ through $G$ are determined by plugging Eq.~\eqref{eq:etsq_xiphiexpand} into Eq.~\eqref{eq:k3pnxiphi}. The result is series expanded.\footnote{This is done by taking $\xi_{\phi} \rightarrow \epsilon^3 \xi_{\phi}$ and $k \rightarrow \epsilon^2 k$ and then series expanding in the small parameter $\epsilon$ (which is then set to $1$ at the end of the calculation). We note that this procedure was also used to check the derivation of \eqref{eq:et_k_xi} at the 2PN level starting from Eq.~\eqref{eq:k3pn}.} Coefficients of the appropriate powers of $\xi_{\phi}$ are then equated to zero, resulting in a system of equations which is solved for the coefficients $A$ through $F$. The result is
\begin{multline}
\label{eq:et_k_xi-phi}
e_t^2 = 1- \frac{3 \xi_{\phi}^{2/3}}{k}  +  \left[ -\frac{35}{4} +\frac{9}{2}\eta + \left( \frac{51}{4}-\frac{13}{2}\eta \right) \left(\frac{\xi_{\phi}^{2/3}}{k} \right) \right] \xi_{\phi}^{2/3}
+  \left\{ \frac{377}{16} + \left( -\frac{367}{12} + \frac{41 }{128}\pi^2 \right) \eta + 9 \eta^2  
\right. \\
+ \left( -\frac{39}{2} + \frac{55}{4} \eta - \frac{65}{8} \eta^2 \right) \left( \frac{\xi_{\phi}^{2/3}}{k} \right)  + \sqrt{3} (10-4 \eta) \sqrt{\frac{\xi_{\phi}^{2/3}}{k}} + \sqrt{3} (-5 + 2\eta) \sqrt{\frac{k}{\xi_{\phi}^{2/3}}}  \\
\left. +  \left[  \frac{77}{48} +  \left( \frac{449}{36} -  \frac{205}{384}\pi^2 \right) \eta + \frac{5}{8} \eta^2 \right] \left( \frac{k}{\xi_{\phi}^{2/3}} \right)  \right\}\xi_{\phi}^{4/3}  + O(\xi_{\phi}^2).
\end{multline}

Using the results above, one could eliminate $e_t$ from our expressions in favor of the gauge-invariant variable $k$. For example, our final expressions for the secular PN approximants depend on an eccentricity $e_0$ at a reference frequency $\xi_{\phi,0}=(\pi M f_0)$. This constant could be replaced with a value $k_0$ corresponding to the periastron advance rate at the reference frequency. However, we feel that it is more sensible to parametrize our waveforms in terms of quantities that have simple physical interpretations in the Newtonian limit. 
\bibliography{maintext_PRDsubmitV3}
\end{document}